\documentclass[twocolumn,prd,superscriptaddress,preprintnumbers,nofootinbib]{revtex4}[11pt]
\pdfoutput=1

\usepackage{amsmath,amssymb,graphicx, color,verbatim}
\usepackage{rotating}
\usepackage{multirow}
\usepackage{lscape}
\usepackage{latexsym}
\usepackage{cancel}
\usepackage{enumerate} 
\usepackage[normalem]{ulem} 
\usepackage {slashed}
\usepackage{epsfig}
\usepackage{url}
\usepackage{hyperref} 
\usepackage{booktabs}
\usepackage{pstricks} 
\usepackage{fancyvrb}
\usepackage{amsthm}
\usepackage{bm}
\usepackage{epstopdf}
\usepackage{wrapfig}
\usepackage{sidecap}
\usepackage{bbold}
\usepackage{fancyhdr}
\usepackage{comment}
\usepackage{array}
\usepackage{titlesec} 
\usepackage{subcaption}
\graphicspath{{figs/}}

\newcommand{\nc}{\newcommand}

\nc{\beq}{\begin{equation}}
\nc{\eeq}{\end{equation}}
\nc{\beqa}{\begin{eqnarray}}
\nc{\eeqa}{\end{eqnarray}}
\nc{\bea}{\begin{eqnarray}}
\nc{\eea}{\end{eqnarray}}
\nc{\ra}{\rightarrow}
\nc{\Tr}{{\rm Tr}}
\nc{\slsh}{\slash\hspace*{-0.22cm}}

\def\be{\begin{equation}}
\def\ee{\end{equation}}
\def\bea{\begin{eqnarray}}
\def\eea{\end{eqnarray}}
\def\bit{\begin{itemize}}
\def\eit{\end{itemize}}

\nc{\barray}{\begin{eqnarray}}
\nc{\earray}{\end{eqnarray}}
\nc{\barrayn}{\begin{eqnarray*}}
\nc{\earrayn}{\end{eqnarray*}}
\nc{\mc}{\mathcal}
\nc{\M}{\mathcal{M}}

\newcommand{\gev}{{\ \rm GeV}}
\newcommand{\tev}{{\ \rm TeV}}

\nc{\h}{$h$}
\nc{\infinity}{\infty}

\def\ben{\begin{enumerate}}
\def\een{\end{enumerate}}

\newcommand{\fref}[1]{Fig.~\ref{f.#1}}
\newcommand{\eref}[1]{Eq.~(\ref{e.#1})}

\newcommand{\aref}[1]{Appendix~\ref{a.#1}}
\newcommand{\sref}[1]{Section~\ref{s.#1}}
\newcommand{\ssref}[1]{Section~\ref{ss.#1}}
\newcommand{\sssref}[1]{Section~\ref{sss.#1}}
\newcommand{\cref}[1]{Chapter~\ref{c.#1}}
\newcommand{\tref}[1]{Table~\ref{t.#1}}

\def\ifb{{\ \rm fb}^{-1}}

\newcommand{\iab}{\,{\rm ab}^{-1}}

\def\to{\rightarrow}

\makeatletter
\def\p@subsection{}
\def\p@subsubsection{}
\makeatother

\addtocounter{page}{2}

\begin{document}

\title{
Towards a No-Lose Theorem for Naturalness
}

\author{David Curtin}
\thanks{dcurtin1@umd.edu}
\affiliation{Maryland Center for Fundamental Physics, University of Maryland, College Park, MD 20742}

\author{Prashant Saraswat}
\thanks{saraswat@umd.edu}
\affiliation{Maryland Center for Fundamental Physics, University of Maryland, College Park, MD 20742}
\affiliation{Department of Physics and Astronomy, Johns Hopkins
  University, Baltimore, MD 21218, USA}

\preprint{UMD-PP-015-013}

\begin{abstract}

We derive a phenomenological no-lose theorem for naturalness up to the TeV scale, which applies when quantum corrections to the Higgs mass from top quarks are canceled by perturbative BSM particles (top partners) of similar multiplicity due to to some symmetry. Null results from LHC searches already seem to disfavor such partners if they are colored. Any partners with SM charges and $\sim$ TeV masses will be exhaustively probed by the LHC and a future 100 TeV collider. Therefore, we focus on neutral top partners. While these arise in Twin Higgs theories, we analyze neutral top partners as model-independently as possible using EFT and Simplified Model methods. We classify  all  perturbative neutral top partner structures in order to compute their irreducible low-energy signatures at proposed future lepton and hadron colliders, as well as the irreducible tunings suffered in each scenario. Central to our theorem is the assumption that SM-charged BSM states appear in the UV completion of neutral naturalness, which is the case in all known examples. Direct production at the 100 TeV collider then allows this scale to be probed at the $\sim$10 TeV level. We find that proposed future colliders probe any such scenario of naturalness with tuning of 10\% or better. This provides very strong model-independent motivation for both new lepton and hadron colliders, which in tandem act as discovery machines for general naturalness. We put our results in context by discussing other possibilities for naturalness, including  ``swarms'' of top partners, inherently non-perturbative or exotic physics, or theories without SM-charged states in the UV completion. Realizing a concrete scenario which avoids our arguments while still lacking experimental signatures remains an open model-building challenge.

\end{abstract}

\maketitle

 \setcounter{equation}{0} \setcounter{footnote}{0}

\section{Introduction}
\label{s.intro}

The discovery of the 125 GeV Higgs boson~\cite{Aad:2012tfa,Chatrchyan:2012ufa} has thrown the hierarchy problem~\cite{Weisskopf:1939zz}  of the Standard Model (SM) into sharp relief. Naturalness suggests that new physics should appear near the TeV scale to stabilize the Higgs mass against UV-sensitive quadratic corrections from top quarks (and, more generally, all SM particles). However, this new physics has so far refused to reveal itself at the LHC. 

There are many natural models which solve the hierarchy problem. Certainly, the LHC has not yet excluded them all, nor is it likely to. However, there are several exciting proposals for future colliders on the horizon. This includes 100 TeV machines like the SPPC~\cite{Tang:2015qga} or FCC-hh~\cite{FCChh}, as well as lepton colliders that can make exquisite precision measurements~\cite{Burdman:2014zta, Dawson:2013bba}. Given the possibility of  such a bright future for high energy experiments, it behooves us to ask: \emph{How can we test the basic hypothesis of naturalness, rather than a particular theory implementation?}

We make progress in tackling this question by focusing on a large class of solutions to the hierarchy problem: theories in which the Higgs is stabilized via some symmetry which guarantees the algebraic cancellation of top loops against loops of some perturbative Beyond-SM (BSM) particles in a 4D effective field theory. 
The most well-known examples are supersymmetry (SUSY)~\cite{Martin:1997ns} and some types of composite Higgs models (CH) (see~\cite{Bellazzini:2014yua} for a classification), including Little Higgs \cite{ArkaniHamed:2001nc, ArkaniHamed:2002qx, ArkaniHamed:2002qy, Schmaltz:2004de}. Slightly more exotic possibilities are models of uncolored naturalness like Folded SUSY~\cite{Burdman:2006tz}, Quirky Little Higgs~\cite{Cai:2008au}, and the Twin Higgs (TH)~\cite{Chacko:2005pe}. 

Given how ubiquitous top partners are in theories of naturalness, it is worthwhile to ask whether \emph{any} theory with partners could be probed, either at the LHC or at proposed future colliders. A \emph{phenomenological no-lose theorem for top partner theories} would go a long way towards addressing the discoverability of naturalness as a general hypothesis. It is our aim to derive such a theorem in this work.

We can classify top partner theories by the gauge quantum numbers of the partners. In SUSY or CH, the top partners carry SM QCD charge. This makes the LHC very well-suited to produce them. However, LHC run~1 searches have so far come up empty~\cite{Chatrchyan:2013xna,CMS:2014wsa,Aad:2014bva,Aad:2014kra}. 
This is not to say that naturalness with colored top partners is excluded. 
Slightly non-minimal models~\cite{Fan:2011yu}, squeezed spectra and kinematic blind-spots~\cite{Martin:2007gf, Martin:2008sv,LeCompte:2011fh,Belanger:2012mk,Rolbiecki:2012gn,Curtin:2014zua,Kim:2014eva,Czakon:2014fka,CMS:2014exa, Rolbiecki:2015lsa, An:2015uwa}, or new states being `just around the corner', can explain the current absence of signal. This leaves open the possibility of discovery at the upcoming 14 TeV LHC run. Looking even further ahead, it is highly unlikely that any colored (or even EW charged~\cite{Low:2014cba, Alves:2014cda}) BSM state with mass around a TeV will escape detection at a 100 TeV collider. We can therefore be confident that naturalness with colored top partners will be discovered.

Theories of \emph{uncolored naturalness} offer motivation for  considering top partners \emph{without} QCD charge.  In these models, the symmetry which protects the Higgs does not commute with color, leading to mirror sectors containing top partners which may carry only electroweak (EW) charge, such as Folded SUSY~\cite{Burdman:2006tz} or Quirky Little Higgs~\cite{Cai:2008au}, or can even be SM singlets, as is the case for Twin Higgs theories~\cite{Chacko:2005pe}. 
These models have undergone a recent revival of interest, which led to group theoretical generalizations of their protection mechanism~\cite{Craig:2014aea, Craig:2014roa}, detailed explorations of electroweak symmetry breaking~\cite{Cohen:2015gaa},  the development of several UV completions necessary at scales of $\sim 5-10 \tev$~\cite{
Batra:2008jy,Barbieri:2015lqa, Low:2015nqa, Geller:2014kta,
Craig:2013fga, Craig:2014fka, Chang:2006ra}, and investigations of their connections to dark matter~\cite{Garcia:2015loa,Craig:2015xla, Garcia:2015toa,Farina:2015uea}, cosmology~\cite{Schwaller:2015tja}, and the neutrino sector~\cite{Batell:2015aha}.

Uncolored top partners lead to phenomenology which is radically different from the usual expectations of naturalness. 
Without colored production of new states around $\mathcal{O}(1 \tev)$, these scenarios can be more difficult to discover than conventional SUSY or CH. 
There are however other avenues for discovery. These have been explored for the various models described above, sometimes leading to exciting, highly exotic signatures.

The LHC has great discovery potential for \emph{electroweak-charged mirror sectors}. In the Folded SUSY and Quirky Little Higgs models, the mirror sector contains its own copy of a confining QCD-like force. Since LEP limits forbid mirror matter with masses below $\sim 100 \gev$, the lightest states of the mirror QCD sector are mirror glueballs~\cite{Morningstar:1999rf, Juknevich:2009ji, Juknevich:2009gg}, which couple to the SM-like Higgs through a top partner loop. This interaction both allows the SM Higgs to decay to mirror glueballs with an appreciable branching ratio, and allows the produced glueballs to decay back in the SM with macroscopic lifetimes $\sim \mu $m to km~\cite{Craig:2015pha}. RG arguments place the glueball mass in the $\sim 10-60 \gev$ window, which allows TeV-scale top partner masses to be probed via the striking signature of exotic Higgs decays with displaced vertices~\cite{Curtin:2015fna, Csaki:2015fba}.

\begin{turnpage} 

\begin{table*}
\vspace*{-5mm}
\begin{center}
\hspace*{-13mm}
\begin{tabular}{|p{2.5cm} ||c|c||c||c||c||c|}
  \hline 
 \begin{tabular}{l}
\emph{Higgs} \\
\emph{coupling}\\
\emph{structure}
\end{tabular} & \multicolumn{2}{c||}{
\begin{tabular}{l}
\emph{Neutral top partner structure}\\
\end{tabular}} &
\begin{tabular}{l}
\emph{Theory}\\
\emph{realization}
\end{tabular} &
\begin{tabular}{l}
\emph{Irreducible low-}\\
\emph{energy probes}
\end{tabular} &
\begin{tabular}{l}
\emph{Irreducible Tunings}
\end{tabular} &
\begin{tabular}{l}
\emph{Section}
\end{tabular} \\
  \hline \hline
  \multirow{4}{*}{
\begin{tabular}{cc}
\\ \\ \\ \\ \\ \\  \\ \\ \\ \\ \\ 
\hspace{-3mm} \includegraphics[width=2.8cm]{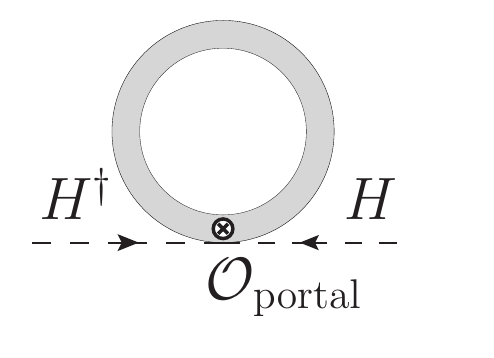} \\
\end{tabular} }
& \multicolumn{2}{c||}{
\begin{tabular}{cc}
\\
Scalar partners ($\phi$) \vspace{1mm}\\  \\ 
$\mathcal{L} \supset \lambda H^\dagger H \phi \phi$ \\ \\ 
\includegraphics[width=3.3cm]{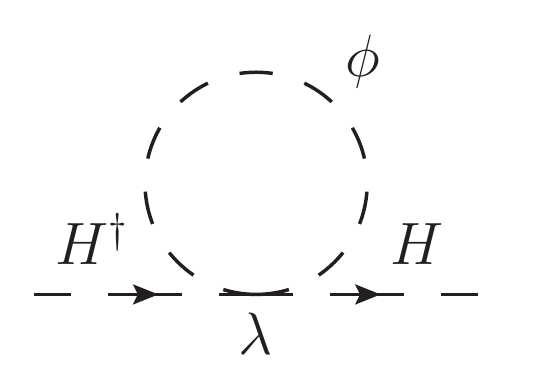} \\ \\ 
\end{tabular}
 } 
&
\begin{tabular}{l}
---
\end{tabular}
&
\begin{tabular}{l}
$\delta \sigma_{Zh}$\\ $\delta \sigma_{hh}$\\ $\sigma_\mathrm{direct}$
\end{tabular}
&
\begin{tabular}{ll}
\\
$\Delta_{h(\phi)}$: &log tuning from incomplete $t-\phi$ cancellation \\
$\Delta_{\phi(h)}$: & quadratic divergence of partner mass from \\ &  Higgs loops\\ \\
\end{tabular}
&
\begin{tabular}{c}
\ref{ss.scalartoppartners}
\end{tabular}
\\ \cline{2-7}
  & \multirow{3}{*}{
\begin{tabular}{cc}
Fermion partners ($T$)\\
(partial UV \\
completion \\
required) \\ \\ \\  \\ 
$\mathcal{L} \supset - \frac{1}{M'} H^\dagger H \bar{T} T $ +  \vspace{1mm}\\
$M_T \bar{T} T$ \\ \\ \\ 
\hspace{-1mm} \includegraphics[width=3.5cm]{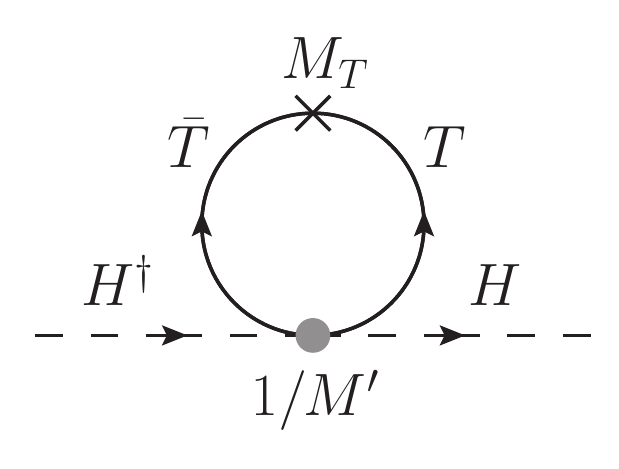} \hspace{-1mm} \\
\end{tabular}
} & \begin{tabular}{cc} 
\\
Strong dynamics \\  \hspace{2.5mm}  \includegraphics[width=2.5cm]{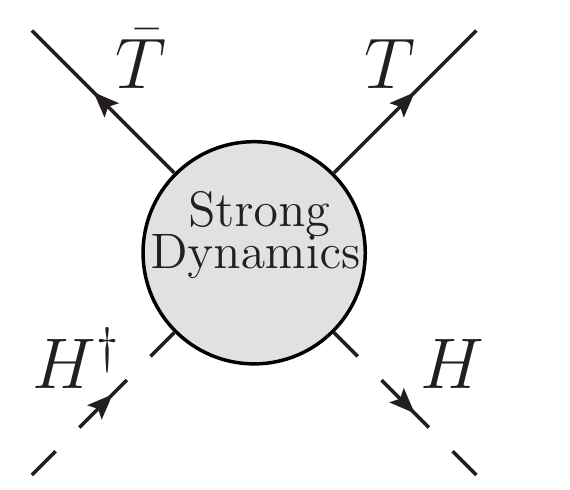} \\ \\ \end{tabular}
   &
\begin{tabular}{l}
Composite/\\
Holographic\\
Twin Higgs\\
\cite{Batra:2008jy, Barbieri:2015lqa, Low:2015nqa, Geller:2014kta}
\end{tabular}
&
\begin{tabular}{l}
Unitarity \\ constraints
\end{tabular}
&
\begin{tabular}{ll}
\\
$\Delta_{h(T)}$: & log tuning from incomplete $t-T$ cancellation \\ \\ 
\end{tabular}
&
\begin{tabular}{c}
\ref{ss.fermionicstrongcoupling}
\end{tabular} \\ \cline{3-7}
  & & \begin{tabular}{cc}
  \\
   Scalar mediators ($S$) \\ \hspace{2.5mm}  \includegraphics[width=2.7cm]{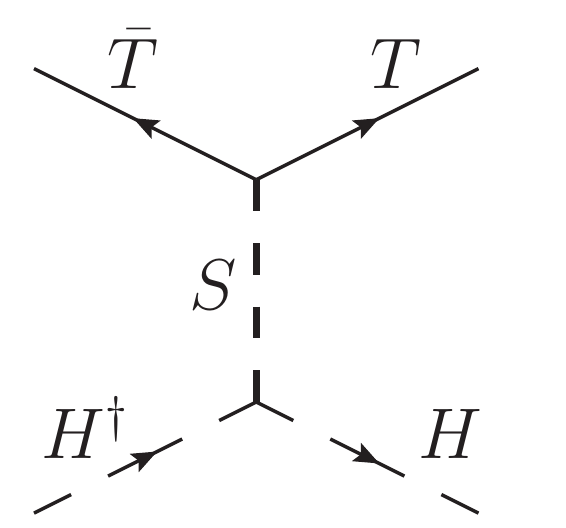} \\ \\ \end{tabular} 
&
\begin{tabular}{l}
4D perturbative \\
Twin Higgs \\
\cite{Chang:2006ra,Craig:2013fga, Craig:2014aea, Craig:2014roa}
\end{tabular}
&
\begin{tabular}{l}
$\delta \sigma_{Zh}$
\end{tabular}
&
\begin{tabular}{ll}
\\
$\Delta_{h(T)}$: & log tuning from incomplete $t-T$ cancellation \\
$\Delta_{h(S)}$: & log tuning from mediator loops\\
$\Delta_{S(T)}$: & quadratic divergence of mediator mass from \\ & $T$ loops, 
  required by Sacrificial Scalar Mechanism \\ \\ 
\end{tabular}
&
\begin{tabular}{c}
\ref{ss.fermionicscalarmediator}
\end{tabular}
\\ \cline{3-7}
  & & \begin{tabular}{cc}
  \\
   Fermion mediators ($F$) \\ \hspace{-1.5mm} \includegraphics[width=3.3cm]{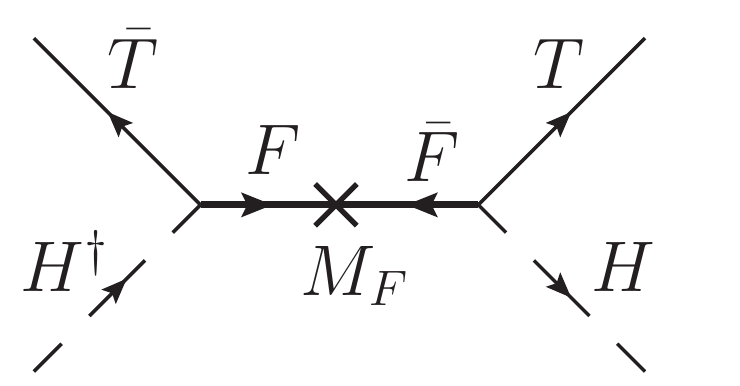} \hspace{-4.5mm} \\ 
  \\
  \end{tabular} 
&
\begin{tabular}{l}
---
\end{tabular}
&
\begin{tabular}{l}\\
EWPO\\
$\delta \sigma_{Zh}$\\
$\delta \sigma_{hh}$\\ \\
\end{tabular}
&
\begin{tabular}{ll}
\\
$\Delta_{h(T)}$: & log tuning from incomplete $t-T$ cancellation \\ \\
\end{tabular}
&
\begin{tabular}{c}
\ref{ss.fermionicfermionmediator}
\end{tabular}
\\ \cline{3-7}
  \hline
\end{tabular}
\caption{
Summary of the neutral top partner structures we examine in this paper, with schematic low-energy effective Lagrangians shown. (Vector top partners are identical to scalar partners for our purposes, see \ssref{vectorpartners}.) For fermion partner scenarios, the $|H|^2 \bar T T$ operator can be generated by non-perturbative physics, scalar mediator exchange, or fermion mediator exchange. The low energy observables we consider are the $Zh$ cross section deviation $\delta \sigma_{Zh}$ and electroweak precision observables (EWPO) at lepton colliders, and the triple Higgs coupling (measured by deviations in di-Higgs production, $\delta \sigma_{hh}$) and direct partner production ($\sigma_\mathrm{direct}$) through the Higgs portal at a 100 TeV collider. \label{t.summary}
}
\end{center}

\end{table*}
\end{turnpage}

Independent of any particular model, if the top partners carry electroweak charge then they are guaranteed to (a) be produced through Drell-Yan type processes at colliders, and (b) modify the electroweak coupling at different scales through RG-running~\cite{Alves:2014cda}.
In the case of a model with mirror QCD, production and decay of these states can result in spectacular glueball-jet signals~\cite{Burdman:2008ek, Kang:2008ea,Harnik:2008ax,Harnik:2011mv, Curtin:2015fna}. In models without a mirror QCD force (see e.g.~\cite{Poland:2008ev}), the top partners may be collider stable or decay electroweakly. In that case, searches in final states with leptons, disappearing tracks or monojets can probe such electroweak states up to masses of a few hundred GeV at the 14 TeV LHC and likely $\sim$ TeV or beyond at a 100 TeV collider~\cite{Low:2014cba}. Regardless of any decay modes, measurements of the Drell-Yan dilepton mass spectrum at a 100 TeV collider with $30\iab$ are sensitive to top partners with masses up to $1-2 \tev$~\cite{Alves:2014cda} or higher (depending on multiplicity). 
Naturalness therefore serves as strong motivation to perform this measurement at current and future colliders.  
The upshot is that electroweak charged top partners, like their QCD-charged cousins, are experimentally discoverable regardless of model details, at current or proposed future colliders.

Discovery is most difficult for models of \emph{neutral naturalness}. This is realized by the Twin Higgs family of theories, where the mirror sector carries no SM charge. It is possible for Twin Higgs models to feature the above glueball signature~\cite{Craig:2015pha}, but the most general Twin Higgs model has light mirror matter and therefore no mirror glueballs. However, in all known concrete theories, tree-level mixing generates Higgs coupling corrections at the level $ \mathcal{O}(v^2/f^2)$, where $f$ is the scale at which an enlarged symmetry containing $SU(2)_L$ is broken. The size of these corrections is also related to the level of tuning in the theory, so naturalness implies sizable deviations. Such coupling shifts are detectable at the $5-10\%$ level at the HL-LHC and at the  $0.5\%$-level at future lepton colliders~\cite{Burdman:2014zta, Dawson:2013bba}. This sensitivity therefore provides a way at least to detect, if not necessarily diagnose, the existence of the Twin Higgs. Detailed exploration may require a 100 TeV collider to probe the UV completion of these models. 

That being said, the argument for discoverability of neutral naturalness is not as tight as for electroweak-charged top partners, where RG effects and direct production guarantee discovery of new states. While the known theory in this category, i.e. Twin Higgs, produces measurable Higgs coupling shifts related to tuning, it is not obvious that this correlation persists in any possible theory with neutral fermionic top partners, which could perhaps have a very different structure. Furthermore, one could also imagine that naturalness is enforced by SM-neutral \emph{bosonic} top partners. (Though no full theory has yet been proposed to realize this via some symmetry, the low-energy consequences of this scenario have been examined in the context of Higgs coupling shifts at lepton colliders~\cite{Craig:2013xia} and direct production of top partners through an off-shell Higgs~\cite{Curtin:2014jma, Craig:2014lda}.)

In order to derive a no-lose theorem for naturalness with top partners, we therefore have to analyze neutral naturalness as model-independently as possible. This will be the focus of the present work. 

Assuming particles of spin of 1 or less, there are four possible \emph{neutral top partner structures} that can cancel the top loop contribution. They are summarized in \tref{summary}. Of these scenarios, only two have been realized in a full theory. Even those which have been realized may arise in completely different theories which have not yet been proposed. 

We take a very conservative approach to our model-independent analysis. In \sref{eft} we construct a low-energy effective field theory (EFT) expansion of Higgs-partner interactions for scalar and fermion partners. It is straightforward to identify the operators which give rise to one-loop quadratically divergent contributions to the Higgs mass. There are many other conditions a natural theory must satisfy, but we only impose the one-loop cancellation of the top quadratic divergence.

These minimal Lagrangians give rise to \emph{irreducible experimental signals} which are sensitive to the low-energy top partner structure. Furthermore, to analyze fermion partners $T$ it is necessary to specify how the $|H|^2 \bar T T$ coupling is generated. This results in three fermion partner scenarios reminiscent of Simplified Models \cite{Alves:2011wf}, which we analyze, along with the scalar partner scenario,  in \sref{simplifiedmodels}. For each case, we also identify a number of \emph{irreducible tunings} $\Delta_i$, which can be used to glean information about the UV completion scale.

\begin{figure}
\begin{center}
\hspace*{-10mm}
\includegraphics[width=7cm]{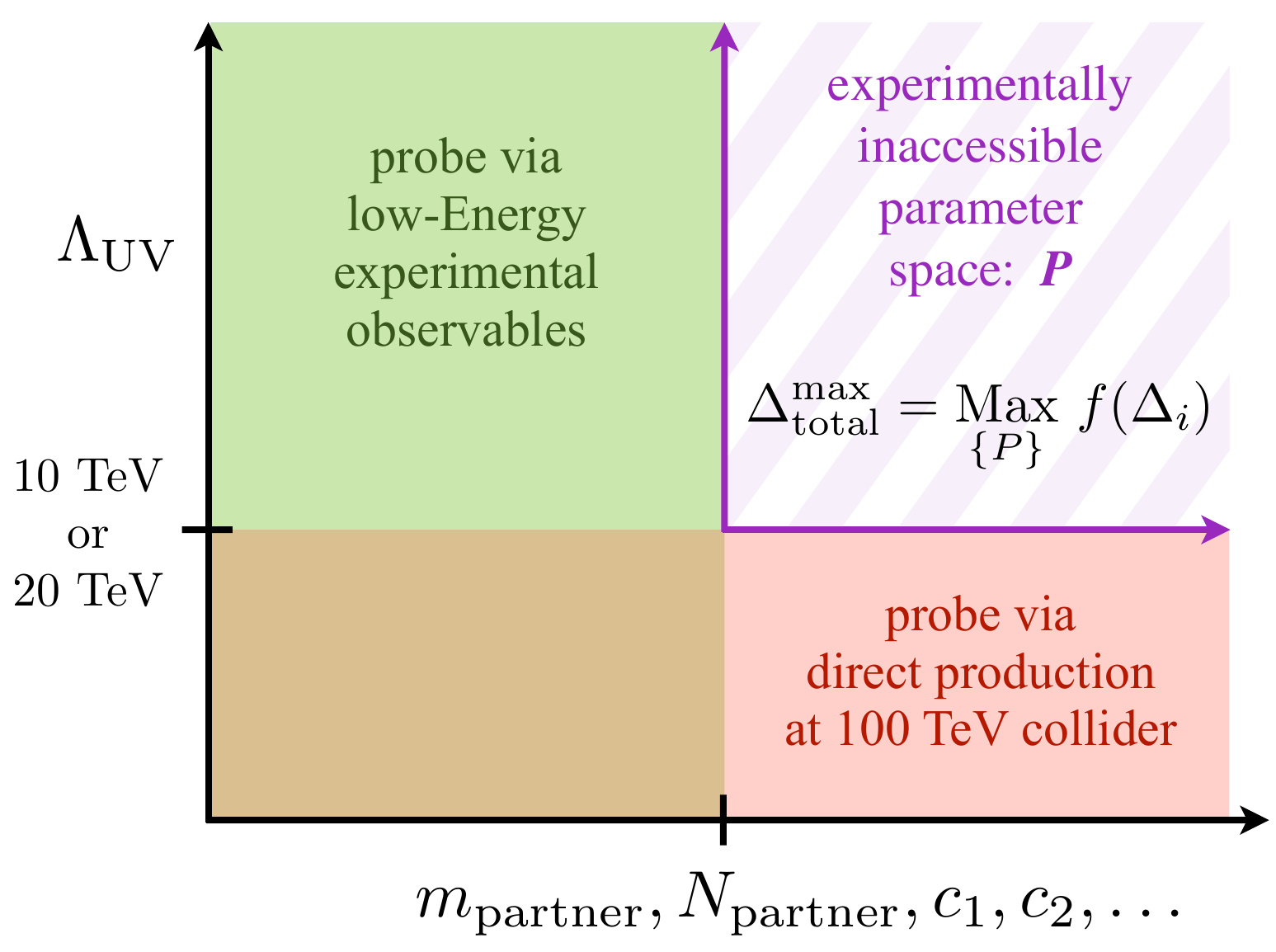}
\end{center}
\caption{
Schematic representation of our analysis strategy. In each neutral top partner scenario, low-energy observations (mostly at lepton colliders) place constraints on some parameters, including partner mass and multiplicity (horizontal axis). Orthogonally, a 100 TeV collider probes UV completion scales as high as 10 or 20 TeV (vertical axis). This defines the \emph{experimentally inaccessible parameter space $P$} for each scenario. The total tuning $\Delta_\mathrm{total} = f(\Delta_i)$ can then be maximized over $P$ to find the guaranteed minimum degree of tuning required to avoid detection.
}
\label{f.strategy}
\end{figure}

The essence of our analysis strategy is schematically represented in \fref{strategy}. Low-energy observables, mostly through measurements at future lepton colliders, constrain some set of each scenario's parameters, including (but not limited to) the top partner mass and multiplicity. 
The various tunings of each scenario must be regulated at some scale $\Lambda_\mathrm{UV}$. We now make an important assumption, namely that new SM-charged BSM states appear at this scale. This is certainly the case in every known UV completion of uncolored naturalness~\cite{Craig:2013fga, Geller:2014kta, Batra:2008jy,Barbieri:2015lqa, Low:2015nqa, Craig:2014fka, Chang:2006ra}, and reflects the expectation that the full symmetry protecting the Higgs is manifest at this scale, so that new particles appear in multiplets which include SM charges. At our level of discussion however we simply take this as an input. In this case, we can expect the 100 TeV collider to exclude any scenario with $\Lambda_\mathrm{UV} \lesssim 10$ or 20 TeV. 

The region not excluded by either low-energy probes or a low value of $\Lambda_\mathrm{UV}$ defines the \emph{experimentally inaccessible parameter space} $P$ of each scenario. Combining the various tunings in the theory as $\Delta_\mathrm{total} = f(\Delta_i)$, either very conservatively or more conventionally (see \ssref{strategy}), we can then define
\begin{equation}
\Delta_\mathrm{total}^\mathrm{max} = \underset{\{P\}}{\mathrm{Max}} f(\Delta_i) \ ,
\end{equation}
which is the total tuning, maximized over the inaccessible parameter space. This represents  the \emph{least severe degree of tuning required in the top partner scenario to avoid detection at proposed future colliders.} 

Fortunately, inverse tunings increase as we move away from the axes in \fref{strategy}, so to speak. 
We will find that $\Delta_\mathrm{total}^\mathrm{max}$ is quite severe for top partner scenarios with multiplicities close to the canonical case (12 real scalar partners or 3 fermion partners). Therefore, such scenarios are expected to be discovered at future lepton and/or 100 TeV colliders, unless they suffer tuning worse than 10\% (in some cases much worse). This result allows us to formulate our phenomenological no-lose theorem for top partner theories in \sref{conclusion}.

In \sref{nolose}, we place our results in context by discussing solutions to the hierarchy problem which might escape our conclusions. First and foremost is the possibility of a ``top partner swarm'', when the number of partner degrees of freedom is very large. One could also imagine UV completing neutral naturalness without new SM-charged particles. No such theory currently exists, but we outline the general features it might have, and argue why some top partner structures might not allow for such a UV completion. Their concrete realization is left as a model-building challenge for future work. Finally, we discuss naturalness in the absence of a perturbative top partner regime. Here we argue  informally that, barring highly exotic possibilities yet to be formulated, such theories should also be discovered at current or proposed future colliders. This allows us to claim that, in fact, our analysis of top partner theories brings us quite close to a true no-lose theorem of naturalness.

\begin{figure*}
\begin{center}
\hspace*{-5mm} 
\begin{tabular}{ccc}
   \includegraphics[width=4cm]{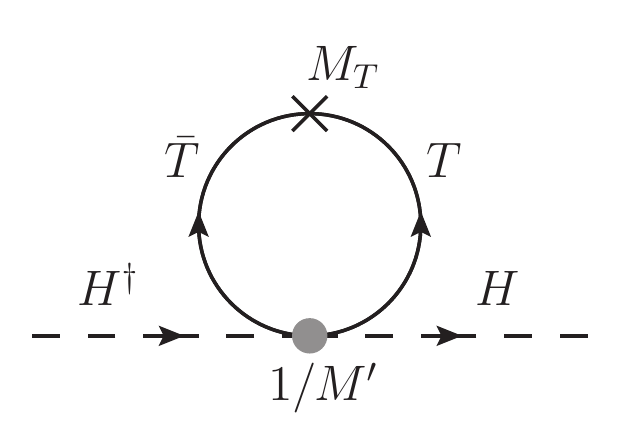}
   &
   \includegraphics[width=3.9cm]{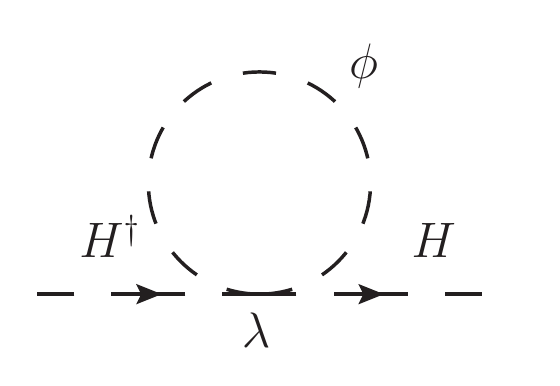}
   &
      \includegraphics[width=3.8cm]{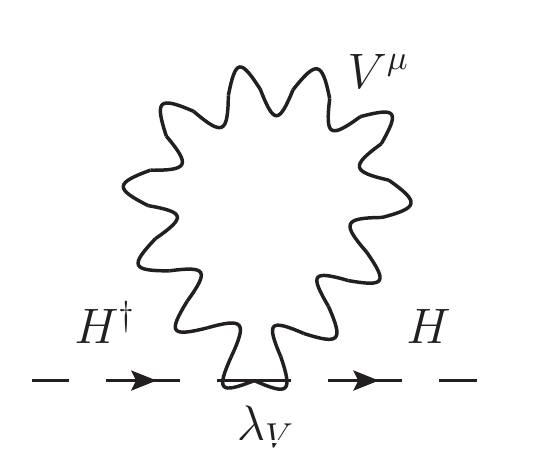}
 \\
 (a) Fermionic top partner
 &
 (b) Scalar top partner
 &
 (c) Vector top partner
 \end{tabular}
 \end{center}
 \caption{
 Neutral top partner loops canceling the quadratically divergent one-loop top contribution to the Higgs mass. }
 \label{f.toppartners}
 \end{figure*}

Overall, our arguments provide significant motivation for the construction of new lepton \emph{and} hadron colliders. Crucially, we show that it is possible for either collider but not the other to see a signal, but unlikely for \emph{neither} to see anything if nature is natural. Therefore, \emph{both} machines will be required to work in tandem as discovery tools of generalized naturalness.

This paper is organized as follows. In \sref{eft}, we establish notation and define the minimal EFT's of neutral bosonic and fermionic top partners. The experimental consequences of these minimal Lagrangians are explored in \sref{simplifiedmodels}. This involves partial UV completions for fermion partners, see \tref{summary}. In each case, we also derive the irreducible tunings, which tie the UV completion scale to the low-energy parameters. This allows us to formulate our no-lose theorem. In \sref{nolose}, we place these results in context by discussing hypothetical theories which avoid our assumptions, as well as solutions to the hierarchy problem without top partners. We sum up our results and explicitly formulate the no-lose theorem in \sref{conclusion}. We make use of experimental sensitivity projections for future colliders, which are summarized in \aref{sensitivities}. Additional technical details of the Sacrificial Scalar Mechanism for neutral fermionic top partners with scalar mediators are collected in  \aref{sacrificialscalar}.

\section{EFT's of Bottom-Up Neutral Naturalness}
\label{s.eft}

In this section we establish notation and define the minimal EFTs of neutral top partner structures.

 We approach naturalness from the bottom-up (as in e.g.~\cite{Farina:2013ssa,Hedri:2013ina}) by considering the SM as a Wilsonian effective theory valid up to some finite momentum cutoff $\Lambda_\mathrm{BSM}$. 
The SM Higgs potential at tree-level is given by
\begin{equation}
V = - \mu^2 |H|^2 + \lambda |H|^4.
\label{e.higgspot}
\end{equation}
Each particle $i$ coupled to the Higgs, with mass $m_i(H) < \Lambda_\mathrm{BSM}$, contributes to the 1PI effective potential at loop order. The one-loop correction is 
\begin{equation}
\begin{array}{l}
\displaystyle 
\delta V_\text{eff}(H) = 
\\ \\ 
\displaystyle  \frac{-i}{2} \sum_i (-1)^{2s_i} g_i \int_{p^2=0}^{p^2=\Lambda_\mathrm{BSM}^2} \frac{d^4p}{(2 \pi)^4} \log \left[p^2-m_i^2(H) + i \epsilon \right]  
\end{array}
\end{equation}
where the $s_i$ is the spin of the $i$th particle , and $g_i$ the associated number of degrees of freedom. After rotating to euclidean momentum space, integrating, and expanding for large $\Lambda_\mathrm{BSM}$:
\begin{equation}
\begin{array}{l}
\displaystyle 
\delta V_\text{eff}(H) = \sum_i \frac{ (-1)^{2s_i} g_i}{64\pi^2}   \ \times \ 
\\ 
\\ \displaystyle 
\ \ \ \ \ \ 
\bigg[2 m_i^2(H) \Lambda_\mathrm{BSM}^2  
- m_i^4(H)  
 \left(
\log\frac{\Lambda_\mathrm{BSM}^2}{m^2_i(H)}+ \frac{1}{2}
\right) \bigg]
\end{array}
\end{equation}

The SM particle with the largest coupling to the Higgs is the top quark, $\mathcal{L} \supset y_t H Q \bar U$. It generates a Higgs mass correction
\begin{equation}
\label{e.deltaVtop}
\delta V_\mathrm{top} = - \frac{3 y_t^2 \Lambda_\mathrm{BSM}^2}{8 \pi^2} |H|^2 + \ldots.
\end{equation}
The quadratic UV sensitivity of this correction to the Higgs mass $\mu^2$ is one way to phrase the well-known Hierarchy Problem. We can quantify the tuning suffered by the Higgs mass as
\begin{eqnarray}
\nonumber
\Delta_{h(t)} &=& \left| \frac{\delta{\mu^2}}{\mu^2_\mathrm{phys}} \right|^{-1}
=
\left| \frac{3 y_t^2 \Lambda_\mathrm{BSM}^2}{8 \pi^2} \frac{1}{\mu^2_\mathrm{phys}} \right|^{-1}
\\ \nonumber \\
&\approx& 
\left\{
\begin{array}{lll}
0.3 &  & \Lambda \sim 800 \gev
\\
0.1 & \mathrm{for} & \Lambda \sim 1.4 \tev
\\
0.01 &  & \Lambda \sim 4.5 \tev
\end{array}
\right. .
\end{eqnarray}
(The subscript denotes the scalar with the tuned mass, with the particle generating the loop correction in brackets.)
A natural theory should not be severely tuned. This is the usual justification for expecting new physics to enter at the TeV scale to cancel the quadratic Higgs mass contributions of the top quark.

We now determine the minimal effective Lagrangians required for \eref{deltaVtop} to be canceled by the loops of perturbative neutral top partners shown in \fref{toppartners}. This stabilization of the Higgs mass will be the only requirement.  As we will see, this is sufficient to extract meaningful information about various tunings and experimental signals. 

It is conservative to assume that electroweak symmetry breaking is entirely due to the vacuum expectation value $v$ of the $SU(2)_L$ doublet $H$ of the Standard Model. Other representations, or additional doublets, would lead to detectable experimental signals like direct heavy Higgs production and Higgs coupling deviations~\cite{Dawson:2013bba}.

In the SM, the physical Higgs boson is $h$, which is a component of the doublet $H$. LHC Higgs coupling measurements suggest that the discovered 125 GeV degree of freedom is \emph{dominantly} $h$-like~\cite{Jez:2015wza}. This implies that  Higgs mass parameter $\mu^2$ is close to the SM value and has to be protected. This justifies imposing the cancellation of \eref{deltaVtop} as our main requirement for neutral top partner structures.

We start by writing the most general interaction Lagrangian between the top quark and the Higgs. 
\begin{eqnarray}
\nonumber
\mathcal{L}_{t} 
&=&
H Q \bar U \left( 
y_t + 
\frac{|H|^2}{\Lambda_{t_{(1)}}^2} + 
\frac{|H|^4}{\Lambda_{t_{(2)}}^4} + 
\ldots
\right)
\\ 
\label{e.Ltgeneral}
&& 
\ + \ (Q \bar U)^2 \left( 
\frac{|H|^2}{\tilde \Lambda_{t_{(1)}}^2} + 
\ldots
\right)
\ \ + \ \ \ldots
\end{eqnarray}
Similarly for a set of general 4-component fermionic top partners $T_i$:
\begin{eqnarray}
\nonumber
\mathcal{L}_{T} &=& \sum_i T_i \bar T_i\left(
 M_{T_i} + 
\frac{|H|^2}{\Lambda_{T_{i(1)}}} + 
\frac{|H|^4}{\Lambda_{T_{i(2)}}^3} + 
  \ldots
  \right)
  \\
\label{e.LTgeneral}
  &&
+
   \sum_{i,j} (T_i \bar T_i)(T_j \bar T_j)
   \left(
   \frac{|H|^2}{\tilde \Lambda_{T_{ij(1)}}^4} +
   \ldots
   \right)
   \ +  \ldots ,
\end{eqnarray}
and for a set of real scalar partners $\phi_i$:
\begin{eqnarray}
\nonumber
\mathcal{L}_\phi &=& - \sum_i \phi_i^2 \left( 
\frac{1}{2} \mu_{\phi_i}^2 + 
\frac{1}{2} \lambda_i |H|^2 + 
\frac{|H|^2}{\Lambda_{\phi_{i(1)}}^2} + 
\ldots
\right)
\\ \label{e.Lphigeneral}
&&
-
\sum_{i,j} \phi_i^2 \phi_j^2 \left( 
\frac{|H|^2}{\tilde \Lambda_{\phi_{ij(1)}}^2} 
+ \ldots
\right)
\ - \ \ldots.
\end{eqnarray}
A discussion of vector top partners is postponed to the end of the section.
As the above illustrates, the number of possible interaction terms is large. However, a \emph{necessary} condition for a theory to be natural from the bottom-up perspective is that the top-loop quadratically divergent Higgs mass contribution is regulated.  Only terms with one or two powers of $H$ and two identical top/partner fields (in the mass basis at $H = 0$)  can contribute to $\delta \mu^2$ at one-loop. Regulation of the one-loop Higgs mass divergence therefore imposes conditions on only the following terms:
\begin{eqnarray}
\label{e.Lt}
\mathcal{L}_{t} 
&\supset&
y_t H Q \bar U 
\\
\label{e.LT}
\mathcal{L}_{T} &\supset& \sum_i T_i \bar T_i\left(
 M_{T_i} - 
\frac{|H|^2}{2 M_i^\prime} \right)
\\
\label{e.Lphi}
\mathcal{L}_\phi &\supset& - \sum_i \phi_i^2 \left( 
\frac{1}{2} \mu_{\phi_i}^2 + 
\frac{1}{2} \lambda_i |H|^2\right)
\end{eqnarray}
where we have renamed one of the scales in $\mathcal{L}_T$ for later convenience. 

The additional terms  in Eqns. (\ref{e.Ltgeneral}) - (\ref{e.Lphigeneral}) can generate, within an effective field theory, $\delta \mu^2$ corrections with quartic or even higher dependence on the cutoff $\Lambda$. However, this can only happen at two-loop or higher order, or due to derivative operators, and these contributions to $\delta \mu^2$ must be subdominant compared to the one-loop contributions as long as the cutoff $\Lambda$ is not too high. Crucially, regulation of these two-loop or higher contributions represent \emph{additional} conditions that a natural theory must satisfy: it might do so by simply not generating (or suppressing) most of the terms in Eqns. (\ref{e.Ltgeneral}) - (\ref{e.Lphigeneral}), or the same symmetry which guarantees cancellation of the one-loop divergence might do the same for higher loop and $\Lambda$ orders (at least approximately, until new protection mechanisms kick in at a higher scale). 

Similarly, some of the terms in Eqns. (\ref{e.Ltgeneral}) - (\ref{e.Lphigeneral}) that are not in Eqns. (\ref{e.Lt}) - (\ref{e.Lphi}) contribute to the quadratically divergent one-loop correction to the Higgs quartic $\delta \lambda$. Taking the cutoff $\Lambda$ to be as high as possible within the realm of validity of the EFT, this would generate $\lambda \sim \mathcal{O}(1)$, corresponding to a $\sim 10\%$ tuning to generate the $\lambda \sim 0.1$ consistent with a 125 GeV Higgs mass (see e.g.~\cite{Bellazzini:2014yua}).  This tuning might be real (and seen as not too serious), or it could be ameliorated with additional symmetry within the theory. The latter would, again, represent an additional requirement that the theory must satisfy. 

It is therefore clear that, by only considering the physical consequences of the \emph{one-loop $\delta \mu^2$ cancellation at quadratic order in the cutoff}, we are deriving the most conservative, necessary signals of neutral naturalness. 
This justifies ignoring the additional terms in Eqns. (\ref{e.Ltgeneral}) - (\ref{e.Lphigeneral}) and only considering Eqns. (\ref{e.Lt}) - (\ref{e.Lphi}). 
In a similar vein, when a top partner scenario necessitates introducing additional fields, the only new tunings we consider will be those computed at one-loop order for scalar masses.
When computing experimental signals and tunings, we will have to take care to ensure that none of these dominantly depend on any operators besides those in in Eqns. (\ref{e.Lt}) - (\ref{e.Lphi}).

As part of this approach we also refrain from making any statements about \emph{tree-level tuning}. 
This is very conservative -- for example, in the Twin Higgs the strongest naturalness constraint in fact arises from such a tree-level tuning, where a $\mathbb{Z}_2$ breaking soft mass has to be adjusted against a tree-level vev to achieve $v \ll f$. 
However, this conservative approach is also necessary, since it is possible to modify or eliminate tree-level tunings without directly impacting the top partner sector. For example in~\cite{Cohen:2015ala} a SUSY model is presented where the Higgs mass is independent of the SUSY $\mu$ parameter, so that the Higgsinos can be heavy without the tuning usually associated with a large $\mu$ parameter. Even so, stops still have to be light for the theory to be natural.

Note that similar arguments can be constructed for quadratic divergences from gauge boson loops, which are subdominant to the top contributions but still have to be regulated at a few TeV to avoid tuning. We conservatively do not consider the consequences of this cancellation, which is an \emph{additional} necessary condition that must be satisfied by a natural theory.

\subsection{Cancellation Condition for Scalar Top Partners}
For real scalar top partners $\phi_i$, the interaction Lagrangian is
\begin{equation}
\label{e.Lscalar}
\mathcal{L}_\phi \supset - \sum_i \phi_i^2 \left( 
\frac{1}{2} \mu_{\phi_i}^2 + 
\frac{1}{2} \lambda_i |H|^2\right),
\end{equation}
giving a physical scalar mass $m_{\phi_i}^2 = \mu_{\phi_i}^2 + \lambda_i |H|^2$.
For $N_r$ identical top partners with $\lambda_i = \lambda_\phi$, $\mu_{\phi_i} = \mu_\phi$, cancellation of the one-loop quadratic divergence from \fref{toppartners}(b) against that from the top quark requires 
\begin{equation}
\label{e.naturalreal}
\lambda_\phi = \frac{12}{N_r} |y_t|^2
 \end{equation}
In the familiar case of the MSSM, this is realized by 12 real (= six complex) scalar tops. The logarithmically divergent correction to the Higgs mass is 
\begin{equation}
\label{e.scalartuning}
\delta \mu^2 = \frac{3 y_t^2}{8 \pi^2} \mu_\phi^2 \log \frac{\Lambda_\mathrm{UV}^2 + \mu_\phi^2}{\mu_\phi^2}
 \end{equation}
 where $\mu_\mathrm{phys}^2 = \mu^2 + \delta \mu^2$ comes with a minus sign in the potential. This corresponds to a log tuning suffered by the Higgs mass:
\begin{equation}
\label{e.DeltaTscalar}
\Delta_{h(\phi)}= \left| \frac{3 y_t^2}{8 \pi^2} 
\frac{\mu_\phi^2}{\mu_\mathrm{phys}^2} 
 \log \frac{\Lambda_\mathrm{UV}^2+\mu_\phi^2}{\mu_\phi^2} \right|^{-1} \ ,
\end{equation}
where the meaning of $\Lambda_\mathrm{UV}$ will be discussed in \sref{simplifiedmodels}.

\subsection{Cancellation Condition for Fermionic Top Partners}

The interaction Lagrangian for fermionic top partner $T_i$,
\begin{equation}
\label{e.Lfermion}
\mathcal{L}_{T} \supset \sum_i T_i \bar T_i\left(
 M_{T_i} - 
\frac{|H|^2}{2 M_i^\prime} \right),
\end{equation}
gives, for finite cutoff $\Lambda$, the following quadratically divergent correction to the Higgs mass term:
\begin{equation}
\delta V_\text{eff.} \supset \sum_i \frac{1}{8 \pi^2}\frac{M_{T_i}}{M_i'}\Lambda^2 H^\dagger H.
\end{equation}
This arises from the loop diagram in \fref{toppartners}~(a). 
For $N_f$ identical fermions with $M_{T_i} = M_T$ etc., the quadratic divergence in the mass term cancels that from the top loop if
\begin{equation}
\label{e.naturalfermion}
\frac{M_T}{M'} = \frac{3}{N_f} y_t^2
\end{equation}
This is of course satisfied in Composite/Little Higgs and standard Twin Higgs models for $N_f = 3$, $M_T = y_t f$ and $1/M' = y_t/f$. 
The physical top partner mass is 
\begin{equation}
\label{e.mTphys}
m_T \equiv M_T - \frac{v^2}{4  M'}.
\end{equation}
The remaining logarithmically divergent correction to the Higgs mass is
\begin{equation}
\label{e.DeltaT}
\Delta_{h(T)} \equiv \left|\frac{\delta\mu^2}{\mu_\mathrm{phys}^2} \right|^{-1} = 
\left|
\frac{3 y_t^2}{8 \pi^2} \frac{M_T^2}{\mu_\mathrm{phys}^2} \log \frac{\Lambda_\mathrm{UV}^2 + M_T^2}{M_T^2}
\right|^{-1} \ .
\end{equation}

 \subsection{Vector Top Partners}
 \label{ss.vectorpartners}
 
A more exotic possibility is that the quadratic divergence of the Higgs mass is canceled by spin-1 (vector) bosons. For example, a model of this type with colored vector top partners arising from a non-abelian gauge symmetry was proposed in~\cite{Cai:2008ss}. More generally, we can consider couplings of the Higgs to a vector field without necessarily imposing gauge invariance, though such models may require UV completion below some scale to ensure unitarity. At the one-loop level, a vector field $X^\mu$ can generate quadratically divergent corrections to the Higgs mass through the operators $H^\dagger D_\mu H X^\mu$ and $|H|^2 X_\mu X^\mu$. The former gives a divergence of the same sign as that from the SM top loop, while the latter gives the opposite sign and is necessary to achieve cancellation. 

The operator $H^\dagger D_\mu H X^\mu$ with coefficient $g_X$ induces mixing between the $X$ field and the Standard Model $Z$ boson at the order $g_X g_2 v^2/M_X^2$ for large $X$ boson mass $M_X$, affecting electroweak precision observables such as the $\rho$ parameter. These constraints then require either large $M_X$, implying large logarithmic contributions to the Higgs mass which strain naturalness, or a small value for the coupling $g_X$, i.e. much smaller than $y_t$, so that the quadratically divergent contribution to the Higgs mass from this operator is well subdominant to that from the top quark. 

To focus on natural theories we can therefore assume that $g_X$ is small and that the operator $|H|^2 X_\mu X^\mu$ exactly cancels the quadratic divergence from the SM top loop. (This operator can exist without the term linear in $X^\mu$ or any other additions if e.g. $X^\mu$ is a Stueckelberg field.) However, if we only consider this operator, then the vector field $X^\mu$ behaves exactly as a collection of three real scalars for all of our purposes; e.g. in temporal gauge the operator becomes simply $|H|^2(X_1^2 + X_2^2 + X_3^2)$. Therefore all experimental signatures we consider for the case of scalar top partners (precision Higgs effects from loops of top partners and instability of the partner mass itself) apply directly to the case of a vector top partner as well, so we will not explicitly discuss this case further.

\section{Irreducible Signatures of Neutral Naturalness}
\label{s.simplifiedmodels}

We now examine the irreducible signatures of neutral naturalness in detail. The aim is to answer the following question:
\begin{center}
\emph{``Are there any natural theories with neutral top partners which would produce no experimental signals at current or proposed future colliders?"}
\end{center}
This of course depends on one's definition of what level of tuning is considered natural, but the results of our analysis can be interpreted according to one's preference in this regard. 
To answer this question we approach the discoverability of neutral naturalness from two different angles, as outlined in \sref{intro} and \fref{strategy}. This strategy is presented in detail in \ssref{strategy}. The individual top partner structures are discussed in Sections \ref{ss.scalartoppartners} - \ref{ss.fermionicfermionmediator}, with simplifying assumptions discussed and justified in \ssref{wrinkles}.

\subsection{Strategy}
\label{ss.strategy}

The first general path to discover theories of neutral naturalness is via probes of the low-energy structure of the top partner sector, which we assume contains no SM-charged states. Of the proposed future measurements, four will be the most useful: (a) precision Higgs coupling measurements, most importantly the $Zh$ production cross section; (b) the Higgs cubic coupling; (c) top partner direct production; and (d) precision electroweak measurements, most importantly the $T$ parameter. In each case we use 95\% CL sensitivity projections of current and future lepton and hadron colliders from the literature, which are summarized in \aref{sensitivities}. 

The only assumptions we make in computing these observables is that the terms in \eref{Lscalar} and \eref{Lfermion} cancel the quadratically divergent top contribution to $\delta \mu^2$. We assume that all top partners have the same coupling to the Higgs, and that there are no off-diagonal couplings between the Higgs and two different top partners. As we explain in \ssref{wrinkles}, this is inherently conservative, essentially because increasing the number of partners reduces all experimental signals and makes all tunings less severe.

These observables offer significant reach into various scalar and fermion neutral top partner scenarios. However, they are insufficient to probe all natural theories by themselves. There will always be some part of each scenario's parameter space, let us call it $P'$, that cannot be excluded using probes of low-energy structure.

The second way to discover theories of neutral naturalness is by probing their UV-completion. One of our basic assumptions is that SM-charged BSM states appear at the UV completion scale. This would allow a 100 TeV collider to discover any such theory by direct production, so long as $\Lambda_\mathrm{UV} \lesssim 10 - 20$ TeV. We leave an attempt to more formally prove this assumption for later work -- for now, it merely represents the expectation that the full symmetry which stabilizes the Higgs is manifest at this higher scale, implying the existence of heavy SM-charged states related to the Higgs and the top quark. 
This is indeed the case in all known UV completions of the Twin Higgs~\cite{Craig:2013fga, Geller:2014kta, Batra:2008jy,Barbieri:2015lqa, Low:2015nqa, Craig:2014fka, Chang:2006ra}, and may be hard to avoid in a general full theory. In fact, for each of the top partner scenarios we examine the implications of this assumption, and what would have to occur in any (as-yet unknown) theory which violates it.

For the purpose of estimating the overall tuning suffered by a scenario, we assume that all divergences are regulated at this scale $\Lambda_\mathrm{UV}$.\footnote{Of course, in particular some logarithmic divergences may be regulated at a much higher scale. In that case, our approach underestimates the level of tuning and is conservative.} The essence of our argument is that an undiscoverable theory has to avoid low-energy signatures (mostly at lepton colliders) and be UV-completed above 10 or 20 TeV (to avoid direct production at 100 TeV). These two requirements push the tuning into the unnatural regime for top partner multiplicities comparable to the canonical 3 fermions or 12 real scalars, as schematically represented in \fref{strategy}.

All the top partner scenarios we consider contain scalars (at least the Higgs) with physical masses $m_j^2$, which include various UV-sensitive loop contributions $\delta m_{j(k)}^2$. This allows us to define a set of independent tunings $\Delta_i \equiv |m_{j}^2/\delta m_{j(k)}^2|$, which depend on the scenario's parameters. $i$ is some arbitrary index to label tunings. Making a very conservative choice to represent simultaneous unrelated tunings by the single most severe tuning, we define an \emph{overall} tuning
\begin{equation}
\label{e.Deltatotal}
\Delta_\mathrm{total} \equiv \underset{i}{\mathrm{Min}}\{  \ 
\Delta_1(\Lambda_\mathrm{UV}) \ , \
\Delta_2(\Lambda_\mathrm{UV}) \ , \
\ldots
\ 
\} \ ,
\end{equation}
where the $i$ below $\mathrm{Min}$ indicates that we minimize with respect to the tuning index $i$, and we have made each tunings' dependence on $\Lambda_\mathrm{UV}$ manifest (the other parameter dependencies are implied). 
We can now define the \emph{least severe possible tuning that could escape experimental detection} in this scenario. Assume that the 100 TeV collider can probe UV-completions up to the scale $\Lambda_\mathrm{UV}^\mathrm{reach}$. In that case, the experimentally inaccessible parameter space $P$ in \fref{strategy} can be schematically represented as
\begin{equation}
P = \{P'\} \cap \{\Lambda_\mathrm{UV} > \Lambda_\mathrm{UV}^\mathrm{reach} \}
\end{equation}
where $P'$, as defined above, is the parameter space that is inaccessible using probes of low-energy structure. 
In order to find out how natural a theory could be while still escaping detection, we maximize the total tuning parameter over all of the scenario's parameter space that is experimentally inaccessible:
\begin{eqnarray}
\label{e.Deltatotalmax}
\Delta_\mathrm{total}^\mathrm{max}(\Lambda_\mathrm{UV}^\mathrm{reach})  &\equiv &\underset{\{P\}}{\mathrm{Max}}
\left[ \ 
\underset{k}{\mathrm{Min}}\{  \ \Delta_k \  \} 
\ 
\right]
\\
\nonumber
&=&
\underset{\{P'\}}{\mathrm{Max}}
\left[ \ 
\underset{k}{\mathrm{Min}}\{  \ \Delta_k(\Lambda_\mathrm{UV}^\mathrm{reach}) \  \} 
\ 
\right]
\end{eqnarray}
The second line follows since all tunings become more severe with increasing $\Lambda_\mathrm{UV}$. 

 The above method of combining tunings is arguably too conservative. Conventionally, if the input parameters of the theory can all be independently adjusted, then tunings of different observables should be multiplied in the overall tuning measure, with $\Delta = f(\{\Delta_i\}) = \Pi_i \Delta_i$, while independent tunings to the same observable should be added in inverse quadrature: $\Delta = f(\{\Delta_i\}) = (\sum_i \Delta_i^{-2})^{-1/2}$ (see e.g.~\cite{Arvanitaki:2013yja} for a discussion).   
The least severe tuning required to avoid experimental detection is then defined as
\begin{equation}
\label{e.DeltatotalmaxMULT}
\tilde \Delta_\mathrm{total}^\mathrm{max}(\Lambda_\mathrm{UV}^\mathrm{reach}) \equiv \underset{\{P\}}{\mathrm{Max}}
\left[ \ f(\Delta_i) \ 
\ 
\right].
\end{equation}
In general, a tilde above a combined tuning will indicate that it has been derived in this less conservative, but more conventional manner.

In the sections below, we compute this irreducible tuning for $\Lambda_\mathrm{UV}^\mathrm{reach} = 10$ and 20 TeV to represent plausible kinematic reaches of a 100 TeV collider.  In all neutral top partner scenarios, in order to escape all experimental detection, a theory has to be either unnatural, or have a large number of top partners (or, in one fermionic top partner scenario, scalar mediators). We call this possibility a ``top partner swarm''. It is also possible to circumvent our arguments if the UV completion introduces no new SM-charged particles. We discuss this in more detail in \sref{nolose}.

\subsection{Scalar Top Partners}
\label{ss.scalartoppartners}

\begin{figure}
\begin{center}
\includegraphics[width=8cm]{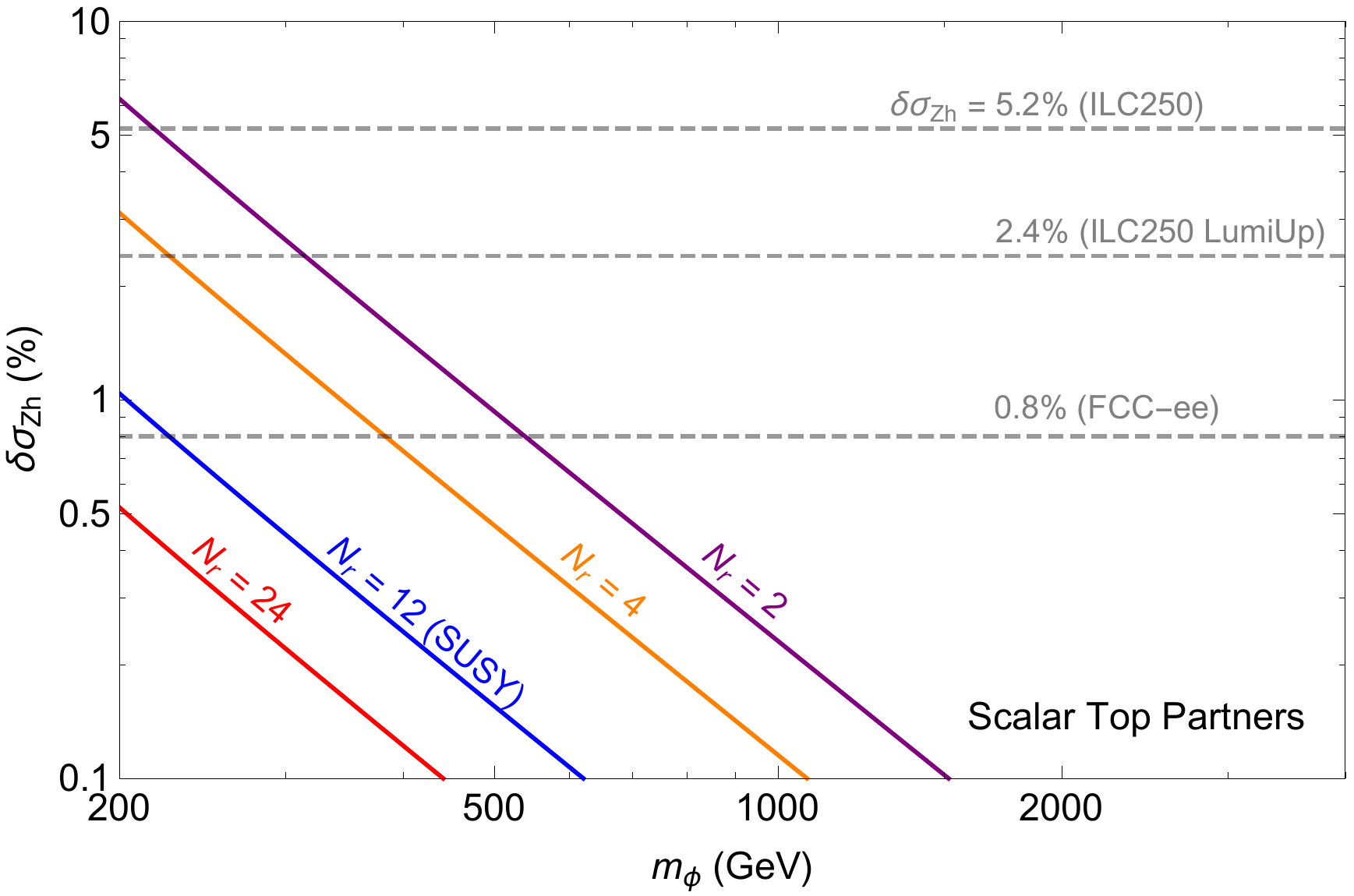} 
\end{center}
\caption{
Deviations in $\sigma_{Zh}$ at lepton colliders with $\sqrt{s} = 250 \gev$ for various multiplicities of real scalar partners, all with equal couplings to the Higgs~\cite{Craig:2013xia}. Dashed lines indicate 95\% CL sensitivities, see \aref{sensitivities}.
}
\label{f.scalarexclusionZh}
\end{figure}

The scalar top partner case is very simple to analyze, since it involves a renormalizable Lagrangian \eref{Lscalar} with a single interaction term. The $\phi^2 |H|^2$ interaction modifies Higgs couplings by generating corrections from closed loops of $\phi$. The resulting $Zh$ cross section shift has been analyzed in~\cite{Craig:2013xia}. We also consider the Higgs cubic coupling shift $\delta \lambda_3$. The same coupling gives rise to direct top partner production through an off-shell Higgs. Finally, since the top partners themselves are scalars, their masses are UV-sensitive and allow us to estimate what level of tuning is required to avoid experimental detection.

Loops of scalar partners rescale all Higgs couplings by contributing to the Higgs wave function renormalization. We use the expressions derived in~\cite{Craig:2013xia}. The most sensitive probe of this coupling shift is the $Zh$ production cross section measurement at future lepton colliders. For $\sqrt{s} = 250 \gev$, the fractional cross section shift, imposing the cancellation condition \eref{naturalreal}, is 
\begin{equation}
\delta \sigma_{Zh} = \frac{9 v^2 y_t^4}{2 \pi^2 m_h^2 N_r} \left[ 1 + F\left(\frac{m_h^2}{4 m_\phi^2}\right) \right],
\end{equation}
where the loop function $F(\tau) = -1 - \frac{2 \tau}{3} - \mathcal{O}(\tau^2)$. In \fref{scalarexclusionZh} we show the $Zh$ cross section shift $\delta \sigma_{Zh}$ for different numbers of top partners, compared to ILC and FCC-ee sensitivities. The reach for $N_r = 12$ (analogous to the MSSM) is about 250 GeV for FCC-ee. Note that the cross section shift scales as $\delta \sigma_{Zh} \propto 1/N_r$ with the number of top partners.

\begin{figure}
\begin{center}
\includegraphics[width=8cm]{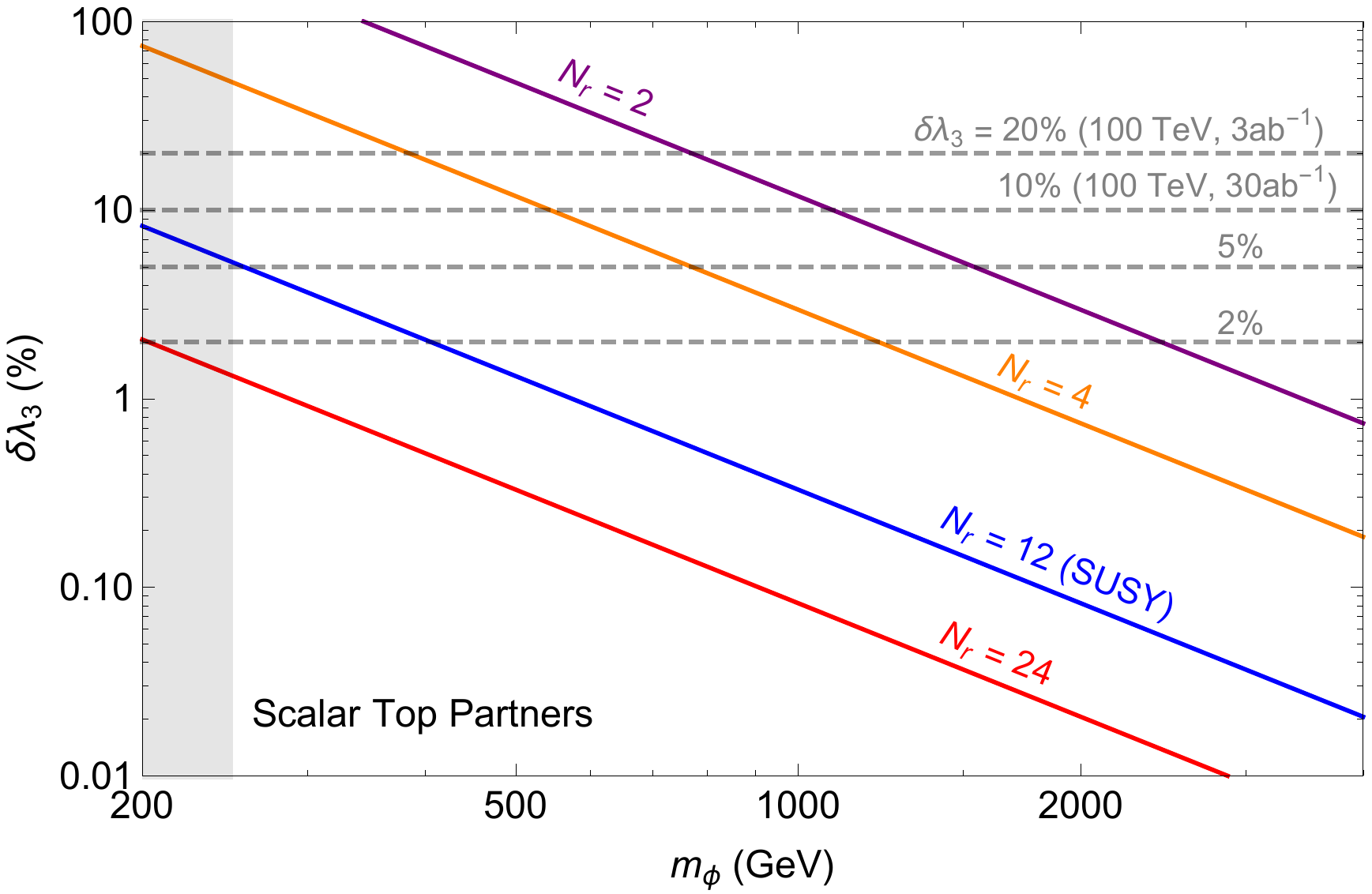}
\end{center}
\caption{
Relative shift in the Higgs cubic coupling  $\lambda_3$ compared to the SM, computed from the one-loop Higgs potential with $N_r$ real scalar top partners. Dashed lines indicate 95\% CL sensitivity projections for measurements at a future 100 TeV collider, as well as the hypothetical reach of more optimistic projections, see \aref{sensitivities}. The vertical gray band indicates where finite-momentum effects have to be taken into account to accurately assess the sensitivity of di-Higgs measurements~\cite{conversationmccullough}. 
}
\label{f.scalarexclusionlambda3}
\end{figure}

The triple Higgs coupling deviation can be parametrically understood as arising from a $|H|^6$ operator generated by scalar partner loops:
\begin{equation}
\delta V_1 \supset \frac{9 y_t^6}{\pi^2 N_r^2  \mu_\phi^2} |H|^6.
\end{equation}
Numerically, however, this is not a great approximation for top partners that are light enough to produce measurable coupling deviations. The relevant non-analytical contributions to the Higgs potential are most conveniently evaluated using the Coleman-Weinberg one-loop potential renormalized in the on-shell scheme~\cite{Anderson:1991zb}, since both top and top partner masses have the form $m^2 =a + b |H|^2$. This gives 
\begin{equation}
\lambda_3 = \frac{1}{6} \left. \frac{\partial^3 V}{\partial h^3} \right|_{h=v} = 
\frac{m_h^2}{2 v} - \frac{v y_t^4}{8 \pi^2} + \frac{9 y_t^6  v^3}{\pi^2 m_\phi^2 N_r^2},
\end{equation}
for the zero-momentum Higgs cubic coupling, with $m_\phi = m_\phi(h=v)$.
The first two terms, designated $\lambda_3^\mathrm{SM}$, come from the tree-level and top one-loop potential respectively. The third term is the scalar top partner one-loop contribution. $\delta \lambda_3 = (\lambda_3/\lambda_3^\mathrm{SM}) - 1$ is shown in \fref{scalarexclusionlambda3} as a function of top partner mass. In comparing to $\delta \sigma_{Zh}$ measurements, two things stand out:
\begin{itemize}
\item Since $\delta \lambda_3 \propto 1/N_r^2$, a $10\%$ measurement is superior to a Higgsstrahlung measurement for fewer top partners than the MSSM-case $N_r = 12$. For example, for four real scalar partners, an FCC-ee $\delta\sigma_{Zh}$ measurement could exclude $m_\phi \lesssim 400 \gev$, while the 100 TeV collider could exclude 550 GeV.
\item The 100 TeV $\lambda_3$ sensitivity projections summarized in \aref{sensitivities} make a few conservative assumptions, and it is possible that the ultimate measurement achieves better $2 \sigma$ precision than 10\%. In that case, the Higgs self-coupling measurement may be superior to $\delta \sigma_{Zh}$ even for $N_r = 12$. This further motivates improved understanding of this important measurement.
\end{itemize}

\begin{figure}
\begin{center}
\hspace*{-1cm}
\includegraphics[width=9.5cm]{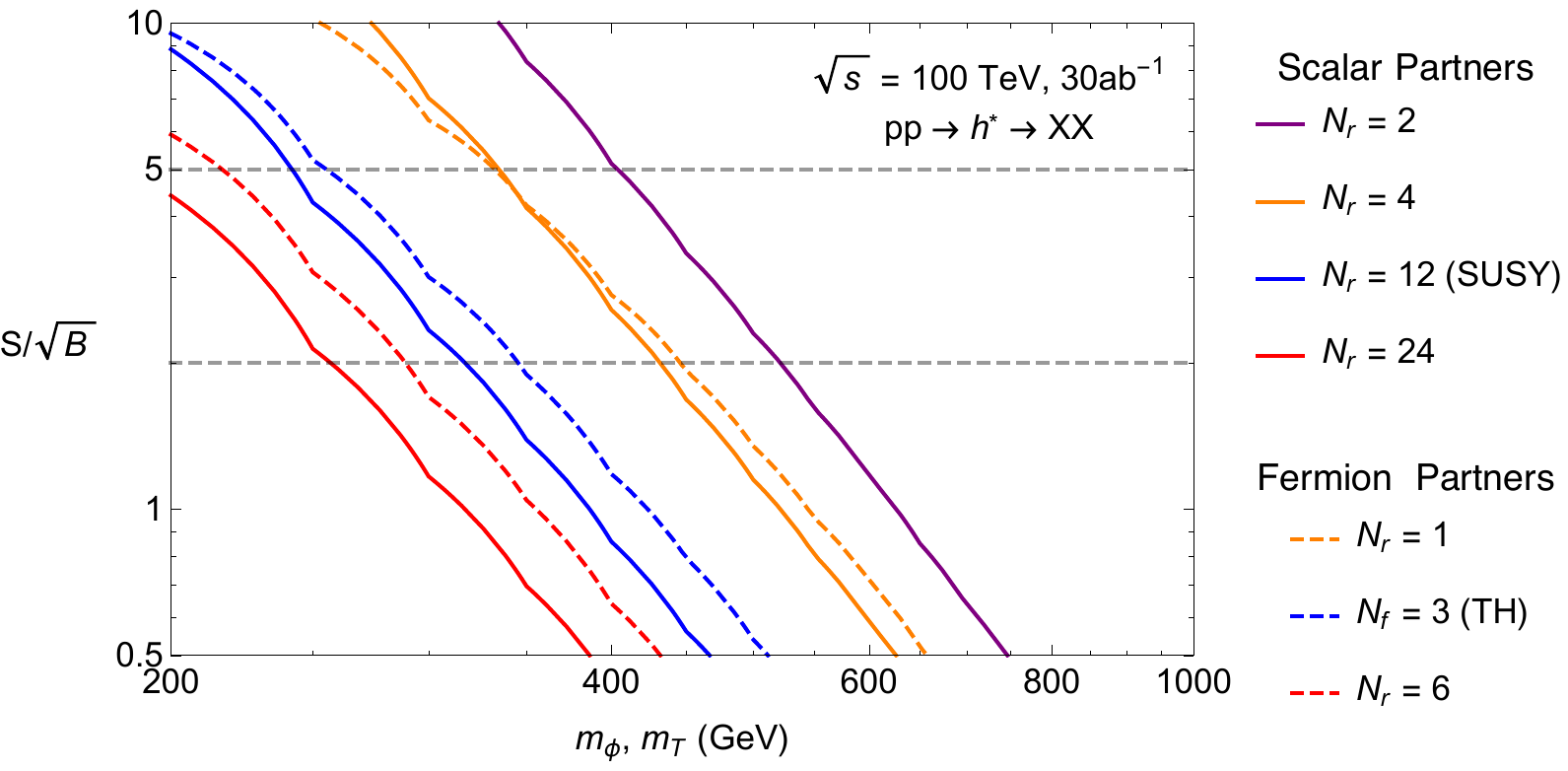}
\end{center}
\caption{Mass reach for direct production of scalar and fermion top partners at a 100 TeV collider with $30\iab$, using rescaled results of the $pp \to h^* \to SS$ analyses in~\cite{Curtin:2014jma}. See \aref{sensitivities} for details.
}
\label{f.directproduction}
\end{figure}

\begin{figure}
\begin{center}
\hspace*{-6mm}
\begin{tabular}{c}
\includegraphics[height=4.7cm]{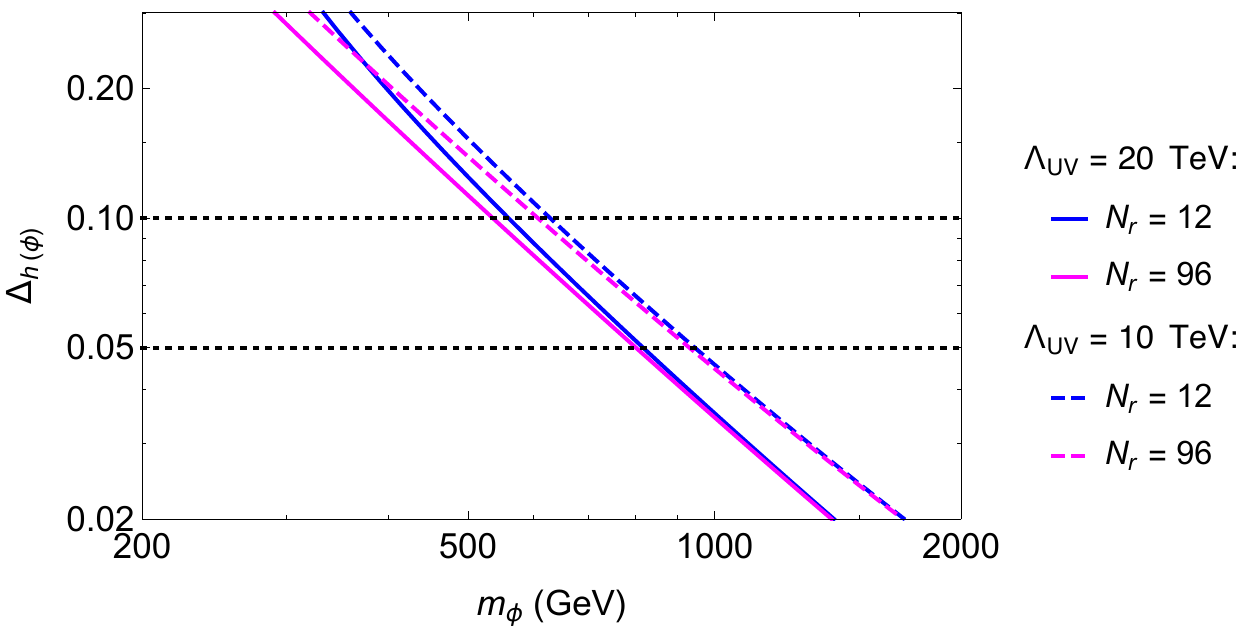}
\\
\includegraphics[height=4.7cm]{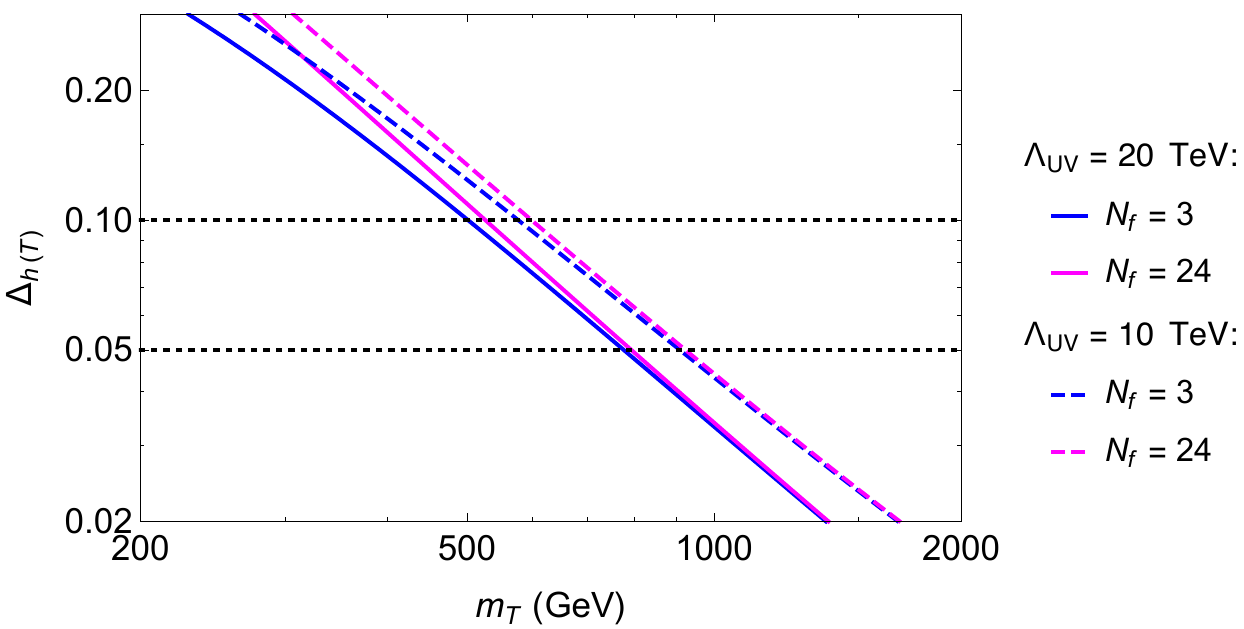}
\end{tabular}
\end{center}
\caption{
The tuning suffered by the Higgs mass due to the incomplete cancellation of top quark and top partner contributions, as a function of the physical partner mass (after EWSB). Assuming the 100 TeV collider has a mass reach of 10 or 20 TeV, any scenario would have to be at least this tuned in order to avoid direct production of SM-charged BSM states. 
The dependence on number of partners asymptotes quickly for moderate $N_r, N_f$.
Top: Scalar top partners, see \eref{DeltaTscalar}. Bottom: Fermion top partners, see \eref{DeltaT}.
}
\label{f.DeltaT}
\end{figure}

Finally, the Higgs portal allows the top partners to be produced directly. Notable sensitivity requires a search in the VBF $h^*\to \phi \phi$ production channel, which was analyzed in~\cite{Curtin:2014jma} in the context of electroweak baryogenesis and~\cite{Craig:2014lda} in the context of neutral naturalness. The two studies agree very well, and we rescale the results of~\cite{Curtin:2014jma} for different number and couplings of the scalars. \fref{directproduction} shows $S/\sqrt{B}$ as a function of top partner mass. For $N_r = 12$, the $2 \sigma$ reach is about 350 GeV with $30\iab$ of data. 

For very light neutral scalar top partners in the few 100 GeV range, there is hope of signals at both future lepton and hadron colliders. Simultaneous measurements could help narrow down the number of scalar top partners, since the signals scale differently with $N_r$. Direct production seems to have the greatest mass reach for the MSSM-like case of $N_r = 12$, but if the ultimate 95\% CL precision of the Higgs cubic coupling measurement can be significantly improved beyond the current estimate of $\sim 10\%$, then this measurement may be competitive or superior.

For heavier scalar top partners, the probes of low-energy structure offer no sensitivity. However, if the UV completion scale is low enough one can expect production of new states at a 100 TeV collider. As outlined in \ssref{strategy}, we now compute the level of tuning required to escape detection by both these experimental probes.

If there are no additional BSM states up to some scale $\Lambda_\mathrm{UV}$, then there are two possible tunings (working always to one-loop order). The first is the logarithmic tuning of the Higgs mass, which for scalar partners was defined in \eref{DeltaTscalar}. Since this affects the Higgs directly, new states with SM charges are strongly expected at scale $\Lambda_\mathrm{UV}$ to regulate the log divergence.

\fref{DeltaT} then shows the least severe level of log tuning suffered by the Higgs mass if $\Lambda_\mathrm{UV}$ is high enough to avoid detection at the 100 TeV collider. If low-energy measurements could probe scalar partners with masses up to $\sim 600 \gev$ regardless of $N_r$, then any theory with tuning better than 10\% could be experimentally probed. This can be seen as significant motivation to improve the low-energy measurements described above. However, it is not clear whether this level of sensitivity is possible.

\begin{figure}
\begin{center}
\includegraphics[height=5cm]{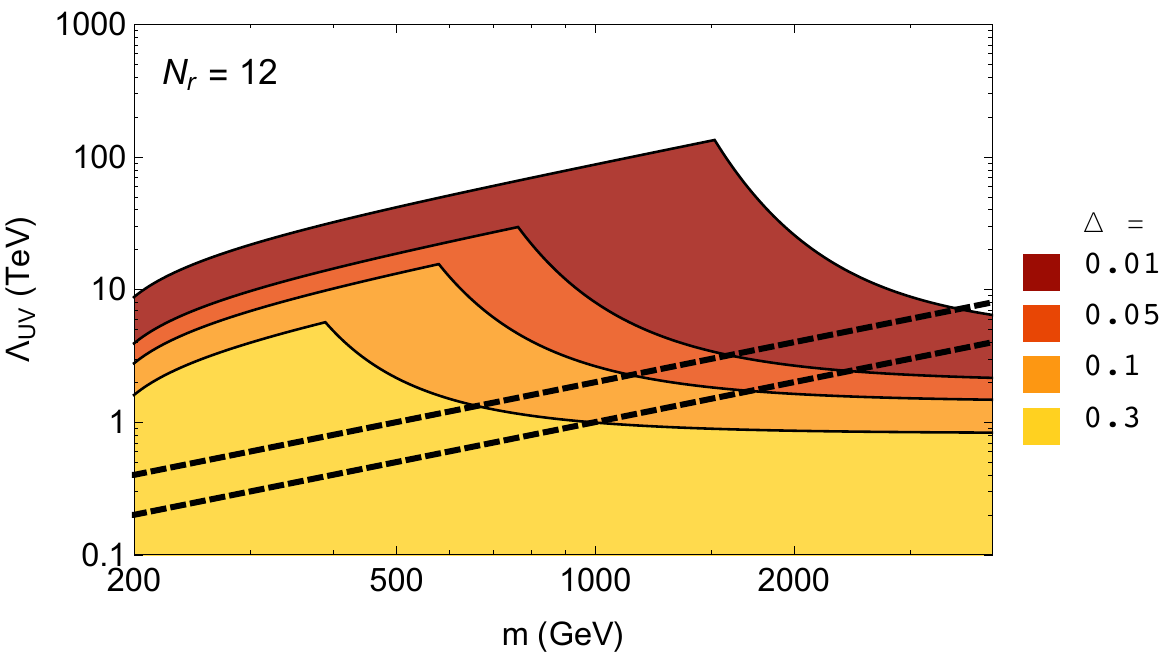}
\end{center}
\caption{
Values of $\Lambda_\mathrm{UV}$ corresponding to a given level of total  tuning $\Delta_\mathrm{total}$ as a function of partner mass (see \eref{Deltatotalscalarpartner}), assuming that corrections to the Higgs mass and the scalar partner mass are regulated at the same UV completion scale. $N_r$ is taken to be 12, analogous to the MSSM. The dashed lines indicate $\Lambda_\mathrm{UV} = m_\phi$ and $2 m_\phi$.}
\label{f.sandduneplot}
\end{figure}

\begin{figure}
\begin{center}
\begin{tabular}{c}
\includegraphics[height=5cm]{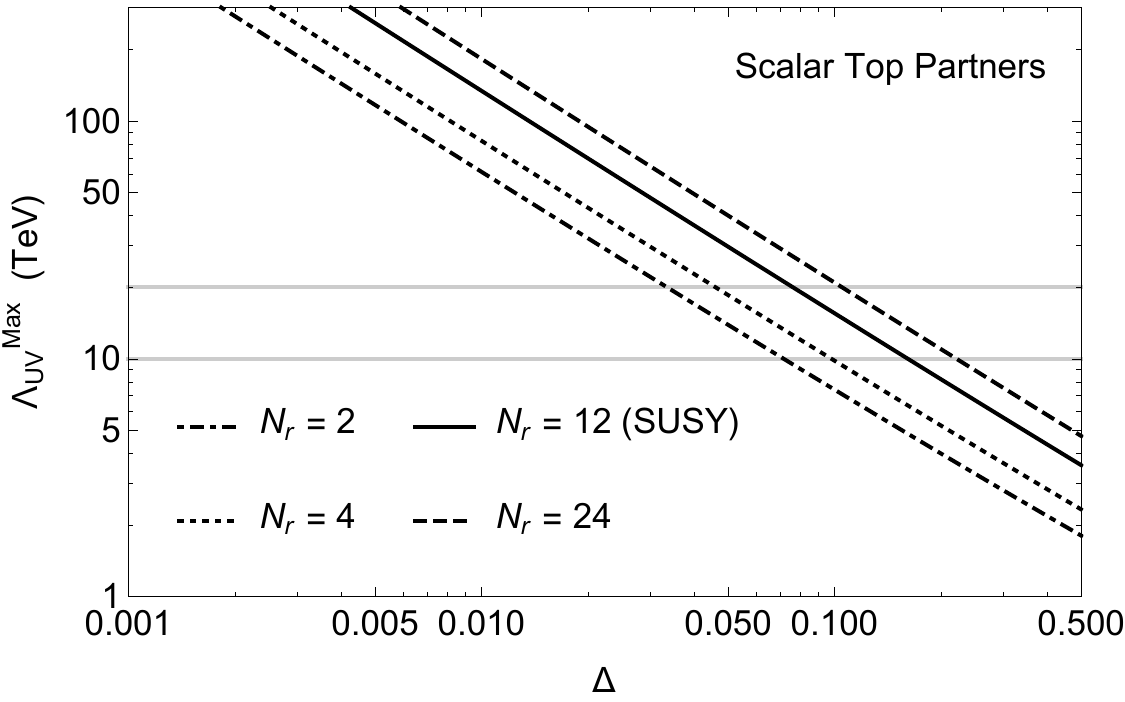}
\\
(a)
\\
\includegraphics[height=5cm]{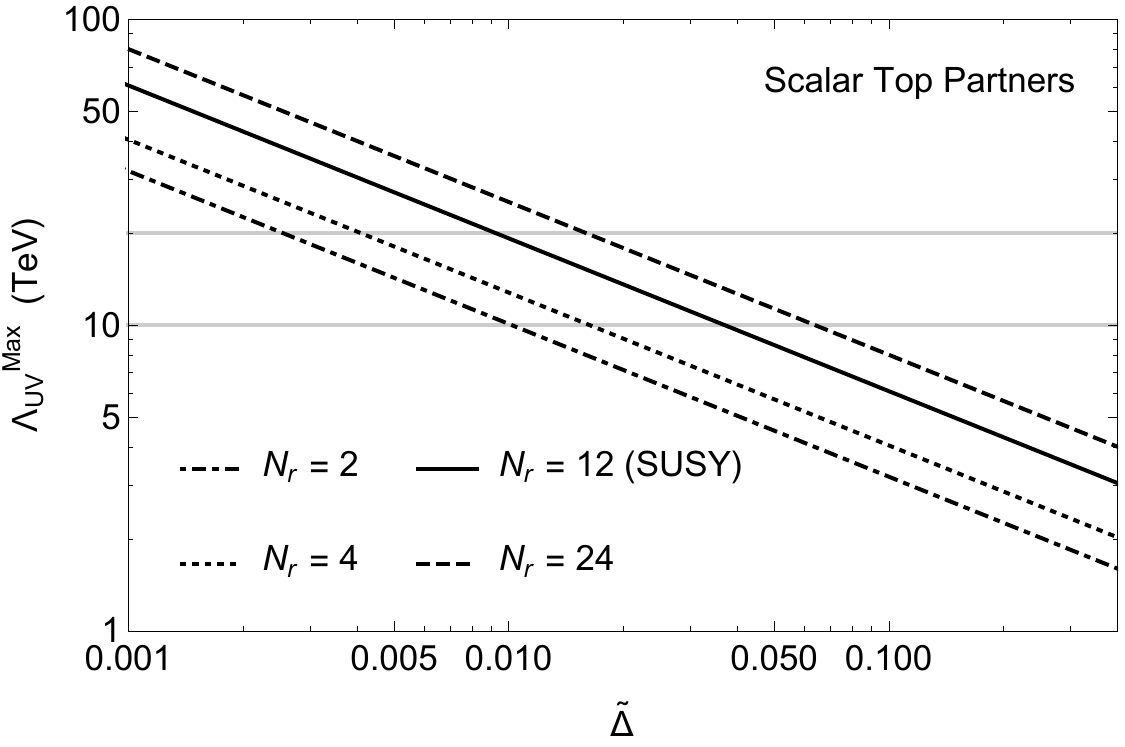}
\\
(b)
\end{tabular}
\end{center}
\caption{
The maximum $\Lambda_\mathrm{UV}$ for different numbers $N_r$ of scalar partners, as a function of allowed tuning $\Delta_\mathrm{total}$, see Eqns.~(\ref{e.Deltatotalscalarpartner}) and (\ref{e.DeltatotalscalarpartnerMULT}).
}
\label{f.scalartuning}
\end{figure}

The Higgs mass if of course not the only scalar mass in this scenario -- the partners themselves are scalars and therefore sensitive to UV physics. By our original assumption, this tuning is regulated at a scale at which new SM charges appear. 
Neglecting self-interactions (which make the tuning more severe), the tuning suffered by the scalar partner masses is
\begin{eqnarray}
\nonumber
\Delta_{\phi(h)} &=& \left(\frac{\delta \mu_{\phi,\mathrm{phys}}^2}{\mu_{\phi,\mathrm{phys}}^2}\right)^{-1} 
= \left( \frac{\lambda_\phi \Lambda_\mathrm{UV}^2}{8 \pi^2} \frac{1}{\mu_{\phi,\mathrm{phys}}^2} \right)^{-1}
\\
\label{e.Deltaphi}
&=&
\left( \frac{3 y_t^2 \Lambda_\mathrm{UV}^2}{2 N_r \pi^2} \frac{1}{\mu_{\phi,\mathrm{phys}}^2} \right)^{-1},
\end{eqnarray}
where we have imposed the naturalness condition \eref{naturalreal} in the last step. 
The overall tuning, combined conservatively as defined in \ssref{strategy}, is then 
\begin{equation}
\label{e.Deltatotalscalarpartner}
\Delta_\mathrm{total}(m_\phi, N_r, \Lambda_\mathrm{UV}) \equiv \mathrm{min}(\Delta_{h(\phi)}, \Delta_{\phi(h)}) \ ,
\end{equation}
where we keep the dependence on all new physics parameters explicit. 

We can now find the the UV completion scale $\Lambda_\mathrm{UV}$ corresponding to a given tuning $\Delta$. This is shown in  \fref{sandduneplot} for $N_r = 12$. For small $m_\phi$, the quadratic tuning of the scalar top partner mass dominates, while for larger $m_\phi$ the logarithmic Higgs tuning dominates. For very large $m_\phi$, new states are required at or below the partner mass (dashed lines), signaling a breakdown of our low-energy description.

The value of $\Lambda_\mathrm{UV}$ for a given $\Delta_\mathrm{total}$ tuning is maximized for the scalar mass where the two tunings are the same (by our conservative tuning combination). If this $\Lambda_\mathrm{UV}^\mathrm{max}$ is below the assumed mass reach $\Lambda_\mathrm{UV}^\mathrm{reach} = 10$  or $20 \tev$ at a 100 TeV collider, then we can expect the scenario to be discovered. This $\Lambda_\mathrm{UV}^\mathrm{max}$  is shown, as a function of tuning for different $N_r$, in \fref{scalartuning}(a). With 20 TeV mass reach, even 24 scalar partners should be discoverable if the tuning is better than $\sim 10\%$.

We can repeat the above exercise with the less conservative combination of tunings by multiplication:
\begin{equation}
\label{e.DeltatotalscalarpartnerMULT}
\tilde \Delta_\mathrm{total}(m_\phi, N_r, \Lambda_\mathrm{UV}) \ \equiv  \ \Delta_{h(\phi)} \ \cdot \ \Delta_{\phi(h)} \ ,
\end{equation}
The resulting maximum UV completion scale as function of $\tilde \Delta$ is shown in \fref{scalartuning}(b). By this measure, 24 scalar partners can be discovered with tuning as severe as 2\%. 

Conversely, rather than showing the \emph{maximum} size of $\Lambda_\mathrm{UV}$ corresponding to a given tuning, we can compute the 
least severe level of tuning $\Delta_\mathrm{tot}$ (or $\tilde \Delta_\mathrm{tot}$) required to avoid detection, assuming $\Lambda_\mathrm{UV}^\mathrm{reach} = 10$ or 20 TeV: see \fref{moneyplot} (top left) in \sref{conclusion}. Scalar partner theories with tuning better than 10\% should be discoverable for $N_r < 20$ $(5)$ for 100 TeV mass reach of $\sim$ 20 (10) TeV with the conservative tuning combination, and $N_r \lesssim 250$ $(40)$ with the conventional tuning combination.

\subsection{Fermionic Top Partners}
\label{ss.fermionictoppartners}

SM-neutral fermionic top partners are realized by Twin Higgs models, but our aim is to consider this scenario in a model-independent way. In the EFT description these partners interact through non-renormalizable operators of the form $|H|^2 \bar T T/2 M'$. However, this EFT requires some completion at scales not too far above $M'$. Considering only the terms in the EFT, \eref{Lfermion}, limits the conclusions we can draw both in terms of probes of low-energy observables and tuning. This motivates us to study partial UV-completions in the next sections.

Like scalars, fermionic partners can be directly produced through the Higgs portal. We can use the Higgs-interaction Lagrangian \eref{Lfermion}, together with the cancellation condition \eref{naturalfermion}, to compute the direct production cross section. Since the experimental signal is nearly identical to the pair production of neutral stable scalars, we can use the results of~\cite{Curtin:2014jma} with a cross-section rescaling to derive the fermionic top partner mass reach of a 100 TeV collider (see \aref{sensitivities} for details). The TH-like case with $N_f = 3$ can be probed in this way for top partner masses up to about 350 GeV.

On the tuning front, information seems similarly sparse. The only computable tuning is $\Delta_{h(T)}$, suffered by the Higgs mass from the incomplete top quark vs partner cancellation,  see \eref{DeltaT}. \fref{DeltaT} shows this level of tuning as a function of top partner mass and $N_f$. 

This allows us to conclude the following: if we assume that a 100 TeV collider can probe UV completion scales up to 10 or 20 TeV, then a natural theory with tuning no more severe than, say, 10\% can only escape detection if the top partner mass is less than about 500 GeV. However, direct top partner production through the Higgs portal is only sensitive to masses of a few hundred GeV. Therefore, we cannot at this point exclude the possibility that natural theories with moderately heavy fermionic top partner masses below $\sim$ 500 GeV escape detection by both low-energy probes and direct production of new states at a 100 TeV collider.

Of course, this minimal conclusion was reached using only a very limited tool-box: a single low-energy observable (direct top partner production) and a single tuning measure. This is very pessimistic compared to what is possible in the explicit Twin Higgs framework, where direct top partner production is not considered as the primary discovery channel, and a more severe tuning arises from having to adjust a soft $\mathbb{Z}_2$ breaking term to achieve $v \ll f$. For the Twin Higgs, Higgs coupling measurements are the smoking gun, giving $\sim 2 \tev$ top partner mass reach at the FCC-ee or CEPC~\cite{preCDR}. Can this be generalized?

\begin{figure}
\begin{center}
\hspace*{-6mm}
\includegraphics[height=4.7cm]{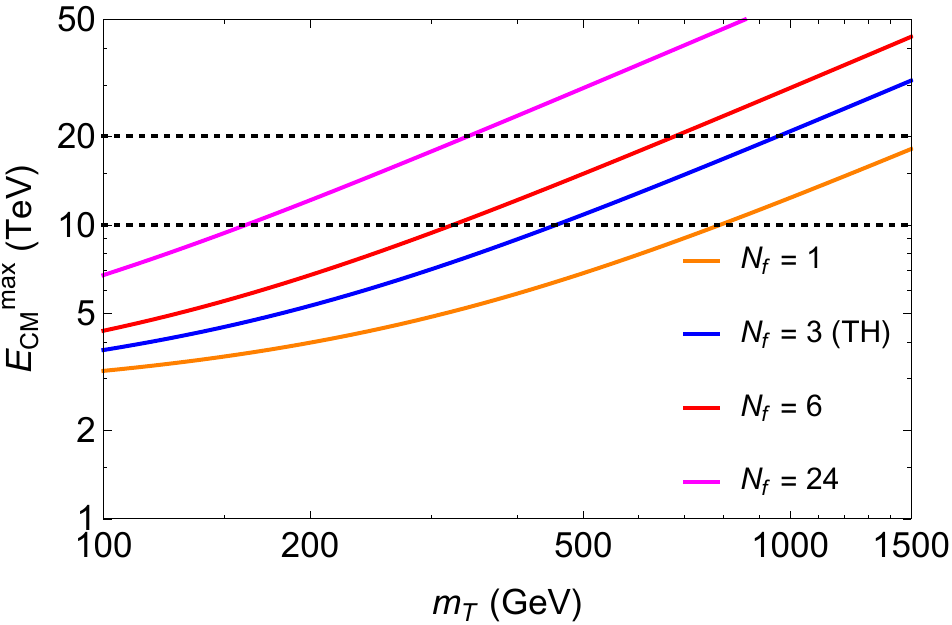}
\end{center}
\caption{
$E_\mathrm{CM}^\mathrm{max}$ as a function of $m_T$ for different numbers of fermionic top partners, see \eref{EcmMax}. New states must appear at or below this scale.}
\label{f.EcmMax}
\end{figure}

To make more progress in constraining fermionic top partners model-independently, we have to examine the nature of the non-renormalizable operator $|H|^2 \bar T T/2 M'$. We can obtain a bound on the scale $\Lambda_\mathrm{complete}$ where the EFT description of this interaction must break down using unitarity arguments. Consider the physical scattering process $h h \to \bar T T$, i.e. pair production of fermionic top partners by colliding physical Higgs bosons. 
In the high-energy limit $E_\mathrm{CM} \gg m_T, m_h$, the total cross section to first order in perturbation theory is
\begin{equation}
\sigma(h h \to \bar T T) = \frac{N_f}{32 \pi} \frac{1}{(M')^2} \ ,
\end{equation}
where we consider the total production cross section of all top partners.

The total scattering cross section cannot exceed the unitarity bound
\begin{equation}
\sigma_J \leq \frac{4 \pi}{E_\mathrm{CM}} (2J + 1) \ ,
\end{equation}
where the cross-section is decomposed into partial wave components $\sigma_J$ of angular momentum $J$. Since the matrix element has no angular dependence we can simply apply this relation to $\sigma(h h \to \bar T T)$ by setting $J = 0$. This yields
\begin{equation}
\label{e.EcmMax}
E_\mathrm{CM} \leq E_\mathrm{CM}^\mathrm{max} = \sqrt{\frac{2}{N_f}}  \ 8 \pi M'
\end{equation}
 This $E_\mathrm{CM}^\mathrm{max}$ is indicative of the highest energy for which this EFT can be valid and perturbative, so that 
\begin{equation}
\Lambda_\mathrm{complete} \leq E_\mathrm{CM}^\mathrm{max}\ .
\end{equation}
$E_\mathrm{CM}^\mathrm{max}$ is shown as a function of top partner mass and multiplicity in \fref{EcmMax}.

\begin{figure}
\begin{center}
\begin{tabular}{c}
\includegraphics[height=3.4cm]{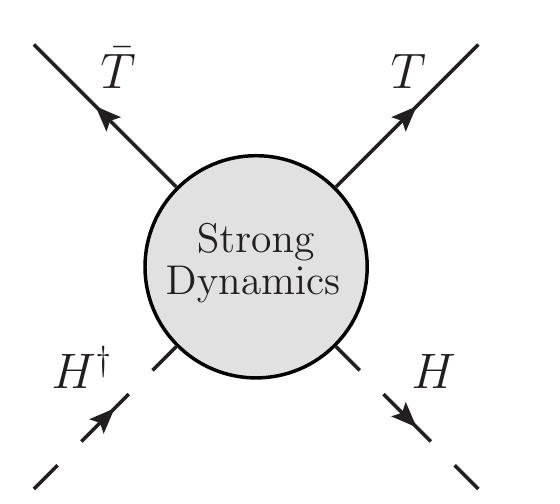}
\\
(a)
\\
\includegraphics[height=3.4cm]{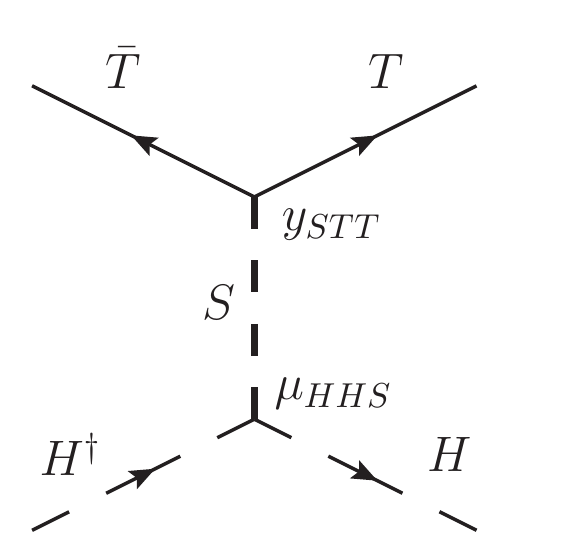}
\\
(b)
\\ \\
\includegraphics[height=2.7cm]{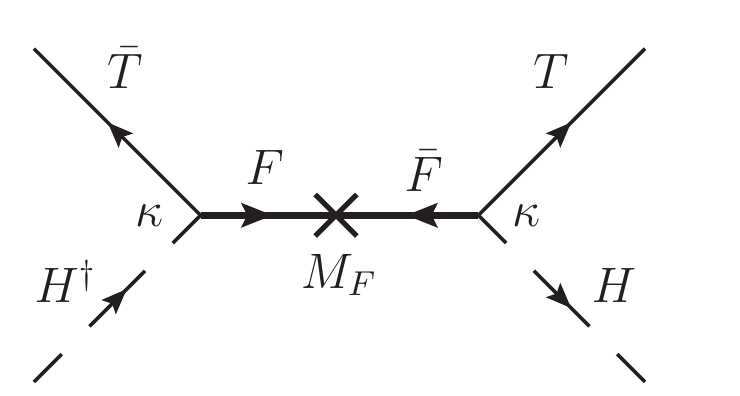}
\\
(c)
\end{tabular}
\vspace*{0cm}
\caption{ The three possibilities for generating the $|H|^2 \bar T T/2 M'$ coupling between the Higgs and neutral fermionic top partners in \eref{Lfermion}: (a) via non-perturbative physics, (b) exchange of SM-singlet scalars $S$, (c) exchange of  $SU(2)_L$ doublet fermions $F$.   
                }
\label{f.fermioncompletion}
\end{center}
\end{figure}

There are various possibilities for completing the EFT operator. One is that non-perturbative physics at or below the scale $\Lambda_\mathrm{complete}$ generates this operator, shown in \fref{fermioncompletion}(a). In this case, arguments akin to those of~\sref{nolose} indicate the presence of new SM charged states. We briefly discuss this case in \ssref{fermionicstrongcoupling}.

Alternatively, perturbative processes could generate the operator. At tree-level, there are just two options: scalar mediators or fermion mediators,  as shown in \fref{fermioncompletion}~(b) and (c). Some Twin Higgs  models realizes the scalar mediator possibility, while the perturbative fermionic mediator has not yet been proposed as part of a full theory of neutral naturalness.
These are not full UV completions, since we do not specify interactions beyond those required for the top partner to cancel the Higgs mass quadratic divergence. 
Even so, this minimal additional structure, what one might call \emph{simplified models of neutral fermionic top partners}, allows us to compute several experimental observables and additional tunings. We consider these cases separately in \ssref{fermionicscalarmediator} and \ssref{fermionicfermionmediator}.

\subsection{Fermionic Top Partner (Non-perturbative Completion)}
\label{ss.fermionicstrongcoupling}

The completion of the operator $|H| \bar T T/2 M'$ in \eref{Lfermion} could involve physics beyond 4D perturbation theory appearing at or below $E_\mathrm{CM}^\mathrm{max}$. This could include strong coupling or 5D physics (possibly being dual descriptions), both involving towers of new states. The Higgs and/or top partner could be realized as composites at this scale; in the latter case we would then expect the top quark to be composite as well, in order for the symmetry protection of the Higgs mass to remain intact. In any case, if the quantum corrections from a non-perturbative top partner sector are related by symmetry to the SM, then the SM states must experience non-perturbative interactions as well.  So in either of these scenarios, SM-charged states appear at a scale below $E_\mathrm{CM}^\mathrm{max}$, which should be discoverable at a 100 TeV collider if $E_\mathrm{CM}^\mathrm{max} \lesssim 10$ or $20 \tev$. 

Using \fref{EcmMax} we can define, for each $N_f$, a minimum top partner mass $m_T^\mathrm{min}$ that is required by unitarity if the UV completion scale is out of reach. $\Delta_{h(T)}(m_T^\mathrm{min}, N_f)$ then defines the minimum severity of tuning that has to be suffered by the theory. This is shown as a function of $N_f$ in \fref{moneyplot} (top right). 
That only way for $N_f$ fermion partners with tuning less severe than $10\%$ to avoid discovery at future colliders is if $N_f > 10$ $(2)$ for 100 TeV mass reach of $\sim$ 20 (10) TeV.

\subsection{Fermionic Top Partners (Scalar Mediator)}
\label{ss.fermionicscalarmediator}

The most obvious perturbative possibility for generating the $|H|^2 \bar T T$ interaction is via one or more scalar mediators, as shown in \fref{fermioncompletion}(a).  This case is realized by theories in which the Higgs is a pseudo-Nambu-Goldstone boson (PNGB) of some linear sigma model, including some UV completions of the Twin Higgs~\cite{Chacko:2005pe} with three partners, and the Orbifold Twin Higgs~\cite{Craig:2014aea, Craig:2014roa} for different numbers of top partners. However, our model-independent approach is even more general than these concrete theories.

\subsubsection{Minimal Lagrangian and the Sacrificial Scalar Mechanism}
\label{sss.minimallagrangiain}

\begin{figure*}
\vspace*{-4cm}
\begin{center}
\hspace*{2mm}
\vspace*{-5cm}
\includegraphics[width=19cm]{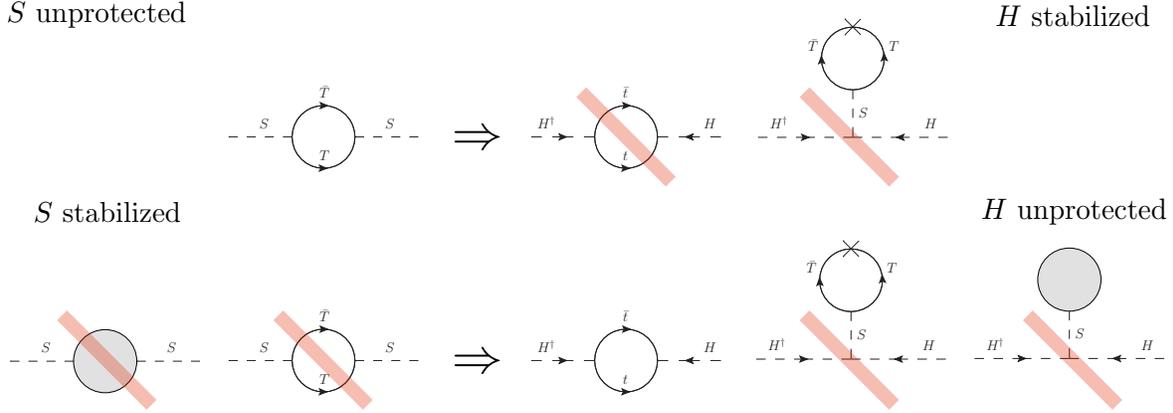}
\end{center}
\caption{
Schematic representation of the Sacrificial Scalar Mechanism in the fermionic top partner scenario with scalar mediators. Red slashes represent cancellation of the quadratically divergent loop contribution. Protecting the scalar mediator negates the cancellation of the top loop in the Higgs sector.
}
\label{f.sacrificialscalar}
\end{figure*}

We consider the general case with $N_f$ top partners $T_j$ and $N_s$ real scalar mediators $S_i$. The minimal interaction Lagrangian is
\begin{eqnarray}
\label{e.Lscalarmediator}
\mathcal{L}_\mathrm{int} &=& y_t H Q \bar U +  (M + y_{STT} S_i) \bar T_j T_j - V(H, S)  \ , \ \ \ 
\end{eqnarray}
where
\begin{eqnarray}
\nonumber
V(H,S) &=& - \mu^2 |H|^2 + \lambda |H|^4 
\\
\nonumber
&&  + \frac{1}{2} \tilde m_S^2 S_i^2  + \mu_{HHS} |H|^2 S_i + \lambda_{HS} |H|^2 S_i^2 + \ldots
\end{eqnarray}
Here we assume that the $S_i$ have been expanded around their respective  vacuum expectation values, so that $\langle S_i \rangle = 0$. We have also made the simplifying assumption that the couplings $y_{STT}, M, \tilde m_S^2, \mu_{HHS}$ and $\lambda_{HS}$ are the same for all top partners and mediators. As for the previous scenarios we discussed, this is conservative, since tunings become less severe and experimental signatures are reduced for increasing $N_f, N_s$, see \ssref{wrinkles}.
In general there will also be the additional allowed interactions like $S^3$, $S^4$, but these affect neither the various tunings we discuss nor the coefficient with which $|H|^2 \bar T T$ is generated.

Generating the $|H|^2 \bar T T$ operator in \eref{Lfermion} by integrating out the scalar gives the naturalness matching condition:
\begin{equation}
\label{e.scalarmediatormatchingcondition}
N_s \frac{\mu_{HHS} y_{STT}}{ m_S^2} = \frac{1}{2 M'}  = \frac{3}{2 N_f} \frac{y_t^2}{M_T}
\end{equation}
where the second step used \eref{naturalfermion}.
Note that $m_S$ in the above expression is the physical scalar mass at $H = 0$, to distinguish it from the tree-level parameter $\tilde m_S$.  The $\lambda_{HS}$ coupling is not fixed by matching to the fermionic top partner theory, but it has to be included in tuning discussions, as we shall see below.

What values are allowed for the physical mass $m_S$ of these singlet mediators? Assume for the moment that $N_s = 1$. (The discussion generalizes to arbitrary $N_s$). 
Intuitively, it makes sense that adding a light scalar with $m_S \sim m_h$ to generate a mechanism of stabilizing the Higgs mass merely pushes the hierarchy problem into a hidden sector. Since we want to find the irreducible experimental consequences of neutral naturalness, we would not have to consider  such a theory with an unprotected light hidden scalar, which allows us to work in the $m_S \gg m_h$ limit.

Naively, there is an obvious loophole in this argument. Why not apply any of the known solutions to the hierarchy problem to stabilize the scalar $S$ at a low mass? In that case we would have to consider the case of $m_S \lesssim m_h$. However, careful examination of the UV-sensitivities reveals that dressing up the hidden sector in such a manner actually prevents it from stabilizing the Higgs. In other words, in order to stabilize the Higgs, the scalar mediator of fermionic top partners cannot itself be stabilized. We call this the \emph{Sacrificial Scalar Mechanism}, represented schematically in \fref{sacrificialscalar}. Any new sources of UV-sensitivity, which one could introduce to cancel quadratic divergences in the singlet sector arising from $T$-loops, are also transmitted to the Higgs sector, where it negates the stabilizing effect of the top partners. This mechanism is elucidated in more detail in \aref{sacrificialscalar}.

Twin Higgs models realize the Sacrificial Scalar Mechanism very transparently. The scalar sector has two degrees of freedom: a pseudo-goldstone Higgs (which is protected by a cancellation of top and top-partner loops), and a mirror-Higgs radial degree of freedom which is unprotected and heavy, analogous to $S$. This realizes the cancellation structure in \fref{sacrificialscalar} (top). It is fascinating to see how this changes once we consider UV completions of the Twin Higgs. In the SUSY Twin Higgs~\cite{Chang:2006ra, Craig:2013fga}, for example, the light SM-like Higgs is protected by cancellations between visible and mirror sector top quark loops in the usual manner. However, at higher scales both the mirror and visible sector stop quarks appear. The visible (mirror) sector Higgs is now protected by visible (mirror) stops canceling visible (mirror) tops. More generally, the protection of the Higgs has to be `handed off' to additional degrees of freedom in the UV, so that the other scalar can be protected from low-energy top partners at higher scales by its own set of additional partners. Invoking our standard assumption, we take these new degrees of freedom to carry SM charge. 

The Sacrificial Scalar Mechanism has important consequences for our low-energy fermionic top partner scenario. 
If the singlet scalars are fairly light -- only slightly above the weak scale, say -- then their lack of protection implies new UV physics within reach of a future 100 TeV collider, or even the HL-LHC. This can be regarded as an experimentally optimistic scenario with clear discovery potential. 

However, the path to discovery is less obvious when the singlets are  significantly heavier than the weak scale, in which case new additional states may or may not be directly accessible. It is this $m_S \gg m_h$ limit which we will focus on in detail below.

\subsubsection{Experimental Signatures and Tuning}
\label{sss.exptuning}

The most important experimental consequence of the scalar mediators is that they mix with the Higgs after the latter acquires a vev. In the small mixing limit, the physical Higgs boson $\rho$ decomposes as
\begin{equation}
\rho =  c_h h - \sum_i s_{\theta_i} S_i
\end{equation}
where 
\begin{equation}
c_h \approx 1 - \frac{1}{2} \sum_i s_{\theta_i}^2
\end{equation}
and the partial mixing angles are 
\begin{equation}
\label{e.scalarmediatormixingangle}
s_{\theta_i}  =  - \frac{\mu_{HHS} \, v}{m_S^2} \  \left[1 + \mathcal{O}\left(\frac{m_h^2}{m_S^2}\right)\right]
\end{equation}
Since $m_S \gg m_h$ we can drop the higher order terms and only use the first term in the above expansion. This allows for the elimination of $\mu_{HHS}$ from \eref{scalarmediatormatchingcondition} to find the partial mixing angle required by naturalness:
\begin{equation}
\label{e.scalarmediatormatchingcondition2}
 s_{\theta_i} \ \approx \  -  \frac{3}{2 N_f N_s} \frac{y_t^2}{y_{STT}} \frac{v}{M_T}
\end{equation}
This mixing is physically observable: it reduces all couplings to the SM particles by a factor of $c_h^2$, thereby leading to a $Zh$ cross section deviation of 
\begin{equation}
\delta \sigma_{Zh} \ \approx \ - \sum_i s_{\theta_i}^2 
\  \approx \ 
-
\frac{9}{4 N_f^2 N_s}
\frac{y_t^4}{y_{STT}^2}
\frac{v^2}{M_T^2}
\end{equation}
This allows us to make definite experimental predictions. For a given physical top partner mass $m_T = M_T  - v^2/4M'$ and multiplicities $N_f, N_s$, the cross section deviation $\delta \sigma_{Zh}$ is fixed up to a choice of the unknown Yukawa coupling $y_{STT}$ of the singlet scalar to the top partner $T$.

\begin{figure*}
\begin{center}
\includegraphics[width=16cm]{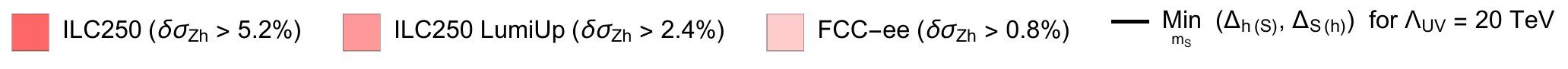}
\hspace*{-8mm}
\begin{tabular}{ccc}
\includegraphics[width=6cm]{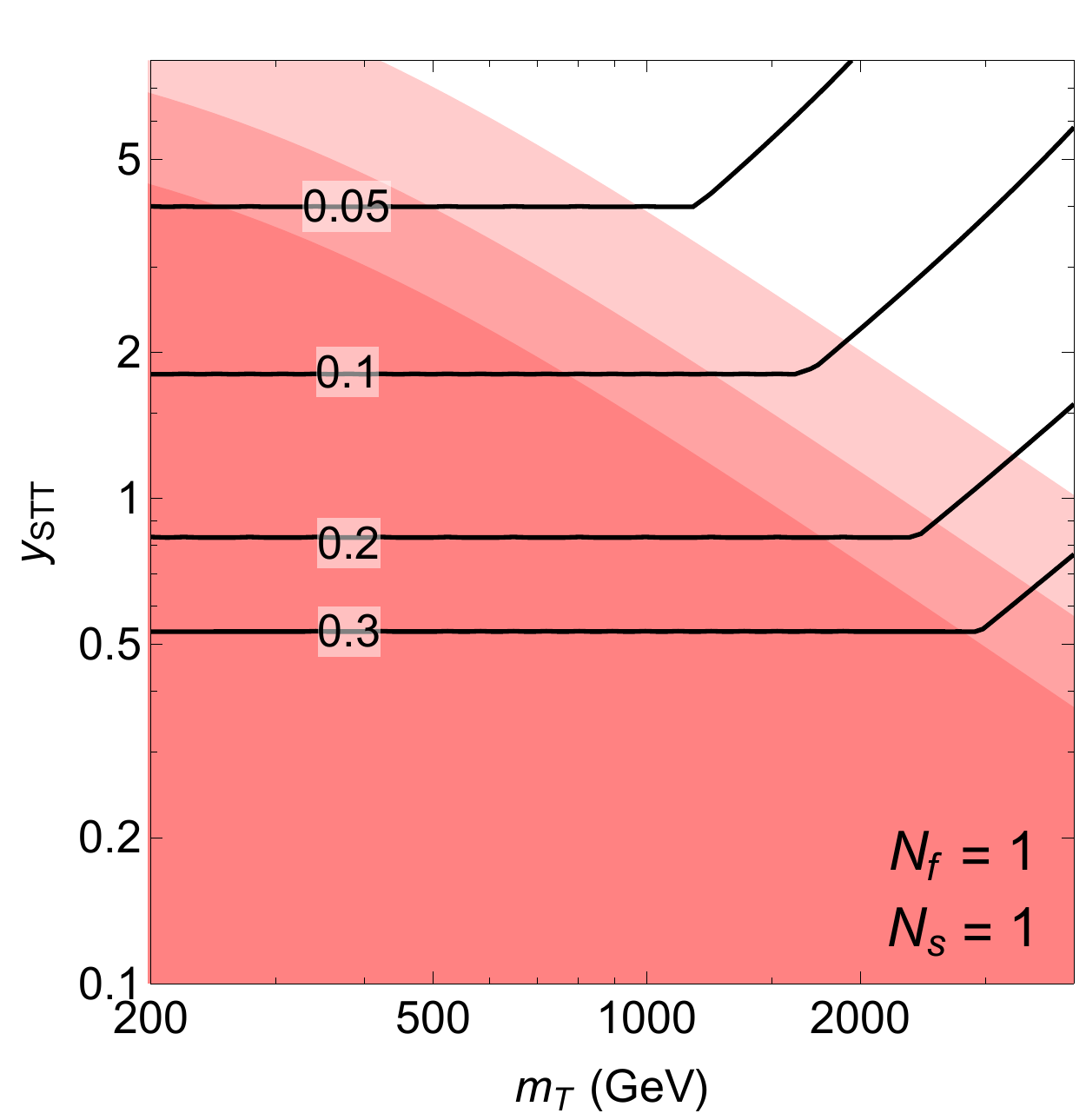}
&
\includegraphics[width=6cm]{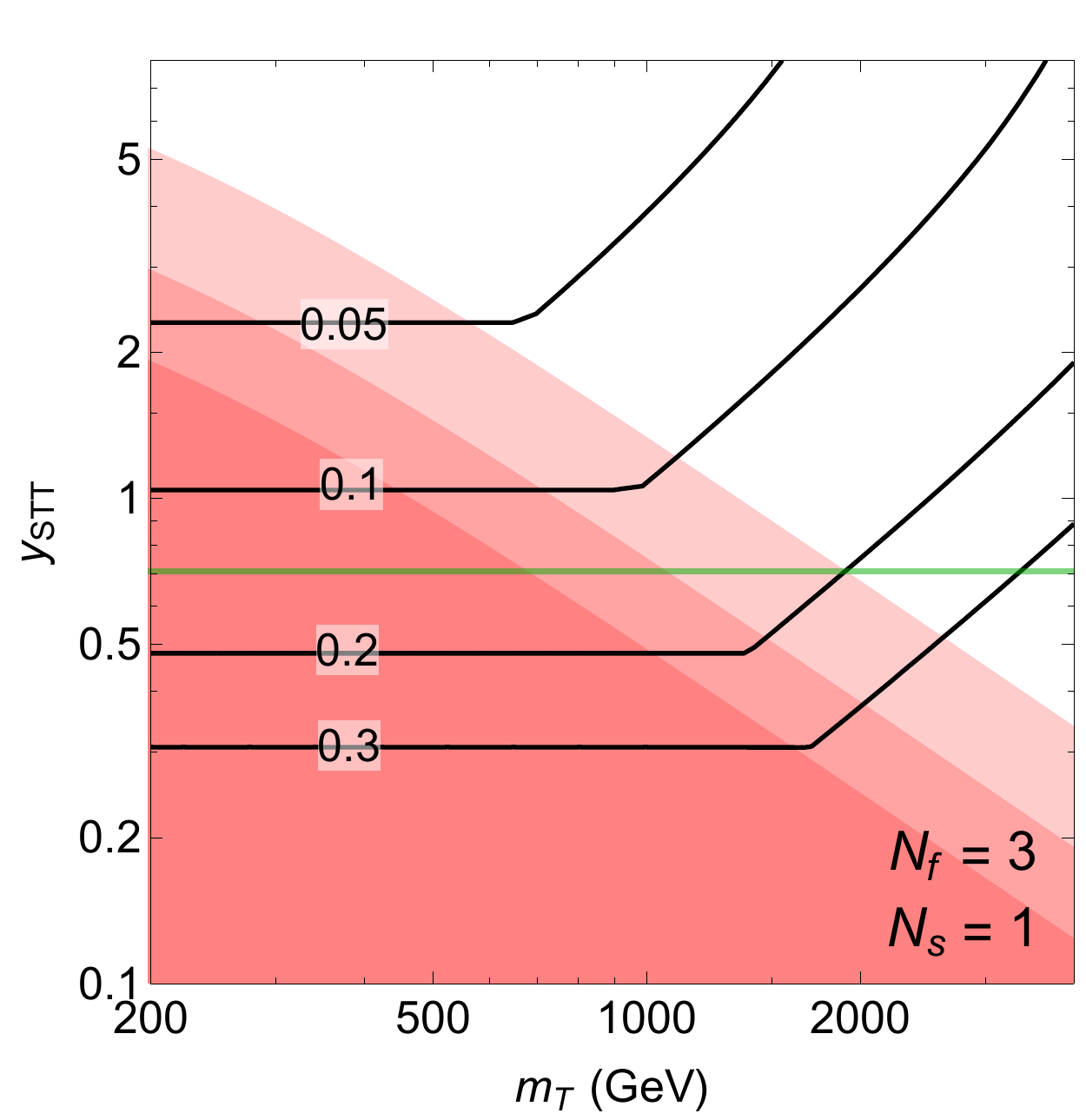}
&
\includegraphics[width=6cm]{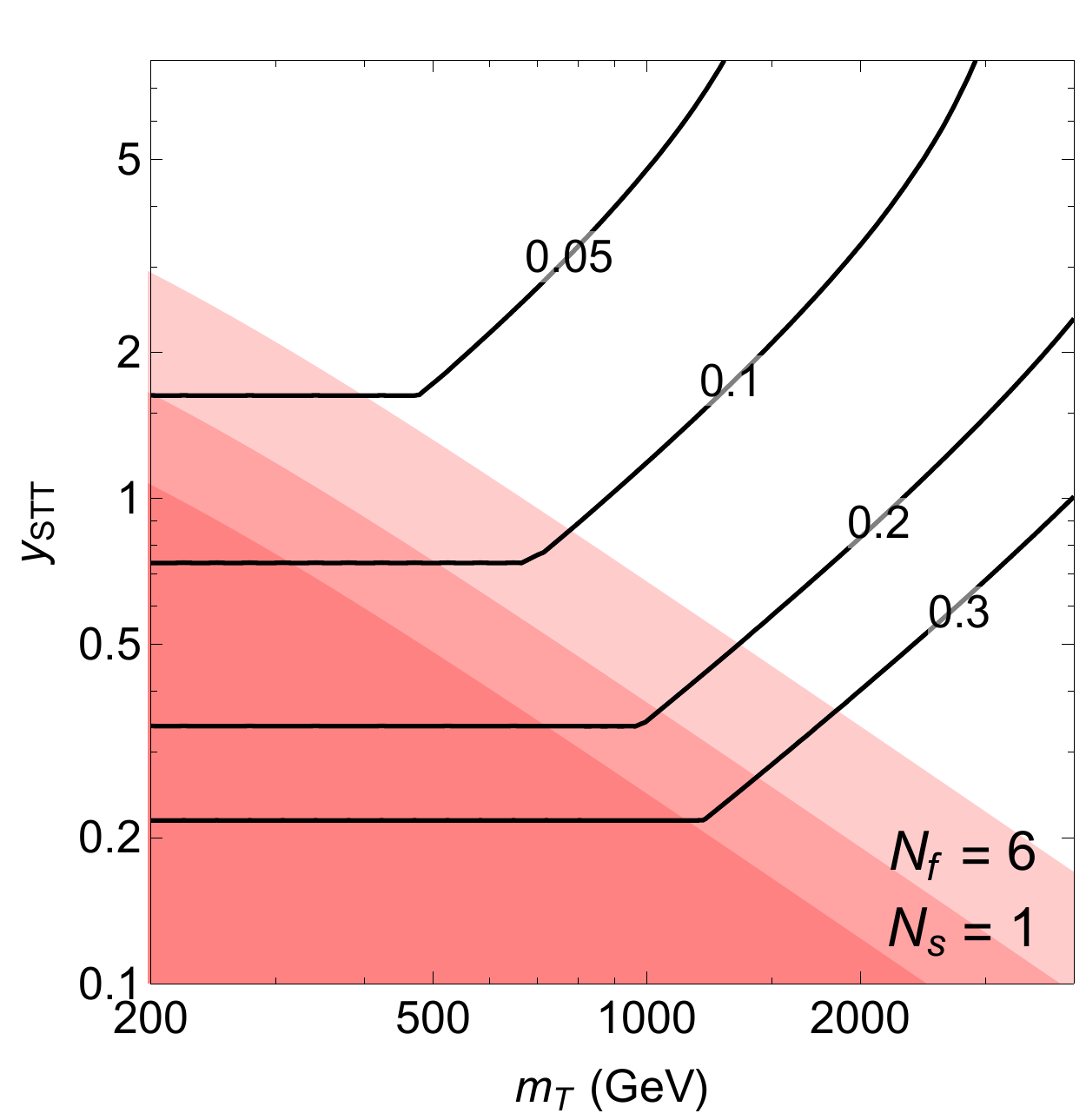}
\end{tabular}
\end{center}
\vspace*{-5mm}
\caption{
The plane of top partner mass $m_T$ and singlet  Yukawa coupling $y_{STT}$ for $N_f = 1, 3, 6$ fermionic top partners with a single real scalar mediator.
Red shaded regions can be probed by measurements of the $Zh$ cross section deviation at future lepton colliders.
Black contours show the value of the tuning associated with only the scalar mediator mass, $\Delta_\mathrm{H, S}(m_T, y_{STT}, N_f, N_s, \Lambda_\mathrm{UV} = 20 \tev)$, see \eref{DeltaHS}. Note that this does \emph{not} include the tuning $\Delta_{h(T)}$ due to the logarithmically divergent mass correction of the top partners $T$ to the Higgs mass. 
The green line in the $N_f = 3$ plot corresponds to $y_{STT} = y_t/\sqrt{2}$, which corresponds to the minimal Twin Higgs scenario. }
\label{f.fermionscalarmediatorexclusion}
\end{figure*}

\fref{fermionscalarmediatorexclusion} shows $\delta \sigma_{Zh}$ in the $(m_T, y_{STT})$-plane for  $N_f = 1,3,6$ with a single mediator $N_s = 1$. The red contours are the $95\%$ CL exclusion limits of proposed future lepton colliders. 
$(N_f, N_s) = (3,1)$ is the canonical Twin Higgs case, which translates to $y_{STT} = y_t/\sqrt{2}$ in our scenario (indicated by a green line). For a Twin Higgs type theory, the $Zh$ cross section measurement probes $\sim 2 \tev$ top partner masses. However, in our model-independent formulation $y_{STT}$ is not fixed. Increasing that Yukawa coupling decreases the observable mixing angle, allowing even relatively light fermionic top partners to escape detection at lepton colliders.

Fortunately, considering this partial UV-completion of the $|H|^2 \bar T T$ operator allows us to make more definitive statements not only about experimental observables, but also about two additional \emph{tunings} that have to be present in this scenario. 

The first tuning arises as a direct result of the Sacrificial Scalar Mechanism, which requires the physical mediator mass to receive quadratically divergent corrections from its coupling to the $T$ fermions:
\begin{equation}
m_S^2 = \tilde m_S^2 - \delta m_S^2 \  \ , \ \ \ \mathrm{where} \ \ \ \ 
\delta m_S^2 = \frac{N_f y_{STT}^2 \Lambda_\mathrm{UV}^2}{4 \pi^2} \ .
\end{equation}
and we ignore terms of lower order than $\Lambda_\mathrm{UV}^2$.
This allows us to define the tuning suffered by the mediator mass:
\begin{equation}
\label{e.DeltaS}
\Delta_{S(T)} \equiv  \left| \frac{\delta m^2_{S}}{m_S^2}\right|^{-1} \ .
\end{equation}
Minimizing the severity of this tuning favors heavy mediators and small Yukawa couplings.

The second tuning is due to the scalar sector contributing to the physical Higgs mass. We differentiate three distinct cases, based on the symmetries in the mediator sector:
\begin{enumerate}

\item The scalar mediators could act in a manner reminiscent of the radial mode in theories where the Higgs is a PNGB of some spontaneously broken symmetry $G$. In that case, $G$-invariant masses and couplings, which would include interactions like $\lambda_{HS}$ and $\mu_{HHS}$, do not generate net corrections to the Higgs potential. 
Of course, the Higgs is not a pure goldstone, and has mass and quartic terms which break $G$ explicitly, but this will not generate loop corrections of $\mathcal{O}(\mu_{hhS}^2, \lambda_{HS}, m_S^2, \ldots)$. 

The $G$-breaking quartic of the Higgs 
generates a quadratically divergent mass correction $\delta \mu^2 = 6 \lambda \Lambda_\mathrm{UV}^2 / 16 \pi^2$. In theories like Twin Higgs, this divergence is canceled by the corresponding $G$-breaking quartic in the mediator sector, which is related to $\lambda$ by a residual symmetry (like $\mathbb{Z}_2$) and is transmitted to the Higgs by integrating out the singlets in the same way as the divergence from $T$-loops. The remaining loop-divergent Higgs mass correction is
\begin{equation}
\label{e.deltamulambda}
\delta \mu^2_{(1)} = \frac{6 \lambda}{16 \pi^2} m_S^2 \log \frac{m_S^2 + \Lambda_\mathrm{UV}^2}{m_S^2} \ .
\end{equation}

\item Alternatively, one could imagine the mediator sector not respecting any symmetry protecting the Higgs mass. 
The $\lambda_{HS}$ coupling would generate a quadratically divergent contribution, so to be conservative we set it to zero. 
In that case, the log-divergent Higgs mass correction is
\begin{equation}
\label{e.deltamumuHHS}
\delta \mu^2_{(2)} =\frac{N_s}{16 \pi^2} \mu_{HHS}^2
\log \frac{m_S^2 + \Lambda_\mathrm{UV}^2}{m_S^2} \ .
\end{equation}
Applying the naturalness matching condition \eref{scalarmediatormatchingcondition} fixes $\mu_{HHS}$:
\begin{equation}
\label{e.deltamumuHHS}
\ \ \ \ \ \ \ \ 
\delta \mu^2_{(2)} =\frac{1}{16 \pi^2}
\frac{m_S^4}{4 y_{STT}^2 N_s  \ (M')^2}
\log \frac{m_S^2 + \Lambda_\mathrm{UV}^2}{m_S^2} , 
\end{equation}
where $M'$ is fixed by the choice of top partner mass and $N_f$.

\item 
One might imagine an exotic  `worst of both worlds' possibility. Start with a scalar sector which contains a \emph{true} goldstone Higgs associated with purely spontaneous breaking of a symmetry $G$. Now add a hard breaking quartic $\lambda = \lambda^{\, \cancel{G}}$ for the Higgs without any residual symmetry that implies a corresponding $G$-breaking quartic to the radial mode(s). (The Higgs mass $\mu^2$ is a soft breaking parameter and does not affect this discussion.) In that case, it seems that the singlet sector neither contributes to nor regulates the quadratic divergence of the Higgs mass $\delta \mu^2 \sim \lambda^{\, \cancel{G}} \Lambda_\mathrm{UV}^2$. One could then add some neutral ``$H$-partners'' (akin to top partners) to cancel this quadratic divergence independently of $m_S$. 

Throughout this paper, we have ignored two-loop divergences under the assumption that they are either subdominant or regulated by the same symmetry which cancels one-loop effects. In this case, however, it appears that the cancellation has to break down due to large two-loop effects.

The goldstone nature of the Higgs requires the $G$-conserving quartic to be $\lambda^G = N_s \mu_{HHS}^2/m_S^2$. Some two-loop diagrams involving both $\lambda^G$ and $\lambda^{\, \cancel{G}}$ cannot be canceled by either the scalar sector nor the $H$-partners. Together with the naturalness condition \eref{scalarmediatormatchingcondition}, this leads to a quadratically divergent two-loop contribution to the Higgs mass:
\begin{equation}
\delta \mu^2_{(3)} \sim \frac{1}{(16 \pi^2)^2} \  \lambda^{\, \cancel{G}} \ \frac{m_s^2}{M_T^2} \ \Lambda_\mathrm{UV}^2
\end{equation}
which, in the relevant parameter space, is either larger or at most comparable to $\delta \mu_{(1)}^2$. We have checked that including this possibility does not change our results. Given this scenario's highly exotic nature, we do not include it in the discussion below.

\end{enumerate}
It is now possible to conservatively define the additional tuning suffered by the Higgs mass:
\begin{equation}
\label{e.natambdaHS}
\Delta_{h(S)} \equiv \left| \frac{\mathrm{Min}\left[\delta \mu_{(1)}^2, \delta \mu_{(2)}^2 \right]}{\mu_\mathrm{phys}^2}\right|^{-1} \ .
\end{equation}
Unlike $\Delta_{S(T)}$, reducing the severity of this tuning favors \emph{lighter} mediators. 

As per our assumptions the mediator mass is not experimentally accessible, but the opposite dependencies of these two tunings on $m_S$ will set a lower bound on the severity of total tuning. Using the conservative tuning combination in \eref{Deltatotal} and making all remaining parameter dependencies explicit, this gives 
\begin{eqnarray}
\label{e.DeltaHS}
\Delta_\mathrm{H, S}(m_T, y_{STT}, N_f, N_s, \Lambda_\mathrm{UV}^\mathrm{reach}) 
\ \ \ \ \ \ \ \ 
&&
\\
\nonumber
\\ \nonumber 
\ \ \ \ \ \ \ \ 
= \underset{m_S}{\mathrm{Max}}\left[ \mathrm{Min} (\Delta_{h(S)}, \Delta_{S(T)}) \right] \ . &&
\end{eqnarray}
For each value of parameters, the mass $m_S$ is chosen to give the least severe tuning, i.e. we maximize $ \mathrm{Min} (\Delta_{h(S)}, \Delta_{S(T)}) $. 
Note that we have named this tuning $\Delta_{H,S}$ instead of $\Delta_\mathrm{total}$ since it does not yet include the log-divergent top and top-partner contribution to the Higgs mass, $\Delta_{h(T)}$ of \eref{DeltaT}. 

Assuming, for the moment, that a UV-completion scale of $\Lambda_\mathrm{UV} = 20 \tev$ can be detected by direct production of new states at a 100 TeV collider, evaluating
 $\Delta_\mathrm{H, S}(m_T, y_{STT}, N_f, N_s, \Lambda_\mathrm{UV}^\mathrm{reach} = 20 \tev)$ gives the minimum level of tuning \emph{from contributions by the scalar sector} that has to be suffered by the theory if it is to avoid detection.  We  show contours of this tuning $\Delta_\mathrm{H, S}$ for $N_s = 1$ and $N_f = 1,3,6$ in \fref{fermionscalarmediatorexclusion}.

From \fref{DeltaT} we know that fermionic top partners heavier than about 500 GeV lead to either tuning $\Delta_{h(T)}$ worse than 10\%, or a UV-completion scale low enough for detection at a 100 TeV collider. \fref{fermionscalarmediatorexclusion} shows that, for $(N_f, N_s) = (3,1)$ for example, top partner masses \emph{lighter} than about 800 GeV lead to tuning worse than 10\% from scalar contributions alone, if the theory is to avoid detection at a 100 TeV collider or through measurable $Zh$ cross section deviations. Therefore, we can immediately see that \emph{in the TH-type scenario of three top partners and one mediator, the theory has to either produce experimental signatures at some future collider, or contain tunings worse than 10\%!}

We now take the program outlined in \ssref{strategy} to its full conclusion and define the least severe \emph{total} tuning possible for this scenario to avoid all experimental detection. The conservative tuning combination gives:
\begin{eqnarray}
\label{e.Deltamaxtotalscalarmediator}
&&\Delta_\mathrm{total}^\mathrm{max}(N_f, N_s, \Lambda_\mathrm{UV}^\mathrm{reach})
= 
 \\ 
 \\
 &&
 \nonumber 
\underset{
\begin{array}{c}
\scriptstyle m_S, \scriptstyle m_T, y_{STT}
\\
\scriptstyle \delta \sigma_{Zh} < 0.8\%
\end{array}
}{\mathrm{Max}} \Big[ \mathrm{Min} (\Delta_{h(S)}, \Delta_{S(T)}, \Delta_{h(T)}) \Big]
\end{eqnarray}
The underscripts below $\mathrm{Max}$ indicate that this is the least severe tuning possible for a given $\Lambda_\mathrm{UV}^\mathrm{reach}$, with respect to all possible choices of $m_S, m_T, y_{STT}$ for which the $Zh$ cross section deviation is less than 0.8\%, i.e. regions of parameter space that \emph{cannot} be probed by FCC-ee. This $\Delta_\mathrm{max}^\mathrm{tot}$ is shown in \fref{Deltatotmaxscalarmediator} (top) as a function of $(N_f, N_s)$ for $\Lambda_\mathrm{UV}^\mathrm{reach} = 10$ and 20 TeV, corresponding to two different assumptions on what scale of UV completion the 100 TeV collider can directly access. 

The entire exercise is easily repeated for the more conventional way of  combining tunings by defining
\begin{equation}
\label{e.DeltamaxtotalscalarmediatorMULT}
\begin{array}{l}
\tilde \Delta_\mathrm{total}^\mathrm{max}(N_f, N_s, \Lambda_\mathrm{UV}^\mathrm{reach})
=
\\ \\
\underset{
\begin{array}{c}
\scriptstyle m_S, \scriptstyle m_T, y_{STT}
\\
\scriptstyle \delta \sigma_{Zh} < 0.8\%
\end{array}
}{\mathrm{Max}} \Big[ 
\left(\Delta_{h(S)}^{-2} + \Delta_{h(T)}^{-2} \right)^{-1/2} \ \cdot \ \Delta_{S(T)}
 \Big]
\end{array}
\end{equation}
This is shown in \fref{Deltatotmaxscalarmediator} (bottom) as a function of $N_f$ and $N_s$.

\begin{figure}
\begin{center}
\begin{tabular}{c}
\includegraphics[width=6cm]{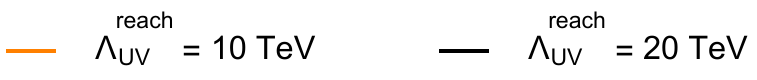}
\\
\includegraphics[width=7cm]{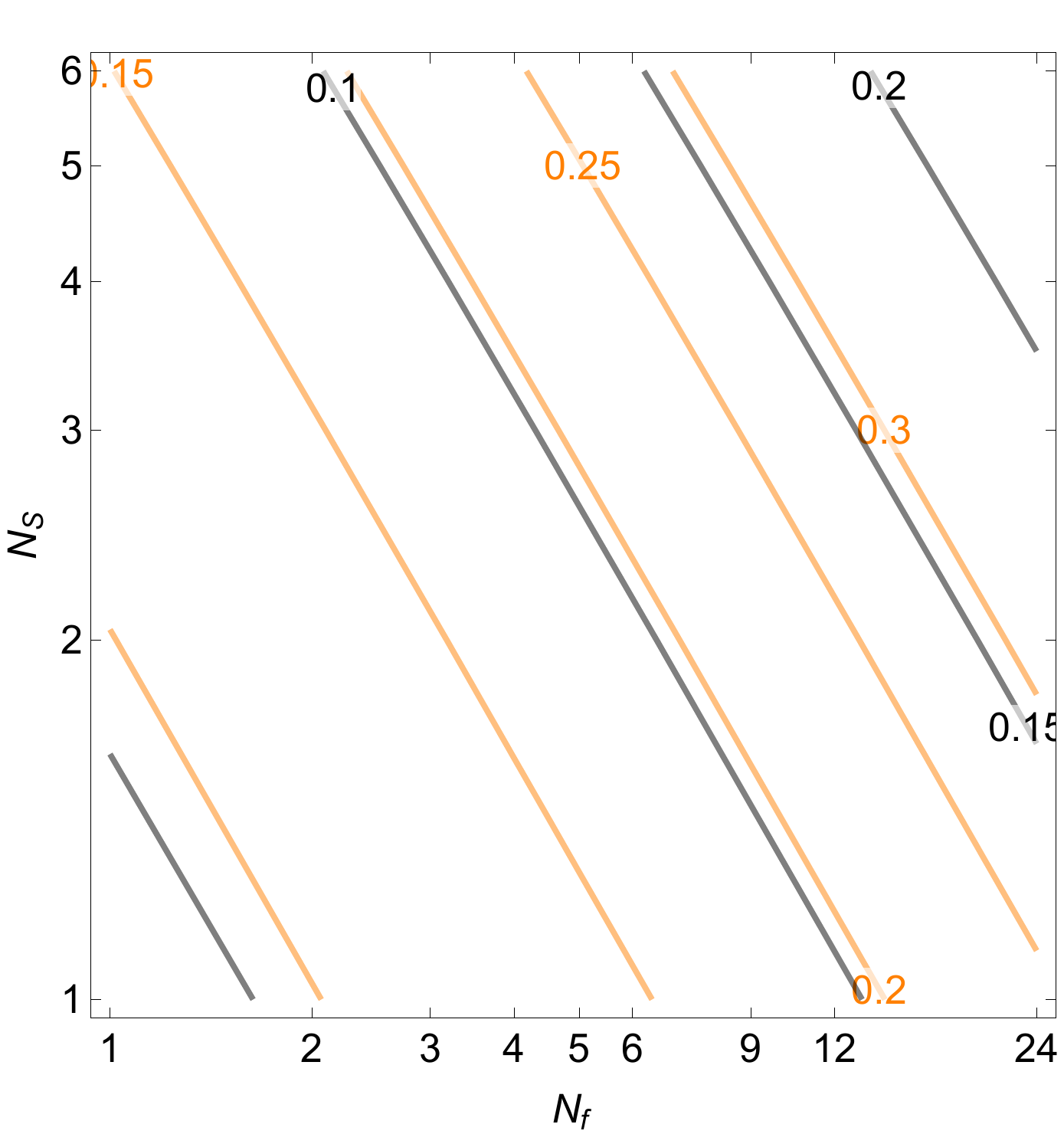}
\\
\includegraphics[width=7cm]{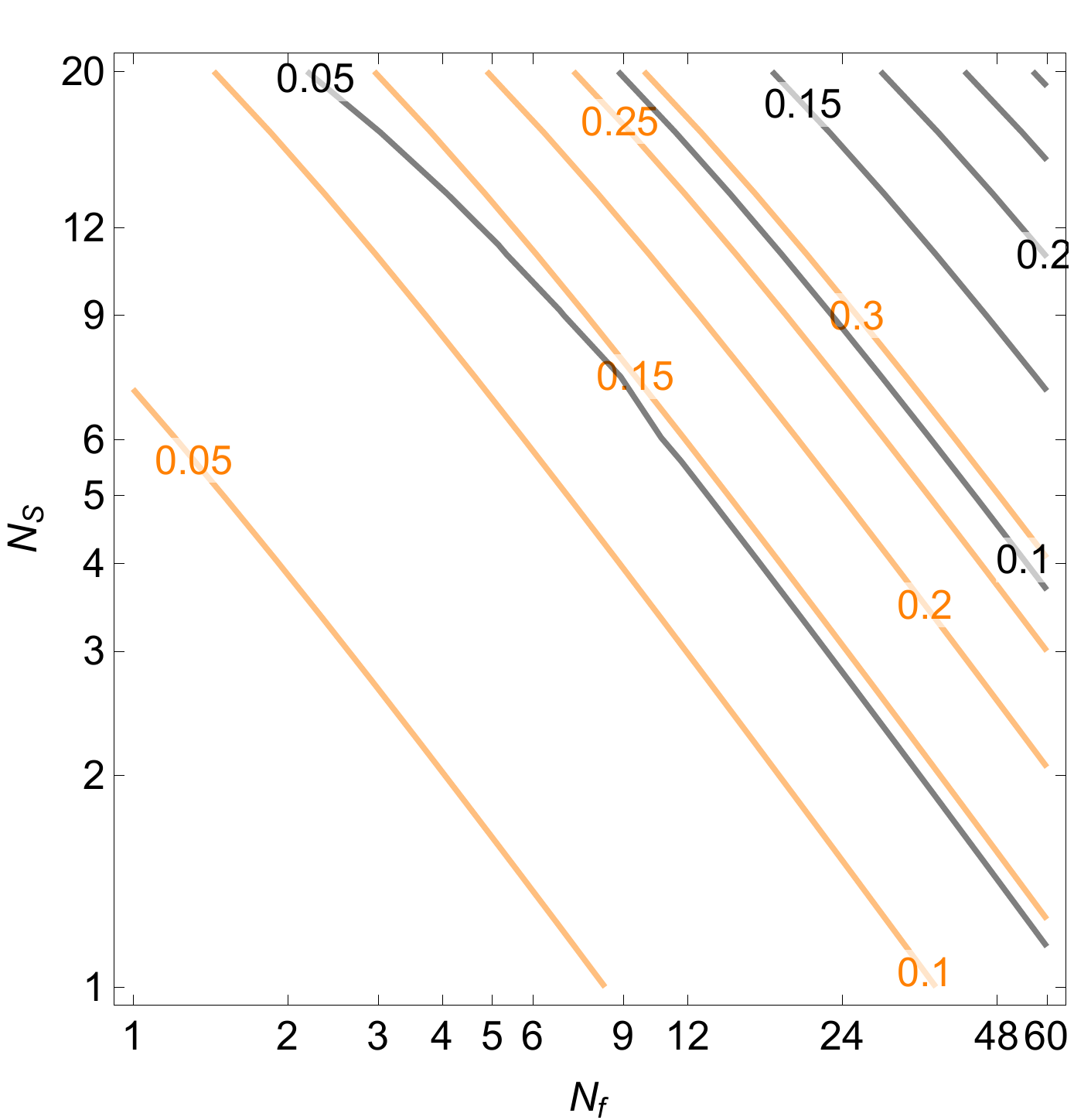}
\end{tabular}
\end{center}
\caption{
Top: Contours of the conservative tuning combination $\Delta_\mathrm{total}^\mathrm{max}$ in the scenario with $N_f$ fermionic top partners and $N_s$ scalar mediators, see \eref{Deltamaxtotalscalarmediator}. 
Bottom: The conventional tuning combination $\tilde \Delta_\mathrm{total}^\mathrm{max}$, see \eref{DeltamaxtotalscalarmediatorMULT}.
Each tuning is computed for $\Lambda_\mathrm{UV} = $ 10 and 20 TeV.
Assuming the 100 TeV collider can probe new SM-charged states at these respective scales, this is the least severe level of tuning that has to be suffered by a theory to avoid experimental detection at both future lepton colliders (through $\delta \sigma_{Zh}$ measurements at FCC-ee) and the 100 TeV collider (through direct production of new states that are part of the UV completion). }
\label{f.Deltatotmaxscalarmediator}
\end{figure}

\fref{Deltatotmaxscalarmediator} suggests that both $\Delta_\mathrm{total}^\mathrm{max}$ and $\tilde \Delta_\mathrm{total}^\mathrm{max}$ depend mostly on the product $N_f \cdot N_s$. Indeed, numerically we find:
\begin{equation}
\Delta_\mathrm{total}^\mathrm{max} \approx 
\left\{
\begin{array}{lll}
0.076 \ (N_f N_s)^{0.37} 
& \mathrm{for}
& \Lambda_\mathrm{UV}^\mathrm{reach} = 10 \tev
\\ \\
0.040 \ (N_f N_s)^{0.37} 
& \mathrm{for}
& \Lambda_\mathrm{UV}^\mathrm{reach} = 20 \tev
\end{array}
\right.
\end{equation}
and
\begin{equation}
\tilde \Delta_\mathrm{total}^\mathrm{max} \approx 
\left\{
\begin{array}{lll}
0.014 \ (N_f N_s)^{0.55} 
& \mathrm{for}
& \Lambda_\mathrm{UV}^\mathrm{reach} = 10 \tev
\\ \\
0.0040\ (N_f N_s)^{0.60} 
& \mathrm{for}
& \Lambda_\mathrm{UV}^\mathrm{reach} = 20 \tev
\end{array}
\right.
\end{equation}
to a good approximation in the relevant range $\Delta \lesssim 0.3$. This is shown in \fref{moneyplot} (bottom left) in \sref{conclusion}.

\begin{figure*}
\begin{center}
\vspace*{-3mm}
\includegraphics[width=3.5cm]{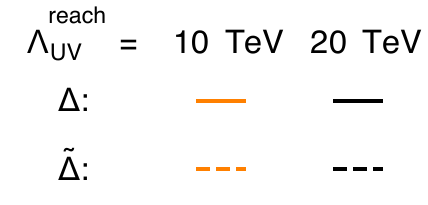}
\vspace{3mm}
\begin{tabular}{cc}
\includegraphics[width=7.6cm]{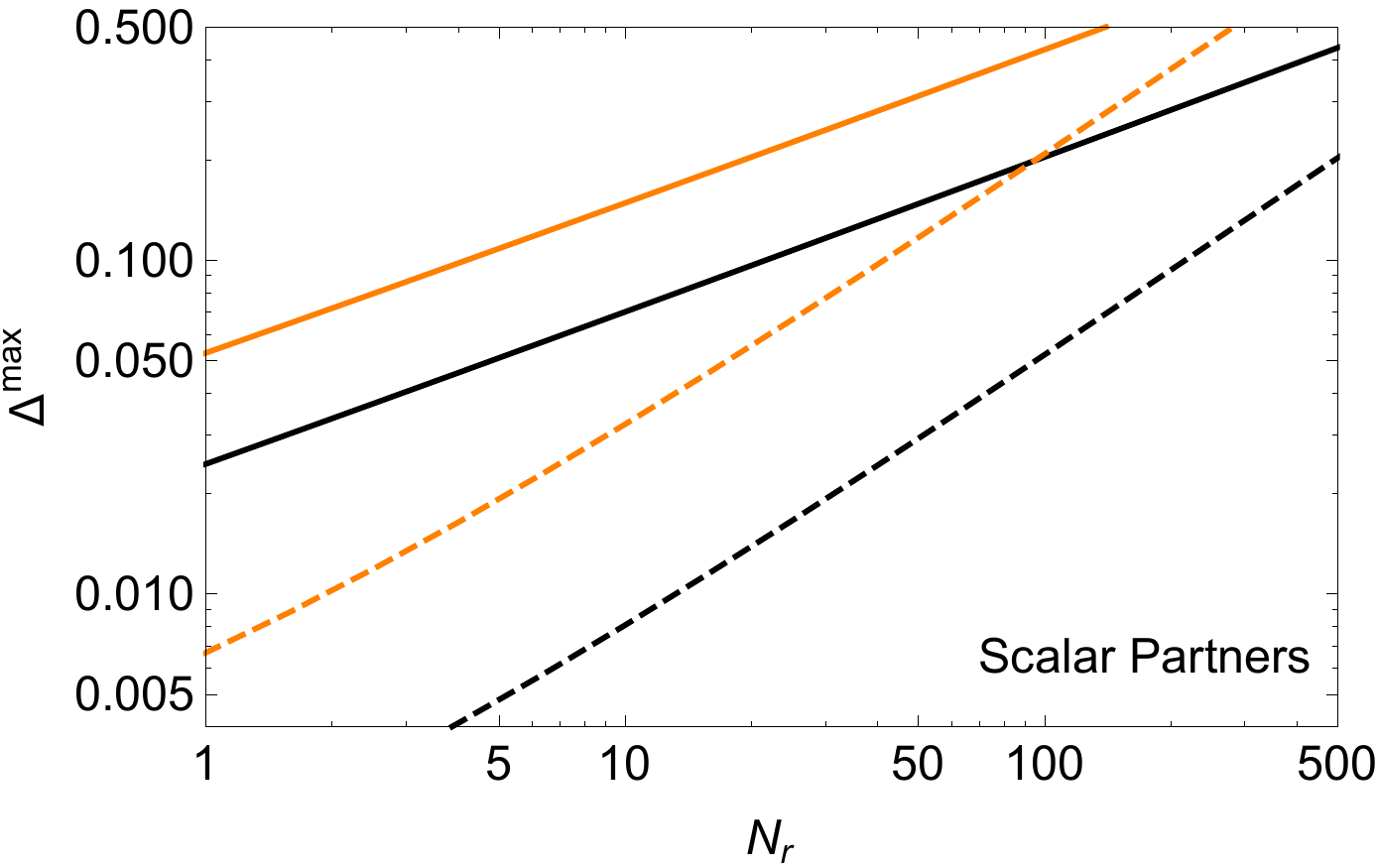}
&
\includegraphics[width=7.6cm]{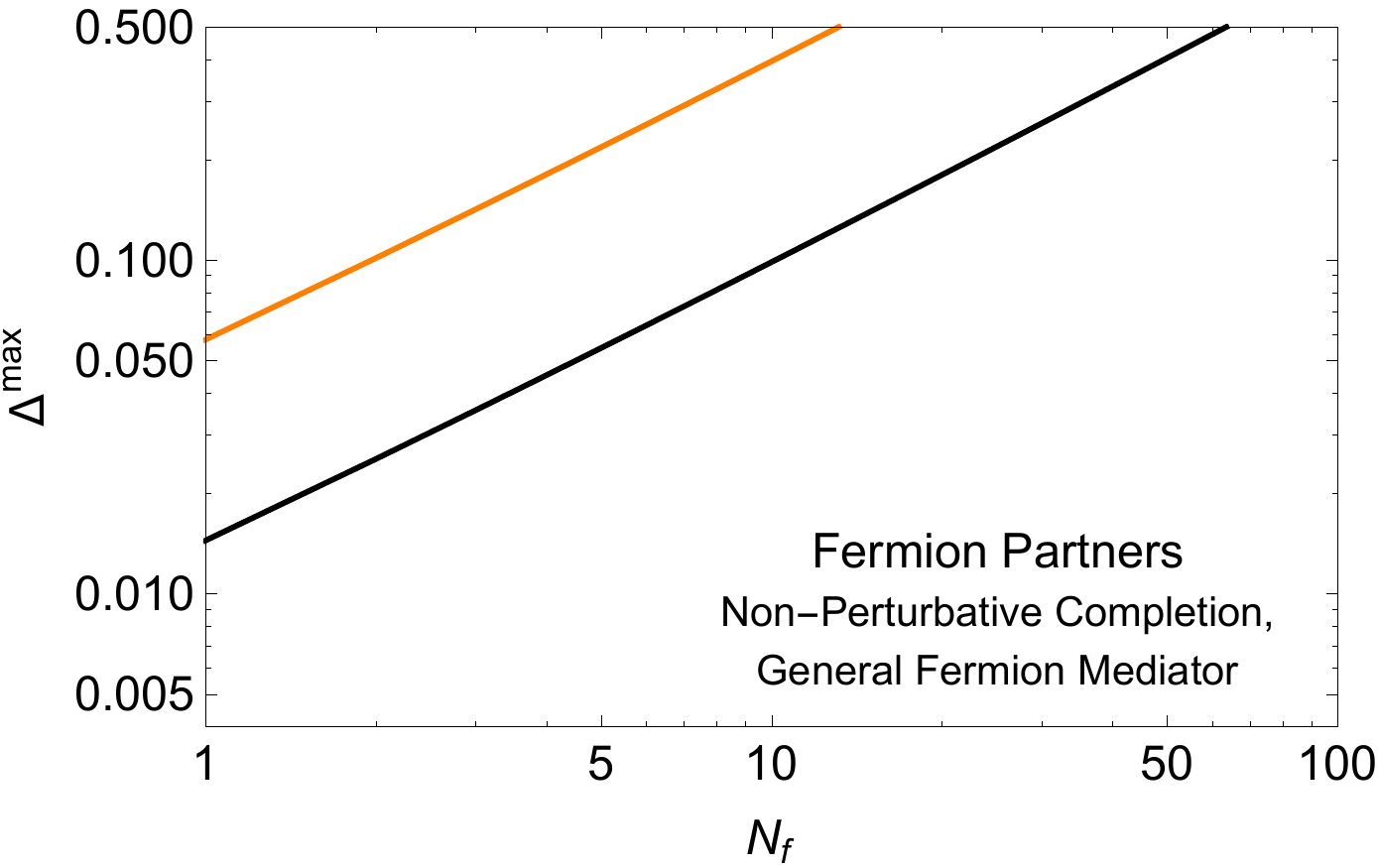}
\\
\includegraphics[width=7.4cm]{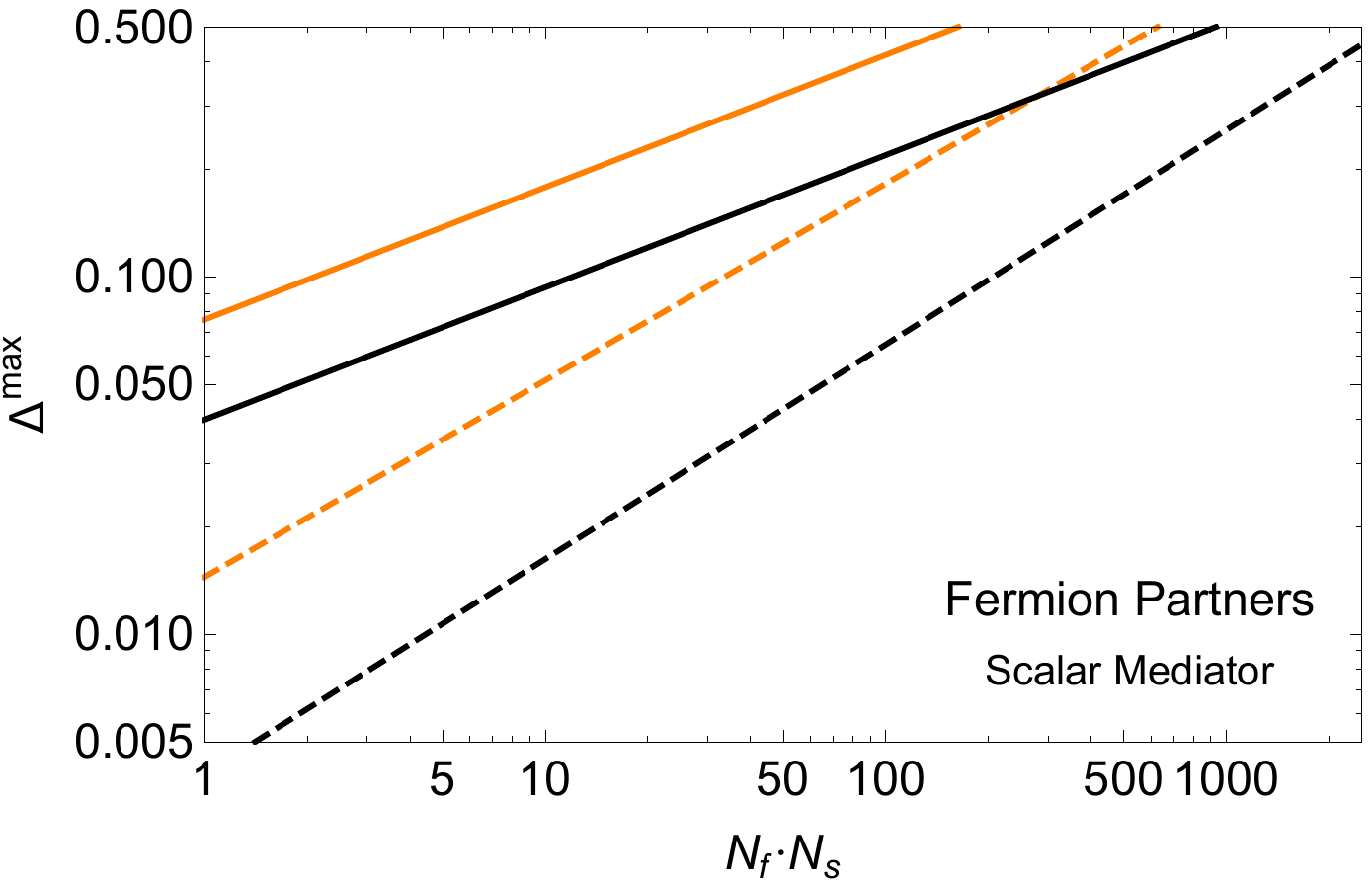}
&
\includegraphics[width=7.6cm]{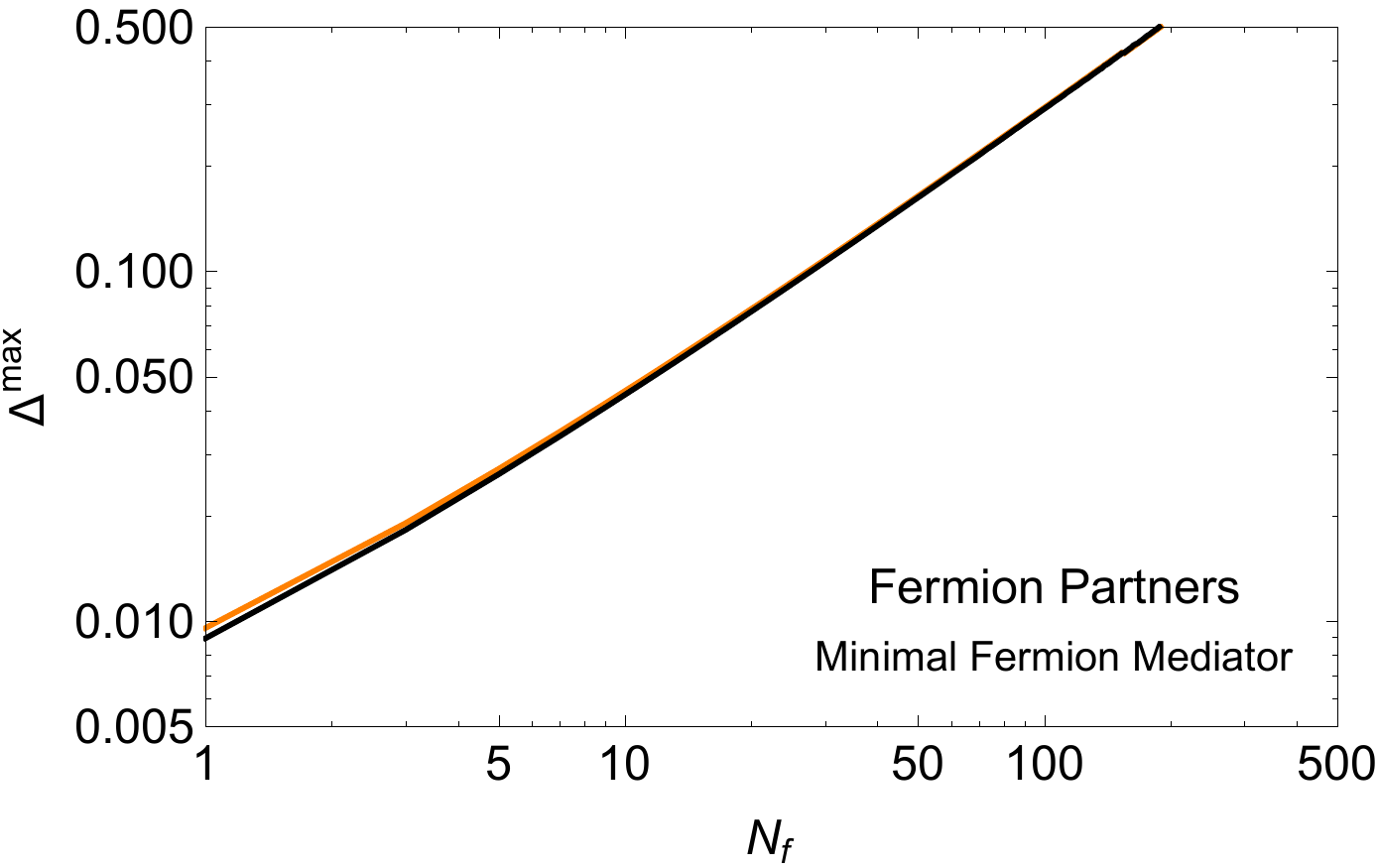}
\end{tabular}
\vspace*{-6mm}
\end{center}
\caption{
Minimum severity of tuning required for a given neutral top partner scenario to escape all experimental probes, i.e. no detectable probes of low-energy structure ($Zh$ cross section deviation, Higgs cubic coupling, direct partner production, and electroweak precision observables) at FCC-ee and a 100 TeV collider, and no direct production of SM-charged BSM states that are part of the UV completion, for assumed 100 TeV collider mass reach of 10 (orange) and 20 TeV (black). The solid lines represent the maximally conservative combinations of tunings $\Delta_\mathrm{tot} = \mathrm{Min}[\Delta_i]$, the dashed lines (for scenarios where more than two tunings exist) represent the slightly less conservative combination of tunings (multiplication and adding in inverse quadrature, as explained in \ssref{strategy}). $N_r$ is the number of real scalar top partners (MSSM $= 12$), $N_f$ the number of four-fermion top partners (TH $= 3$), and $N_s$ the number of real scalar mediators in the scalar mediator scenario (TH~$=1$). All couplings to Higgs and mediators are taken to be identical, which is conservative. 
}
\label{f.moneyplot}
\end{figure*}

Figs.~\ref{f.Deltatotmaxscalarmediator} and~\ref{f.moneyplot} allow us to reach a very strong conclusion. 
Assume the 100 TeV collider can probe  20 (10) TeV SM-charged states. In that case, the only way for $N_f$ fermion partners and $N_s$ scalar mediators with tuning less severe than $10\%$ to avoid discovery at future colliders is for $N_f \cdot N_s > 12$ $(2)$ with the conservative tuning combination, and $N_f \cdot N_s \gtrsim 210$ $(30)$ with the conventional tuning combination.

The scalar mediator scenario serves as a powerful demonstration that \emph{both} kinds of future colliders, lepton and 100 TeV, are required to discover generalized scenarios of neutral naturalness. Consider $m_T \lesssim 500 \gev$ in  \fref{fermionscalarmediatorexclusion}, so $\Delta_{h(T)}$ is not too severe. Then for small $y_{STT}$, Higgs mixing is sizable but the UV completion scale could be higher than 20 TeV without severe tuning. In this case, a 100 TeV collider alone would not make the discovery, but a lepton collider would detect the mixing. Conversely, if $y_{STT}$ is large, Higgs mixing might be undetectable but the UV completion scale has to be low to avoid tiny $\Delta_{H,S}$. In that case, a lepton collider sees no signal, but a 100 TeV collider would produce new states. Both machines have to work in tandem to probe the space of generalized naturalness!

In this scenario we have not explicitly calculated the Higgs cubic coupling as a potential low-energy observable. Depending on the specific mediator sector, detectable cubic coupling deviations $\delta \lambda_3$ may be generated, either through sizable loop contributions or through mixing effects. These measurements are therefore motivated within fermionic top partner theories as well. However, a precise determination of the Higgs cubic coupling at one-loop requires a more complete model than our simplified mediator Lagrangian \eref{Lscalarmediator}, and would be sensitive to terms that are not directly related to the cancellation of quadratic divergences in $\delta \mu^2$ or the unavoidable level of tuning in the theory.

\subsubsection{Matching to Orbifold Higgs}

It is interesting to map our simplified model description to a concrete theory realization. The generalized Twin Higgs models  called ``orbifold Higgs'' theories~\cite{Craig:2014aea, Craig:2014roa} can realize top partner multiplicities beyond the canonical 3. Since this is a full theory, the equivalent of $y_{STT}$ is determined by symmetries. For the abelian $\mathbb{Z}_n$ orbifold, the following mapping applies:
\begin{equation}
N_s = 1 \ \ , \ \ \ 
N_f = 3(n-1) \ \ , \ \ \ 
y_{STT} = \frac{y_t}{\sqrt{2(n-1)}} \ .
\end{equation}
So for $N_f = 3, 6, 9$, $Zh$ cross section measurements at FCC-ee give $m_T$ reaches of 1.9, 1.4, 1.1 TeV. This is obviously much better than the model-independent reach we derive in \sssref{exptuning}. It also suggests that the most pessimistic limit possible in our model-independent approach -- large $y_{STT}$ and large $N_f, N_s$ -- could be difficult to realize in a full theory.

\subsection{Fermionic Top Partners (Fermionic Mediator)}
\label{ss.fermionicfermionmediator}

The previous examples, where fermionic top partners couple to the Higgs through scalar mediators or a non-perturbative sector, represent all known theories of neutral naturalness. However, from a bottom-up perspective, there is one alternative, which to our knowledge has never been explored in this context. The operator  $|H|^2 \bar T T$ could be generated within 4D perturbation theory by integrating out a heavy EW-charged fermion $F$, as shown in \fref{fermioncompletion}~(c). We distinguish this from composite/holographic TH models~\cite{Batra:2008jy, Barbieri:2015lqa, Low:2015nqa, Geller:2014kta} in which an infinite tower of states generate the operator, requiring strong coupling or extra dimensions. Taking the top partner fields $T$ to be  SM singlets requires the heavy states $F$ in \fref{fermioncompletion}~(c) to be Dirac fermions transforming as doublets under $SU(2)_L$. 

In such a model the cancellation of quadratic divergences breaks down at scales above the mediator mass $M_F$, since $F$ loops start contributing to the Higgs mass. 
(Contrast to the scalar mediator case, where the Higgs mass protection from the top quark loop is still realized above the singlet mass.) Therefore this scenario requires further new physics at or below the scale $M_F$  to contribute to the cancellation of quadratic divergences:
\begin{equation}
\Lambda_\mathrm{UV} \lesssim M_F.
\end{equation}
As we have done throughout, we assume that at the scale $\Lambda_\mathrm{UV}$ additional SM-charged states appear, such that $\Lambda_\mathrm{UV} < 10 - 20\tev$ can be probed by a 100 TeV collider. (This is explicitly satisfied here, at the very least, by the presence of electroweak charges with mass $M_F$.) 
In the following we make the worst-case assumption $\Lambda_\mathrm{UV} = M_F$ when computing projected sensitivities for the model.

As discussed in~\ssref{fermionictoppartners}, the mass $M_F$ at which the top partner EFT is completed cannot be too high if perturbation theory is to remain valid (\fref{EcmMax}). Therefore, for a given top partner mass and multiplicity one obtains a bound on $\Lambda_\mathrm{UV} < M_F < E_\mathrm{CM}^\mathrm{max}$, giving constraints that are identical to those obtained in \ssref{fermionicstrongcoupling}, where strong coupling generates the fermionic top partner EFT. In particular, the tuning suffered by this scenario can be no better than that case, see \fref{moneyplot} (top left).

However, in this partial UV completion we can make more definitive statements beyond the unitarity arguments. For different fermion mediators, one can compute  low-energy observables including Higgs coupling deviations and electroweak precision observables. This may sharpen the tuning statements we can make.

The low-energy phenomenology of the most minimal fermionic mediator was recently explored, in a general BSM context, by the authors of~\cite{Fedderke:2015txa}. Their results are easily adapted to our specific application of neutral naturalness. In addition to the observables discussed there, we also consider deviations in the Higgs cubic coupling. The strongest constraints arise from measurements of the Peskin-Takeuchi $T$ parameter at lepton colliders. 

It is possible to go beyond this minimal scenario by custodially completing the fermion mediators and eliminating deviations to the $T$-parameter at leading order. In that case the strongest low-energy constraint is derived from $Zh$ cross section measurements. 
While these measurements serve as a useful diagnostic, the strongest restrictions on the minimal level of tuning in an undiscoverable custodial mediator scenario arise from the simple unitarity arguments of \ssref{fermionicstrongcoupling}.

\subsubsection{Minimal mediator}
\label{sss.minimalfermionmediator}

\begin{figure*}
\begin{center}
\hspace{-8mm}
\begin{tabular}{ccc}
\hspace*{8mm}
\includegraphics[height=2cm]{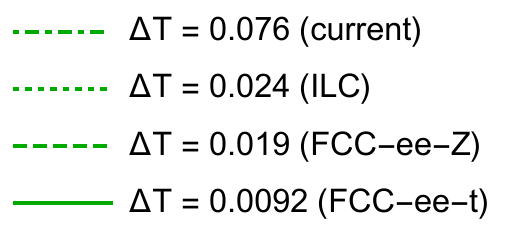}
&
\hspace*{8mm}
\includegraphics[height=2cm]{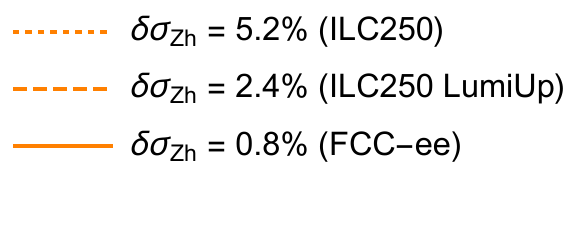}
&
\hspace*{7mm}
\includegraphics[height=2cm]{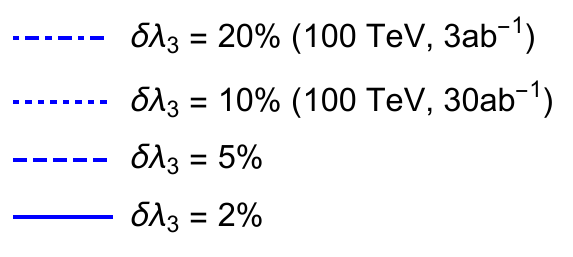}
\\
\includegraphics[width=6cm]{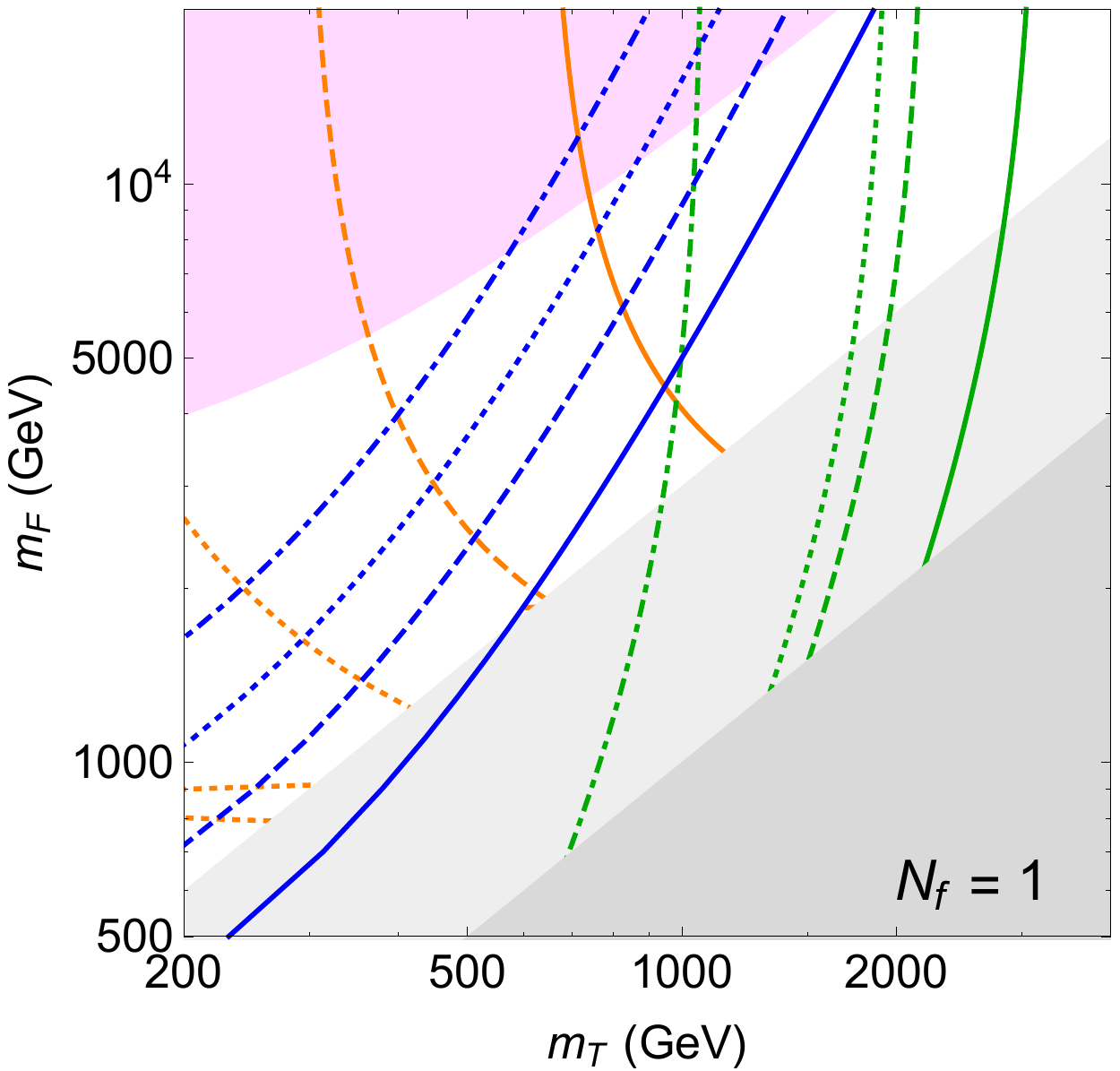}
&
\includegraphics[width=6cm]{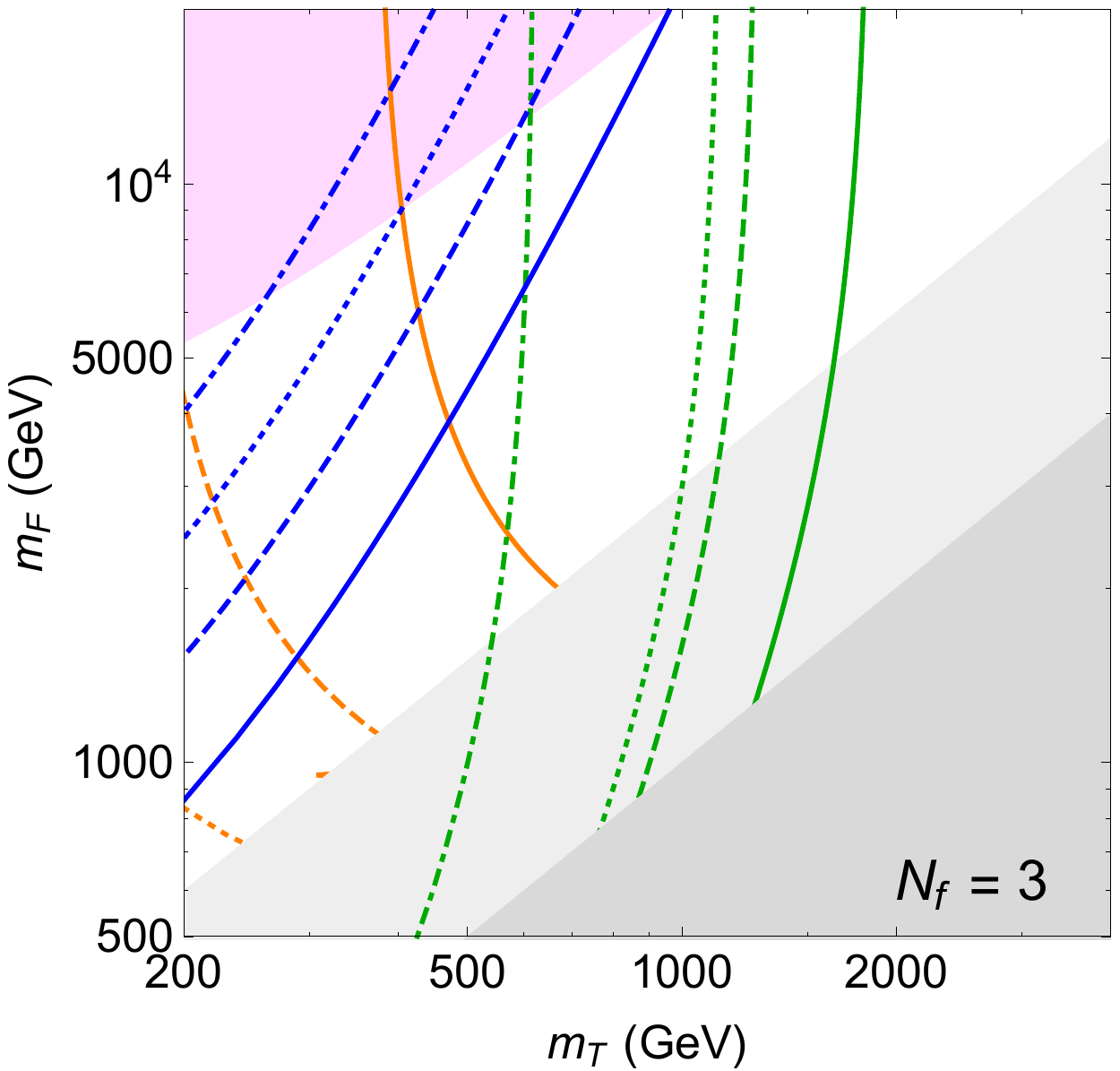}
&
\includegraphics[width=6cm]{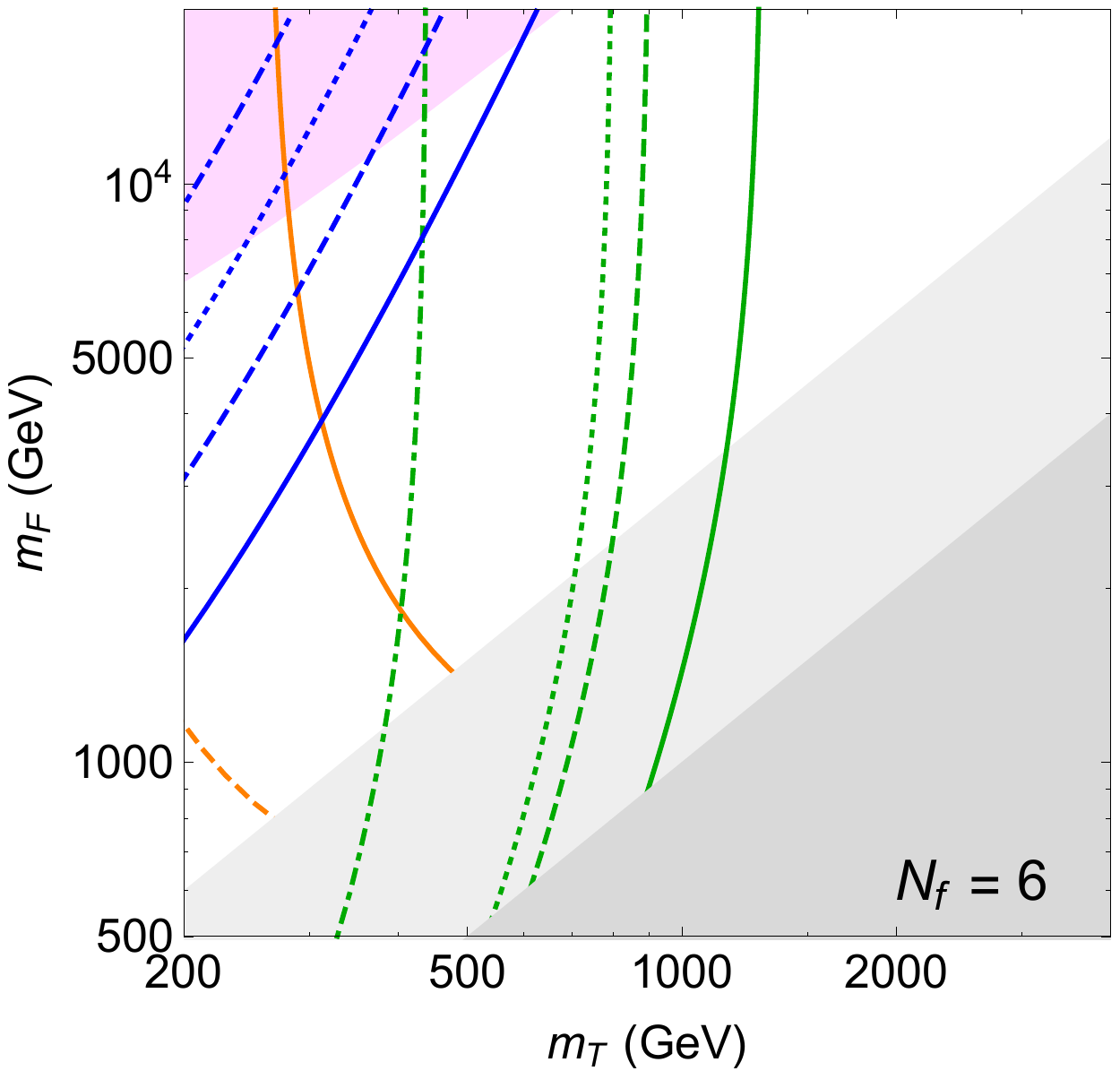}
\end{tabular}
\end{center}
\vspace*{-5mm}
\caption{
Exclusion potential of  future lepton and hadron colliders for fermionic top partners with a fermionic mediator, in the minimal model without a custodial symmetry (\sssref{minimalfermionmediator}). $Zh$ cross section and $T$-parameter deviations computed using expressions in~\cite{Fedderke:2015txa}. The light gray area indicates  $M_F < 3 m_T$ where the EFT calculation of $\delta \sigma_{Zh}$ is likely unreliable. The dark gray area indicates $M_F < M_T$ where the theory ceases to qualify as a scenario of neutral naturalness. Unitarity arguments forbid the pink shaded region in the top left corner, see \ssref{fermionictoppartners}. In the model with a custodial symmetry (\sssref{custodialfermionmediator}), similar results apply except that the $T$ parameter constraints are eliminated, and the $\delta \sigma_{Zh}$ contours are shifted by a negligible amount.
}
\label{f.fermionfermionmediatorexclusion}
\end{figure*}

A minimal set of interactions to generate the desired top partner operator is as follows (assuming same coupling for all top partners):
\begin{equation}
\mathcal{L} \supset - M_T \bar{T}_i T_i  - M_{F} \bar{F}_i F_i  - \kappa \bar{F}_i H T_i - \kappa \bar{T}_i H^\dagger F_i 
\label{e.fermmodel}
\end{equation}  
We show the top partner index $i$ explicitly to emphasize that each $T$ couples to its own $F$, so there are as many doublet mediators as there are partners, $N_f$. (This is to be experimentally conservative and avoid off-diagonal terms $|H|^2 T_i T_j$, see \ssref{wrinkles}.) We have taken $F$ to transform under the SM as $(\mathbf{1},\mathbf{ 2})_{1/2}$. For simplicity and to minimize additional experimental signals we have also assumed $CP$ symmetry.  Integrating out $F_i$ generates the operator $|H|^2 \bar T_i T_i$ with coefficient 
\begin{equation}
\frac{1}{2M'} = \frac{\kappa^2}{M_{F}}
\end{equation}  
Recalling the cancellation condition \eref{naturalfermion}, we find that 
\begin{equation}
\label{e.kappa}
\kappa^2 = \frac{3}{2 N_f} \frac{M_{F}}{M_T} \ y_t^2
\end{equation}
is required for naturalness. 

As discussed in~\cite{Fedderke:2015txa}, the Yukawa-like coupling $\kappa$ violates custodial symmetry and thus gives a contribution to the Peskin-Takeuchi parameter $T$. 
 Within an EFT approximation (valid in the limit $M_{F} \gg M_T \gg v$), integrating out the $F_i$ and $T_i$ generates the operator $|H^\dagger D_\mu H|^2$, which determines the $T$ parameter. The one-loop diagram contributing to this operator involves a loop of two $T$ particles and two $F$ particles with four insertions of the coupling $\kappa$, so that $T$ scales as $\sim N_f \kappa^4/M_{F}^2$. Imposing \eref{kappa} yields
\begin{eqnarray*}
T &\approx& \frac{3 v^2 y_t^4}{128 \pi^2 \alpha_e N_f M_T^2 } \left[ 5 - (2 M_F + 15 M_T) \frac{M_T}{M_F^2} \right] 
\\ \nonumber
&=&  \frac{15 v^2 y_t^4}{128 \pi^2 \alpha_e N_f M_T^2} + \mathcal{O}\left(\frac{M_T}{M_F}\right)
\end{eqnarray*} 
We see that $T \sim 1/(N_f M_T^2)$ to first order. 
\fref{fermionfermionmediatorexclusion} shows green contours of the $T$ parameter, corresponding to the sensitivities of future lepton colliders, as a function of the top partner mass $m_T$ and the heavy fermion mass $M_F$. 
This is computed using the full loop expressions for $T$ in~\cite{Fedderke:2015txa}.
Unless $N_f$ is taken to be very large, electroweak precision measurements have at least TeV reach in the top partner mass.

The interactions of~\eref{fermmodel} also correct the Higgsstrahlung cross-section $\sigma_{Zh}$. This was computed in~\cite{Fedderke:2015txa} using the results of~\cite{Craig:2014una} in an EFT approach. Contributing low-energy operators include $\left( \partial_\mu |H|^2 \right)^2$  (as for scalar top partners), but also other operators such as $|H|^2 W_{\mu \nu}^a W^{a,\mu \nu}$.  Orange contours of $\delta \sigma_{Zh}$ corresponding to various projected future sensitivities are also shown in \fref{fermionfermionmediatorexclusion}. While the $T$ parameter generally provides a more sensitive probe of this model, the correlated signature of $\delta \sigma_{Zh}$ could be visible at future colliders as well, allowing partial diagnosis of the hidden sector. Furthermore, this $\delta \sigma_{Zh}$ may be the superior signature if unrelated additional contributions \emph{accidentally} cancel the heavy doublet fermion's deviation in $T$.

In addition to the signatures considered in~\cite{Fedderke:2015txa}, there are potentially observable deviations to the triple Higgs coupling, which we compute from the Coleman-Weinberg potential. The interactions of \eref{fermmodel} generate an effective $|H|^6$ operator 
\begin{equation}
\delta V_1 \supset \frac{ C_{60}}{M_F^2} |H|^6
\end{equation}
The triple-Higgs coupling is then
\begin{equation}
\lambda_3 = \frac{m_h^2}{2 v} + \frac{C_{60} v^3}{M_F^2} + \ldots
\end{equation}
with the first term being the SM value at tree-level, and the second term being the contribution from the $H \bar F T$ interaction. The coefficient of the operator is
\begin{eqnarray*}
C_{60} &=& \frac{N_f \kappa^6 M_F^2}{12 \pi^2 (M_F - M_T)^5} \ \times \ 
\\ \nonumber
&&
 \bigg( 
(M_F - M_T) (M_F^2 + 10 M_F    M_T + M_T^2)
\\ \nonumber
&& 
 + 6 M_F M_T (M_F + M_T)\log \frac{M_T}{M_F} \bigg)
\\
&=& \frac{1}{N_f^2} \frac{M_F^3}{M_T^3} \left[ \frac{9 y_t^6 }{32 \pi^2} + \mathcal{O}\left(\frac{M_T}{M_F}\right) \right]
\end{eqnarray*}
where in the second line we have applied \eref{kappa} and expanded for large $M_F$. The reach of various proposed or potential $\lambda_3$ measurements are shown as blue contours in \fref{fermionfermionmediatorexclusion}. The best possible Higgs self coupling measurements can only compete with even current electroweak precision bounds if $M_F$ is very heavy. However, since additional new physics could affect $\lambda_3, \delta \sigma_{Zh}$ and $T$ differently, all three measurements can be important to pin down the hidden sector.

Finally, the heavy doublet fermion $F$ could be directly pair produced at the LHC or a 100 TeV collider. In the minimal model, the relevant process is $p p \to \bar F F \to 2h + X$. Since the mass of $F$ is not fixed by the naturalness requirement \eref{kappa}, we do not pursue this signature in detail, beyond our stated assumption that $\Lambda_\mathrm{UV} \leq M_F$ can be probed at the 10 or 20 TeV level at a 100 TeV collider. However, we point out that unlike top partner direct production, this signal grows with $N_f$ instead of being suppressed by the number of partners.

\begin{figure}
\begin{center}
\includegraphics[width=8cm]{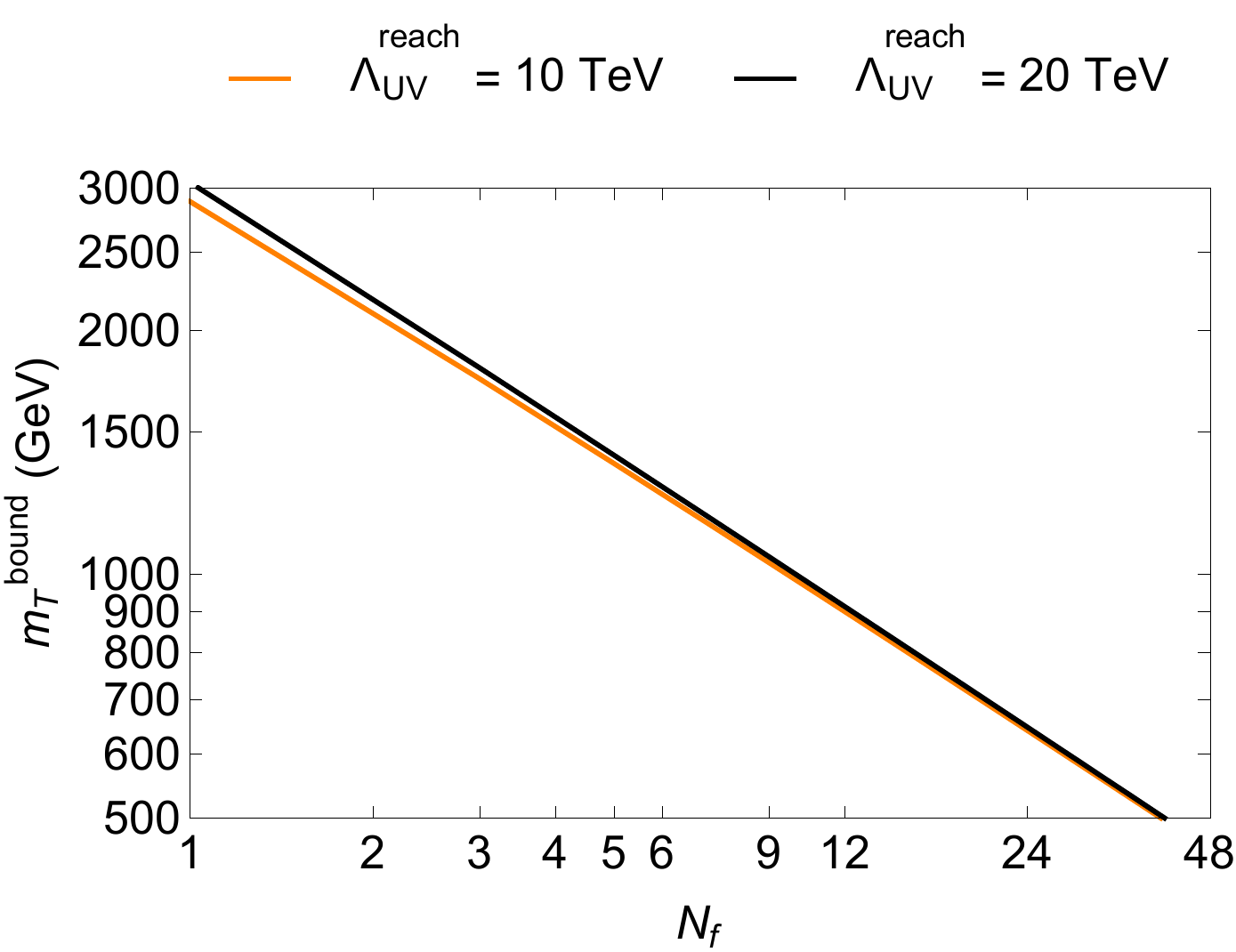}
\end{center}
\caption{
$m_T^\mathrm{bound}$, the smallest top partner mass that can always be excluded for fermionic top partners completed with minimal fermionic mediators, either by direct production at a 100 TeV collider ($M_F$ mass reach assumed to be up to $\Lambda_\mathrm{UV} = 10$ or 20 TeV) or by probing deviations in the $T$-parameter at FCC-ee. 
}\label{f.minexcludablemT}
\end{figure}

The small dependence of the $T$-parameter deviation on $M_F$ allows us to robustly define a minimum top partner mass $m_T^\mathrm{bound}(N_f, \Lambda_\mathrm{UV}^\mathrm{reach})$ which can always be excluded,  for some assumption of direct UV physics reach $\Lambda_\mathrm{UV}^\mathrm{reach}$, by either directly producing the charged fermion $F$ or by probing the $T$-parameter. This is simply derived by solving  $T(m_T, M_F = \Lambda_\mathrm{UV}^\mathrm{reach}) = \Delta T_\mathrm{FCC-ee}$ for $m_T$, and is shown in \fref{minexcludablemT}. The $m_T$ reach depends very little on the assumed $\Lambda_\mathrm{UV}^\mathrm{reach} = M_F$ reach. Up to $\sim 9$ top partners can be excluded with masses up to a TeV.

We can now again ask how tuned this scenario would have to be in order to escape all detection, either by direct production of new states at $\Lambda_\mathrm{UV} = 10$ or 20 TeV, or by probes of low-energy structure. There is only one tuning, $\Delta_{h(T)}$, from the log-divergent top partner contributions to the Higgs mass. $\Delta_{h(T)}$ is maximized by choosing the smallest non-excludable top partner mass, i.e. $m_T^\mathrm{bound}$. The resulting $\Delta_{h(T)}^\mathrm{max}$ is shown in \fref{moneyplot} (bottom right). A natural theory with tuning no worse than 10\% will produce experimental signals unless it has more than 30 - 40 top partners, a quite baroque scenario.

\subsubsection{Custodially symmetric mediator}
\label{sss.custodialfermionmediator}

Although the above model represents a minimal realization of the fermionic mediator scenario, one can introduce additional particle content to restore the custodial $SU(2)_L \otimes SU(2)_R$ symmetry of the SM Higgs sector. This suppresses contributions to the $T$-parameter which constitute the most sensitive probe of the minimal fermionic mediator.

There are two possible ways to restore custodial symmetry: by embedding $T$ into an $SU(2)_R$ doublet, and by embedding $F$ into an $SU(2)_L \otimes SU(2)_R$ bi-doublet.\footnote{We thank Kaustubh Agashe for bringing the latter possibility to our attention.}

The first possibility is incompatible with the requirements of neutral naturalness, since the additional charged particle would have to be close in mass to the top partner. This would be easily observable through direct production and Drell-Yan measurements~\cite{Alves:2014cda}, potentially even at the LHC. 

On the other hand, embedding $F$ in a bi-doublet does not have to lead to additional light states. We call this the custodially symmetric fermion mediator. 

The Lagrangian of the custodially symmetric fermion mediator can be written down by extending the previous model with an additional doublet $F^C$, transforming as $(\mathbf{1},\mathbf{2})_{-1/2}$ (i.e., opposite hypercharge compared to $F$):
\begin{align}
\mathcal{L} \supset - M_T \bar{T}_i T_i  - M_{F} \bar{F}_i F_i - \kappa \bar{F}_i H T_i &- \kappa \bar{T}_i H^\dagger F_i \nonumber \\
-M_{F} \bar{F}^C_i F^C_i  - \kappa \bar{F}^C_i \varepsilon H^* T_i &+ \kappa \bar{T}_i H^T \varepsilon F^C_i 
\label{e.custodialmodel}
\end{align}  
where $\varepsilon$ is the two-index antisymmetric tensor. This model realizes custodial symmetry, with the Higgs embedded in a real $SU(2)_L \times SU(2)_R$ multiplet as $\Phi \equiv \left( \varepsilon H^*, H \right)$ and the mediators combining into a multiplet $\mathbb{F} \equiv \left(F^C, F \right)$. 

In this extended model, the matching condition for naturalness is slightly modified compared to \eref{kappa}:\begin{equation}
\label{e.kappa2}
\kappa^2 = \frac{3}{4 N_f} \frac{M_{F}}{M_T} \ y_t^2 .
\end{equation}
Contributions to the $T$ parameter vanish to leading order in the hypercharge coupling. 
It remains straightforward to calculate the corrections to the triple Higgs coupling (from the operator $|H|^6$), and they are practically identical as a function of $(m_T, M_F)$ to the minimal mediator case. To compute the $Zh$ cross-section deviation, one must in principle  repeat the one-loop EFT analysis of~\cite{Fedderke:2015txa} to find the coefficients of all the operators which contribute. 

Fortunately, we can bootstrap our way to an estimate for $\delta \sigma_{Zh}$ in the custodial model using the existing results in~\cite{Fedderke:2015txa}. In the minimal fermionic mediator model, $\delta \sigma_{Zh}$ is very well approximated by only including operators that scale as $\mathcal{O}(\kappa^4)$ for $N_f \lesssim 20$, which have dimension six with four Higgs bosons and two derivatives. Amplitudes for diagrams with four external Higgses and $F^C$'s in the loop are related to those with only $F$ loops by a permutation of external momenta. This allows us to relate the $\mathcal{O}(\kappa^4)$ operators in the custodial mediator scenario to those in the minimal mediator scenario. In the notation of~\cite{Fedderke:2015txa}:
\begin{eqnarray}
\nonumber
C_{(4,2),A}^\mathrm{custodial} &=& 4 \, C_{(4,2),A} +  2 \, C_{(4,2),B}
\\
C_{(4,2),B}^\mathrm{custodial} &=& 0
\\ \nonumber
C_{(4,2),C}^\mathrm{custodial} &=& 4 \, C_{(4,2),C}
\end{eqnarray}
Due to the modified naturalness condition \eref{kappa2}, the final $\delta \sigma_{Zh}$, as a function of $(m_T, M_F)$, is only slightly larger than in the minimal mediator, increasing top partner reach by $\sim 10 - 20 \gev$ compared to what is shown in \fref{fermionfermionmediatorexclusion}. This fails to probe the perturbative region with $M_F > 10 \tev$. Therefore, the minimum level of tuning that must be suffered by an undiscoverable theory is still given by the unitarity constraints of \ssref{fermionictoppartners}, identical to the strong coupling case of \ssref{fermionicstrongcoupling}. This is shown as a function of $N_f$ in \fref{moneyplot} (top left).

\subsection{Discussion of Simplifying Assumptions}
\label{ss.wrinkles}

We have assumed throughout, for both scalar and fermion partners, that the couplings of those partners to the Higgs (and of scalar mediators to partners and to the Higgs) are all identical, quantifying different scenarios simply by their number $N_r$ and $N_f$ ($N_s$). Of course, it is possible for the couplings to be unevenly distributed amongst the partners, as explicitly realized in~\cite{Craig:2014aea, Craig:2014roa}. However, since all experimental signatures and inverse tunings depend on negative powers of $N_r, N_f$ ($N_s$), making some couplings bigger essentially decreases the effective number of important top partners, increasing the size of all experimental signals and making all tunings more severe. This makes our analysis conservative.

In general, one could also consider off-diagonal Higgs-top partner interactions like  $|H|^2 \phi_i \phi_j$ for scalars and $|H|^2 \bar T_i T_j$ for fermions. However, working in the basis where the scalar or fermion mass terms $\mu_{\phi}^2$, $M_T$ are diagonal, these interactions will not contribute to the cancellation of quadratic divergences. They are therefore not connected to the question of naturalness. It is of course possible for judiciously chosen large mixings to cancel low-energy signatures if the top partners are light compared to the Higgs, but this cancellation might be regarded in itself as a form of tuning.

For the fermionic mediator scenario of fermionic top partners we made the assumption of one mediator (or one $SU(2)_R$ doublet of mediators, in the custodially symmetric case) per partner. If each mediator coupled to more than one partner (in the basis where $M_T$ is diagonal) or if there were other mixing effects, the result would be  additional off-diagonal terms $|H|^2 \bar T_i T_j$ which we ignore as discussed above.  On the other hand, we could couple \emph{each} partner to $N_F > 1$ mediators. However, all the observables we consider are loops of $F$ and $T$ fermions with $2n$ insertions of the $\kappa$ coupling ($n = 1, 2, \ldots$), or pure gauge loops of only $F$ fermions. The former amplitudes scale as $\kappa^{2n} N_f N_F^n \propto N_f^{n-1}$, i.e. are independent of $N_F$, and the latter will be enhanced by $N_F$. The direct production cross section of mediators would be enhanced by the same factor. Therefore, setting $N_F = 1$ is the minimal assumption with the smallest possible experimental signals.  Our assumption of CP symmetry for $F$ and $T$ is similarly conservative, since additional vertices unconstrained by naturalness yield additional signals.

\section{Towards a No-Lose Theorem for Naturalness}
\label{s.nolose}

In \sref{intro}, we argued that TeV-scale perturbative top partners will be discovered at current or proposed future colliders if they have any SM gauge charge. The results derived in \sref{simplifiedmodels}, and summarized in \fref{moneyplot}, prove that neutral top partners can also not escape detection, unless the theory is tuned worse than 10\%, or the top partner multiplicity is much higher than the SM top. Together, this amounts to a phenomenological no-lose theorem for naturalness with general top partners.

We now place our no-lose theorem in context by discussing solutions to the hierarchy problem which lie explicitly \emph{outside} of the scope of our arguments: theories of neutral naturalness \emph{without} SM-charged new states at the UV-completion scale, and theories without top partners. Taken together, it is clear that our results make significant progress towards a truly general no-lose theorem for the hypothesis of naturalness, while also sharpening those questions which remain unsolved.

\subsection{SM charges at the UV completion scale?}
\label{ss.loopholes}

In determining whether a given top partner scenario would lead to production of new states at the 100 TeV collider, we assumed that at whatever scale the most severe tuning is regulated, new SM-charged BSM states should appear. This can simply be taken as an input assumption for our analysis, with attempts to more formally prove it left for future work. 

That being said, this assumption is very reasonable from the expectation that the full symmetry protecting the Higgs mass must become manifest in the UV. Assuming the Higgs and top to reside in larger multiplets of this symmetry, new charges should appear. This is certainly the case in every known UV completion of uncolored naturalness~\cite{Craig:2013fga, Geller:2014kta, Batra:2008jy,Barbieri:2015lqa, Low:2015nqa, Craig:2014fka, Chang:2006ra}.

However, we can in principle imagine several ways in which this assumption might fail. We now outline the features such a theory would have to have for each top partner structure. None of these possibilities have been realized in the literature, so our results articulate a clear model-building challenge for future work.

In the case of scalar top partners (\ssref{scalartoppartners}),  the tuning $\Delta_{\phi(h)}$ in \eref{Deltaphi} could be eliminated if the mass of scalar partners were to be stabilized by SM-neutral ``$H$-partners'' that couple to $\phi$ to cancel the Higgs contribution. This would not be allowed to affect the Higgs mass, otherwise new sources of tuning would be introduced. In that case, scalar partners below about 600 GeV (so that the Higgs log tuning is not too severe) but heavier than a few hundred GeV (beyond the reach of low-energy probes) could escape detection while remaining natural.

For fermion partners with scalar mediators, the Sacrificial Scalar Mechanism (see \sssref{minimallagrangiain}) requires new partner states to ``take over'' the protection of the Higgs mass at whatever scale the singlet mediator mass becomes protected. As we have discussed, to ensure that the tuning  $\Delta_{S(T)}$  in \eref{DeltaS} does not become too severe, the scale $\Lambda_\mathrm{UV}$ at which these states appear must not be too low, such that they would be visible if they carried SM charge.  However, one could in principle imagine an \emph{epicyclic neutral top partner scenario}. For example, one could introduce SM-neutral scalars $\varphi,  \phi$, where the former protects the singlet against $T$ loops above scale $m_{\varphi}$ and the latter takes over protection of the Higgs against $t$ loops above scale $m_\phi$. In that case, the Higgs mass correction (for small $m_\varphi^2 - m_\phi^2$) is
\begin{eqnarray}
&& \delta \mu^2 = \frac{3 y_t^2}{8 \pi^2} \ \times   
\\ \nonumber 
&& 
 \bigg[
M_T^2 \log \frac{\Lambda_\mathrm{UV}^2}{M_T^2} 
+ 
(m_{\phi}^2 -  m_\varphi^2) \left( \log \frac{\Lambda_\mathrm{UV}^2}{m_\varphi^2} - 1\right)
\bigg]  \ ,
\end{eqnarray}
where we have taken the large-$\Lambda_\mathrm{UV}$ limit for clarity. The first term is the usual log divergence giving rise to the $\Delta_{h(T)}$ tuning. The second term includes the additional log divergences from the thresholds at $m_\phi$ and $m_{\varphi}$, but in the limit $m_{\phi} \to  m_\varphi$ this cancels and there is no additional tuning suffered by the Higgs mass, even if the singlet is stabilized at scale $m_{\varphi} \ll \Lambda_\mathrm{UV}$. However, even a small fractional mass splitting between $\varphi$ and $\phi$ will reintroduce large loop corrections, and hence make the tuning on the Higgs much more severe than the $\Delta_{h(T)}$ estimate. Even if this loophole were realized, light mediators generally lead to larger Higgs mixing, and depending on the size of the Yukawa coupling $y_{STT}$ may lead to Landau poles at $\mathcal{O}(10 \tev)$. 

A significantly more contrived possibility would be if the mediator did not protect the Higgs mass but had large and negative $\lambda_{HS}$. This could reduce the size of the log-divergent correction \eref{deltamumuHHS}, but would only lead to less overall tuning if somehow the resulting Higgs mass contribution $\delta \mu^2 \propto (6 \lambda + N_s \lambda_{HS}) \Lambda_\mathrm{UV}^2$ was canceled by some alternative mechanism, which, apart from the danger of potential runaways, seems very strange indeed. 

Conversely, it seems that our assumption is quite robust for fermion partner scenarios with non-perturbative or fermionic mediators. In the former case, maintaining the symmetry which stabilizes the Higgs would seem to force new SM charges to appear at $\Lambda_\mathrm{UV}$. In the latter case, EW-charged fermions are explicitly a part of the top partner structure.

\subsection{Theories without top partners}
\label{ss.notoppartners}

Our approach to a general no-lose theorem for naturalness has a useful historical analogue in the ``no-lose theorem'' for discovery of the mechanism of electroweak symmetry breaking~\cite{Chanowitz:1998wi}. This was the argument that $WW$ scattering must be unitarized either by a perturbative Higgs sector, or by more exotic non-perturbative physics appearing at scales $\lesssim$ TeV (e.g. technicolor). The former case was systematically analyzed~\cite{Espinosa:1998xj} to show that discovery of Higgses at (proposed) linear colliders was guaranteed.  Our no-lose theorem for TeV-scale naturalness with perturbative top partners can be seen as the analogue of this claim.

Non-perturbative or exotic solutions to the hierarchy problem are, by their very nature, much more difficult to study and exhaustively classify. Even so, we still expect new QCD or at least EW-charged physics at the TeV scale, which can in principle be probed by the LHC and future colliders. We now perform an informal survey of the known mechanisms of solving the hierarchy problem without top partners to illustrate this point.

One possibility to cut off corrections to the Higgs mass is that the Higgs itself ceases to be a sensible degree of freedom. This is its ultimate fate in composite Higgs models including their 5D realizations (e.g.~\cite{Bellazzini:2014yua, ArkaniHamed:2001nc, ArkaniHamed:2002qx, ArkaniHamed:2002qy, Schmaltz:2004de, Contino:2003ve,Agashe:2004rs, Contino:2006qr}, though many of these models also have a perturbative top partner regime at low energies). Since the Higgs has electroweak charge, this would seem to require the existence of new electroweak-charged states at scale $\Lambda_\mathrm{BSM}$. (In most such theories, there are new colored states as well.) These states can be directly produced with sizable cross section at the LHC or a future 100 TeV machine, which should  lead to discovery.

\begin{figure}
\begin{center}
\begin{tabular}{ccc}
\includegraphics[width=4cm]{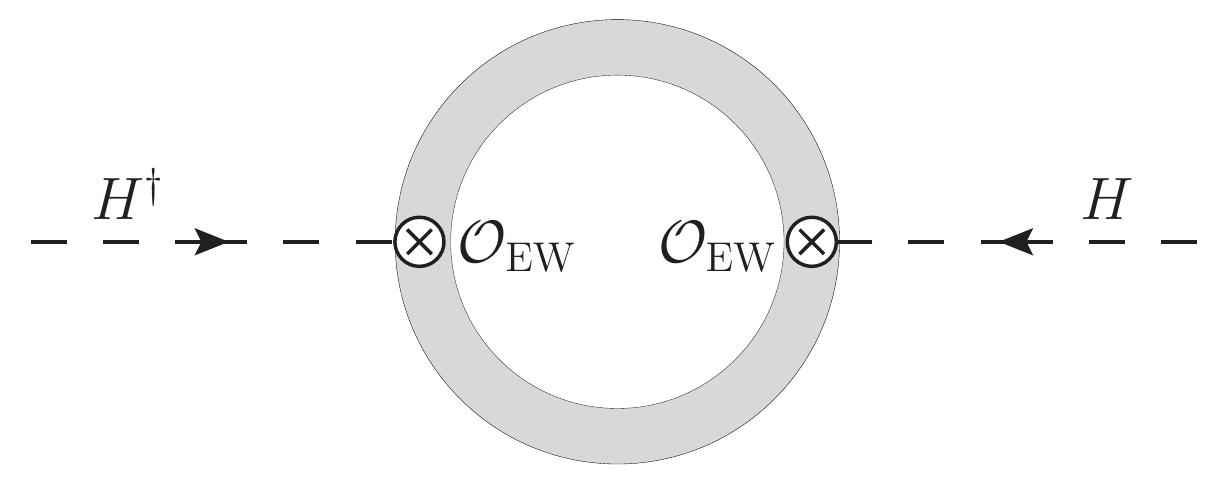}
& \phantom{blabla}  &
\includegraphics[width=2.8cm]{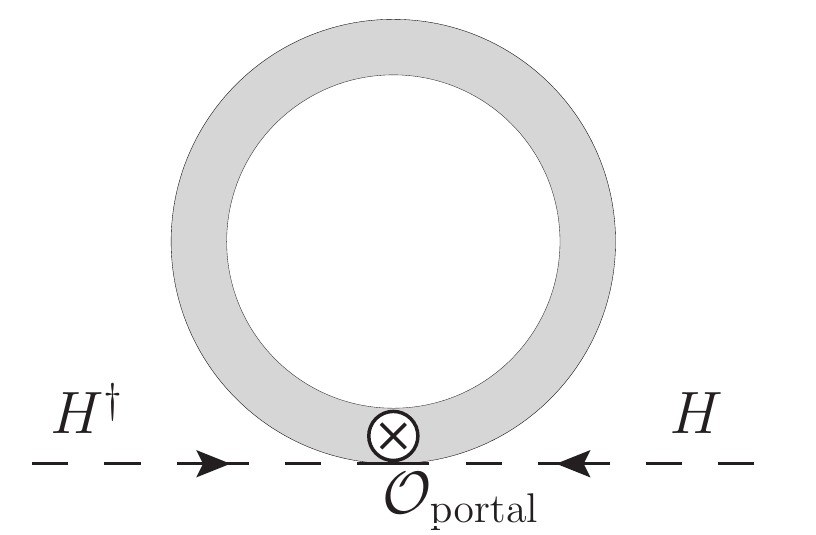}
\end{tabular}
\end{center}
\caption{
The two forms of quantum corrections which could cancel the top loop, featuring operators $\mathcal{O}_\mathrm{EW}$ (left) and $\mathcal{O}_\mathrm{portal}$ (right).
}
\label{f.OewOportal}
\end{figure}

What if the Higgs remains as an (approximately) elementary degree of freedom at the scale where SM top loops are regulated? Assuming this cancellation arises due to some symmetry we know this cancellation occurs loop-order by loop-order. There are now, in principle, two possibilities for how the Higgs mass could be stabilized. One possibility is that the Higgs couples to an $SU(2)_L$ doublet operator $\mathcal{O}_\mathrm{EW}$ (i.e. $\mathcal{L} \supset \mathcal{O}_\mathrm{EW} H$), and the relevant quantum correction is generated by two insertions of $ \mathcal{O}_\mathrm{EW}$, see \fref{OewOportal} (left). Here the central blob represents a set of quantum corrections involving the fields of $\mathcal{O}_\mathrm{EW}$. Since the new physics operator $\mathcal{O}_\mathrm{EW}$ will involve new electroweak states, this scenario should again be probed by direct production at future colliders.\footnote{The exception is the possibility that the necessary electroweak charge of $\mathcal{O}_\mathrm{EW}$ is entirely due to SM fields in the operator, such as $L$, $D_\mu H^\dagger$, etc. This however implies mixing between the SM neutrinos, gauge bosons etc. and some new physics operators, which is strongly constrained, so that such interactions cannot give large contributions to the Higgs mass.}  

The other possibility is that the Higgs couples to new quantum loops through a SM-singlet ``portal'' operator
$\mathcal{L} \supset |H|^2 \mathcal{O}_\mathrm{portal}$, generating a Higgs mass contribution shown in \fref{OewOportal} (right). This is the only possibility which naively does not seem to explicitly guarantee the existence of new SM-charged states at the TeV scale, and of course includes the neutral top partner scenarios we have discussed.

Suppose now that the new physics quantum corrections which cancel against the top contribution to the Higgs mass in \fref{OewOportal} (right)  are non-perturbative. For the cancellation to proceed by an explicit symmetry then requires top physics to \emph{also} become non-perturbative at $\Lambda_\mathrm{BSM} \sim \tev$.  This is reminiscent of top compositeness models~\cite{Georgi:1994ha, Kumar:2009vs, Agashe:2005vg, Lillie:2007hd}, and implies the existence of new physics charged under QCD (related to the top) at the TeV scale. This is a very favorable scenario for discovery. Similar reasoning applies to extra-dimensional models as well (which can often be considered dual to strong coupling): if the sector acting to cancel the top divergence is described by a 5D theory, then the top quark itself must become a 5D field.

We are therefore left with perturbative new physics sectors, coupled to the Standard Model through $|H|^2 \mathcal{O}_\mathrm{portal}$. Considering particles of spin 1 or less, these are the top partner scenarios shown in \fref{toppartners} and \tref{summary}, which are subject to our no-lose theorem. At this level of rigor, the argument for the discoverability of general naturalness seem very promising. However, it would obviously be extremely interesting to find natural models which evade the arguments we have presented here. One example is the recently proposed ``cosmological relaxation'' mechanism~\cite{Graham:2015cka} (see~\cite{Kobakhidze:2015jya, Espinosa:2015eda, Batell:2015fma, Gupta:2015uea, Jaeckel:2015txa,Antipin:2015jia,Hardy:2015laa,  Patil:2015oxa} for related work), in which a small Higgs mass is selected by dynamics rather than symmetry, so that tree-level terms dynamically adjust to cancel any radiative corrections. Another possibility is an explicit failure of Wilsonian effective field theory, where an apparently tuned theory at low energies actually arises from a natural UV-complete theory. This could arise due to a non-decoupling of heavy states~\cite{Dienes:1994np, Dienes:1995pm, Dienes:2001se}. The arguments of~\cite{Dong:2014tsa, Dong:2015gya} might point in a similar direction. However, it is important to keep in mind that even within the realm of symmetry-based low-scale solutions to the hierarchy problem, unexpected and  exotic possibilities may still exist.

\section{Conclusion}
\label{s.conclusion}

The main motivation for our work is the question: \emph{"Can naturalness be tested as a general hypothesis?"} We make significant progress towards such a general no-lose theorem by analyzing a well-defined class of theories in which the Higgs is stabilized below the TeV scale by perturbative top partners. In light of proposed future colliders, any such partners with SM gauge charge will be discovered~\cite{Alves:2014cda}. We therefore focused our attention on neutral top partners, analyzing them as model-independently as possible.

We considered all possible new interactions of the Higgs with SM-neutral bosonic and fermionic states which could give rise to one-loop quadratically divergent corrections to the Higgs mass, and derived conditions on the couplings for these divergences to cancel that from the SM top quark (\sref{eft}). While there are other requirements for a fully natural theory (e.g. cancellation of divergences from gauge loops, smallness of two-loop divergences, lack of tunings between tree-level parameters, and regulation of gauge boson loops), the former represent a set of necessary conditions for naturalness and allow us to determine the minimal experimental consequences thereof (\sref{simplifiedmodels}). The experimental observables that can be sensitive to the new Higgs interactions include the $Zh$ production cross-section at lepton colliders ($\sigma_{Zh}$), double Higgs production ($\sigma_{hh}$) or direct top partner production ($\sigma_{\mathrm{direct}}$) at a 100 TeV hadron collider, and in some cases electroweak precision measurements at lepton colliders.  

We also identify the irreducible tunings that must be present in each top partner scenario. These tunings have to be regulated at some scale $\Lambda_\mathrm{UV}$, which cannot be too high for the theory to be natural. We then make the central assumption that SM-charged BSM states appear at this scale, and that therefore any theory with $\Lambda_\mathrm{UV} < \Lambda_\mathrm{UV}^\mathrm{reach} = 10$ or 20 TeV can be discovered at a 100 TeV collider by direct production.
We explicitly outline in each case what would be required of a full theory to satisfy or violate this condition, but we can simply accept this assumption as an input to our analysis. It represents a reasonable expectation~\cite{Craig:2013fga, Geller:2014kta, Batra:2008jy,Barbieri:2015lqa, Low:2015nqa, Craig:2014fka, Chang:2006ra} about the full symmetry of the theory which should be apparent at higher scales. Its proof (or the formulation of theories which violate it) is left to future work.

The tunings and most important low-energy observables for each top partner scenario are summarized in \tref{summary}.

We first examined the case where the top partners are $N_r$ \emph{neutral scalars} (which for our purposes also covers the case of vector top partners), with $N_r = 12$ being analogous to the MSSM. Both $\sigma_{\mathrm{direct}}$ and the corrections to $\sigma_{Zh}$ scale as $1/N_r$; these signatures have been previously explored in the literature~\cite{Craig:2013xia, Craig:2014lda}. In this work we additionally considered the correction to $\sigma_{hh}$, which scales as $1/N_r^2$ and is in fact the most sensitive probe for low $N_r$ (possibly as high as $N_r \approx 12$, depending on the ultimate precision of this measurement at a 100 TeV collider). 
Unfortunately, these precision measurements are only sensitive to top partner masses of a few hundred GeV for MSSM-like values of $N_r$. 
However, the presence of scalar partners introduces an additional tuning, since their mass is quadratically sensitive to the scale of new physics through Higgs loops. Together with the log-tuning of the Higgs mass, naturalness then forces the UV completion scale to be relatively low:
Assuming a mass reach of $\sim 20$ $(10)$ TeV for the production of new SM-charged BSM states at a 100 TeV collider, \emph{a natural scalar partner theory with less than 10\% tuning will be discovered at future lepton or hadron colliders} provided $N_r < 20$ $(5)$ if tunings are combined very conservatively, or $N_r \lesssim 250$ $(40)$ if tunings are combined more conventionally, see \fref{moneyplot}~(top left).

We next considered $N_f$ \emph{fermionic top partners}, which must couple to the Higgs through the non-renormalizable operator $|H|^2 \bar T T$. The detailed experimental predictions and tuning depend on how this operator is generated, see \fref{fermioncompletion}. If non-perturbative physics generates the operator directly, symmetry arguments imply that SM-charged non-perturbative new physics should show up coupled to the top or Higgs sector, which would be discoverable at a 100 TeV collider (or the HL-LHC). Unitarity arguments can then be used to show that \emph{for any natural theory with less than 10\% tuning}, production of SM-charged states is expected at the 100 TeV collider for $N_f < 10$ $(2)$, assuming a 100 TeV mass reach of $\sim$ 20 (10) TeV, see \fref{moneyplot}~(top right)

The operator can also be perturbatively generated by exchange of either a SM-neutral scalar mediator or an electroweak doublet fermion mediator. Each of these cases leads to very different phenomenology and tuning.

The case of $N_s$ singlet scalar mediators represents the known models of fermionic top partners including Twin Higgs and Little Higgs theories (in which $N_s = 1$), where the Higgs emerges as a PNGB. Taking a model-independent approach, we identify the quadratic divergence of the Higgs from the new sector as arising directly from uncanceled divergences of the scalar mediator potential, so that the mediator mass is necessarily larger than the weak scale in a natural theory (and hence related to the ultimate UV completion scale). We dub this the ``Sacrificial Scalar Mechanism,'' and it is transparently manifest in the Twin Higgs model and its UV completions. The mixing of the Higgs with the heavy singlet, observable through $\sigma_{Zh}$, is predicted as a function of the top partner mass $m_T$ and the singlet-top partner Yukawa coupling $y_{STT}$. Future lepton colliders will be able to explore significant parts of the $(m_T, y_{STT})$ parameter space. The $Zh$ cross section deviation decreases with increasing $y_{STT}$, but this also increases the severity of several new tunings that can be identified in this scenario (in part as a result of the Sacrificial Scalar Mechanism). Assuming a 20 TeV mass reach for the UV completion at a 100 TeV collider with less than 10\% tuning arising from the mediator sector, all top partners lighter than $\sim 800 \gev$ lead to some kind of experimental signature if one takes $(N_f, N_s) = (3,1)$ analogous to the TH case. As for the scalar top partner case, all experimental signatures and inverse tunings are diluted for large $N_f, N_s$. 
Even so, assuming a mass reach of $\sim 20$ $(10)$ TeV for the production of new SM-charged BSM states at a 100 TeV collider, \emph{a natural fermionic partner theory with scalar mediators and less than 10\% tuning will be discovered at future lepton or hadron colliders} provided $N_f \cdot N_s < 12$ $(2)$ if tunings are combined very conservatively, or $N_f \cdot N_s \lesssim 210$ $(30)$ if tunings are combined more conventionally, see \fref{moneyplot}~(bottom left).

Finally, the fermionic top partner operator $|H|^2 \bar{T} T$ could also in principle be generated perturbatively by a heavy EW-doublet fermion mediator. (As with neutral scalar top partners, this has yet to be realized in a complete theory of neutral naturalness.) 
However, the cancellation of the Higgs mass then fails above the mass of the mediator fermion, so the mediator mass again points towards the necessary UV completion scale and its associated SM-charged BSM states.
The charged mediator has to couple strongly to the Higgs to give the necessary coefficient for $|H|^2 \bar{T} {T}$, generating sizable deviations in $\sigma_{Zh}$ and $\sigma_{hh}$. It can also be produced directly at the 100 TeV collider. 
For custodially symmetric fermionic mediators, the strongest tuning constraints are then derived from unitarity bounds, as for the non-perturbative case: see \fref{moneyplot} (top left). The minimal fermionic mediator breaks custodial symmetry, generating large deviations to the $T$-parameter. 
Assuming a mass reach in the range of $10 - 20$ TeV for the production of new SM-charged BSM states at a 100 TeV collider, \emph{a natural fermionic partner theory with minimal fermion mediators and less than 10\% tuning will be discovered at future lepton or hadron colliders} provided $N_f  \lesssim 35$, see \fref{moneyplot}~(bottom right).

Taken together, the above results allow the formulation of a  \emph{phenomenological no-lose theorem for the detectability of TeV-scale naturalness with perturbative top-partners}:
\begin{enumerate}

\item TeV-scale top partners with SM charges will be discovered by direct production and their effect on the DY spectrum~\cite{Alves:2014cda}, either at the LHC or a 100 TeV collider.

\item Our work identifies the minimal low-energy signatures of neutral top partners, showing that current and future colliders will be able to set model-independent lower bounds on both scalar and fermion top partner masses, as a function of their multiplicity. 

\item The low-energy scenarios we describe all have to be UV-completed at some scale $\Lambda_\mathrm{UV}$. The minimal motivation for this arises from the logarithmic tuning of the Higgs mass, due to the incomplete cancellation of top and top partner contributions. However, all of the top partner scenarios also require a sufficiently low UV completion scale to regulate other tunings and/or maintain the cancellation of Higgs quadratic divergences. At that scale, \emph{we make the assumption} that new SM-charged states appear as the full symmetry protecting the Higgs mass becomes manifest.

We have then shown that, if this UV completion scale is within the $\sim 10 - 20$ TeV kinematic reach of a 100 TeV collider, then any \emph{discoverable} neutral top partner theory is \emph{either} more natural than (say) $10\%$ tuning, \emph{or} it has to feature a large number of degrees of freedom in the partner sector (see \fref{moneyplot}): what we might call a \emph{``swarm of top partners''}.

\end{enumerate}

In \sref{nolose}, we place this result in theoretical context by discussing solutions to the hierarchy problem which do not fall under its scope: theories of neutral naturalness \emph{without} SM charges in the UV completion, and theories without top partners. Examples of the former have yet to be formulated, and present an open model-building challenge. The latter possibility, we argue, should also lead to discovery at the LHC or future colliders, but one cannot exclude the existence of exotic exceptions. This shows how the remaining questions are sharpened by our results. Generically, however,  there is  strong reason to be optimistic about the discoverability of general naturalness.

It should be pointed out that, since most of our low-energy probes are precision observables derived from loop corrections, it is also possible for \emph{unrelated} new physics to accidentally cancel these observables. Whether this constitutes an unnatural tuning is a matter of interpretation, but it would seem strange for nature to feature both top partners and unrelated new physics which cancel each other's experimental traces exactly.

The particular loophole of top partner swarms might be addressed if some experimental probe could be devised which does not lose sensitivity with increasing number of partner degrees of freedom. This, while challenging, would highly strengthen any no-lose theorem. 

The conservative and model-independent nature of our analysis implies that we would typically expect many \emph{more} experimental signals than those we examined here. Furthermore, since our predictions for the required tuning to escape detection in \fref{moneyplot} represent a minimization of inverse tuning over the experimentally inaccessible parameter space, any additional feature of the full theory that restricts which part of our simplified model parameter space it can map on to will obviously tighten our conclusions. This is, for example, the case for Twin Higgs model, where top partner mass reach through $Zh$ cross section deviations is about 2 TeV rather than our conservative estimate of 800 GeV. 

Our work therefore raises important model-building questions even within the constraints of perturbative top partner theories. Is it actually possible for any complete theory based on symmetries to realize the experimental worst-case scenarios we have defined here? For example, is there some version of Twin Higgs which could realize the Scalar Mediator scenario with large $N_f, N_s$ \emph{and} large mediator-partner Yukawa coupling to avoid all low-energy probes? This is not realized in the Orbifold Higgs~\cite{Craig:2014aea, Craig:2014roa}, where the Yukawa coupling decreases with $N_f$.

Our work supports the idea that naturalness can be model-independently tested as a general hypothesis. While generalized theories of neutral naturalness are currently unconstrained, and will remain largely so by the end of the LHC program, they are still amenable to detection: a future lepton collider is needed for low-energy precision measurements, a 100 TeV collider for direct production of heavy SM-charged states. Our work provides strong new motivation for the construction of these two machines. Crucially, \emph{both} kinds of new colliders are required. It is possible for a neutral top partner structure to generate signals at one kind of machine only, but our analysis shows that no such natural theory can avoid discovery at both machines unless it is quite baroque. Barring great unkindness on behalf of nature, these future colliders working in tandem should therefore reveal the new physics which stabilizes the electroweak scale.

\subsection*{Acknowledgments}
The authors are very grateful to 
Kaustubh Agashe,
Zacharia Chacko,
Timothy Cohen,
Nathaniel Craig,
Jamison Galloway,
Matthew McCullough, and
Raman Sundrum
 for valuable comments on the manuscript. We also thank 
Kaustubh Agashe,
Zacharia Chacko,
Spencer Chang,
Timothy Cohen,
Nathaniel Craig,
Keith Dienes,
Jamison Galloway,
Howard Haber,
Roni Harnik,
Simon Knapen,
Tongyan Lin,
Markus Luty,
Matthew McCullough,
David Pinner,
Lisa Randall,
Yael Shadmi,
Raman Sundrum,
Ofri Telem,
John Terning,
Yuhsin Tsai, and
Lian-Tao Wang
for useful conversations. 
The research of DC and PS was supported by the National Science Foundation  Grant No. PHY-1315155 and by the Maryland Center for Fundamental Physics. The research of PS was also supported by National Science Foundation grant PHY-1214000. 
The research of DC was performed in part at the Munich Institute for Astro- and Particle Physics (MIAPP), part of the DFG cluster of excellence ``Origin and Structure of the Universe". 
DC also thanks the Galileo Galilei Institute for Theoretical Physics for its hospitality, and the INFN for partial support during the completion of this work.
  The work of PS was performed in part at the Aspen Center for Physics, which is supported by National Science Foundation grant PHY-1066293.

\appendix

\section{Experimental Sensitivities}
\label{a.sensitivities}

\begin{table}
\begin{center}

\begin{tabular}{ccc}

\begin{tabular}{|l|r|c|}
\hline
Experiment & $\delta \sigma_{Zh}$ (95\%CL) & Reference
\\ \hline
ILC250  ($250\ifb$) & 5.2\% & \multirow{2}{*}{\cite{Asner:2013psa, Dawson:2013bba}}
\\ \cline{1-2}
ILC250LumiUp  ($1150\ifb$) & 2.4\% & 
\\ \hline
FCC-ee & 0.8\% & \cite{Gomez-Ceballos:2013zzn}
\\ \hline
\end{tabular}

 \\ \\

\begin{tabular}{|l|r|c|}
\hline
Measurement & $\Delta T$ (95\%CL) & Reference
\\ \hline
current & 0.076 & \multirow{4}{*}{\cite{Fan:2014vta}}
 \\ \cline{1-2}
ILC & 0.024 &
 \\ \cline{1-2} 
FCC-ee-Z & 0.019 &
 \\ \cline{1-2}
FCC-ee-t & 0.0092 &
\\ \hline
\end{tabular}

\\  \\ 

\begin{tabular}{|l|r|c|}
\hline
Experiment & $\delta \lambda_3$ (95\%CL) & Reference
\\ \hline
HL-LHC ($3 \iab$) & 60 - 100\% & \cite{ATLAS:lambda3, Baglio:2012np, Goertz:2013kp, Barger:2013jfa,Yao:2013ika}
\\ \hline
ILC (1 TeV, $2.5 \iab$) & 26\% & \cite{Asner:2013psa, Tian:2013yda} 
\\ \hline
100 TeV ($3 \iab$) & $\sim$ 20\% & \multirow{2}{*}{\cite{Barr:2014sga}}
\\ \cline{1-2}
100 TeV ($30 \iab$) & $\sim$ 10\% & 
\\ \hline
\end{tabular}

\end{tabular}

\end{center}
\caption{
Projected 95\% exclusion sensitivities of proposed future lepton and hadron colliders to  $Zh$ cross section deviations (top), $T$-parameter deviations if $\Delta S = 0$ (middle), and triple Higgs coupling shifts $\delta \lambda_3$ (bottom). The CEPC reach for $\delta \sigma_{Zh}$ is very similar to FCC-ee~\cite{preCDR}.
}
\label{t.sensitivities}
\end{table}

Here we briefly summarize projections of experimental sensitivities for the irreducible observables of neutral naturalness we explore in this work. In all cases we quote projected $95\%$ CL (2 sigma) exclusion sensitivities. 

\subsection{Higgs Coupling Shifts}
\label{sec:hshift}

Future lepton colliders will be the best probe of precision Higgs coupling measurements. In all the models we consider, the main BSM effect on Higgs couplings arises as an overall coupling rescaling relative to the SM. Therefore, the most sensitive probe will be Higgsstrahlung measurements of the $Zh$ production cross section $\sigma_{Zh}$, as originally suggested by~\cite{Craig:2013xia}. We will consider three representative sensitivity projections on $\delta \sigma_{Zh} = \Delta\sigma_{Zh}/\sigma_{Zh}$, shown in \tref{sensitivities} (top).

\subsection{Electroweak Precision Observables}

The sensitivity of future lepton colliders to electroweak precision observables (EWPO) has recently been estimated in~\cite{Fan:2014vta}. We follow the convenient parameterization of these results in~\cite{Fedderke:2015txa}. In the models we consider, the most important sensitivity comes from contributions to the $T$-parameter. Assuming no significant $S$-parameter shift, we can extract the exclusion sensitivities shown in \tref{sensitivities} (middle).

\subsection{Direct Top Partner Production}

The only guaranteed direct production channel for neutral top partners is the Higgs portal. This was recently examined by  the authors of~\cite{Curtin:2014jma} and~\cite{Craig:2014lda}, who studied pair production of invisible real scalars $S$ coupled to the Higgs via $\mathcal{L} \supset - \lambda_\mathrm{HS} |H|^2 S^2$ via the process $p p \to h^* + X \to S S + X$ at a 100 TeV collider. (Lepton colliders have very limited reach~\cite{Chacko:2013lna}.) The latter analysis was more sophisticated but the results agree very closely. Both analyses found that vector boson fusion production of the off-shell Higgs was the best channel to look for this challenging signature, and a 100 TeV collider is required for meaningful sensitivity. 

The results of~\cite{Curtin:2014jma} are shown in \fref{soversqrtb}, corresponding to the production of scalar top partners with Lagrangian \eref{Lscalar} and $N_r = 1$.  
For a single real top partner, the cancellation condition \eref{naturalreal} implies $\lambda_1 = 12 y_t^2$. The sensitivity for greater $N_r$ can be read off by computing the ``effective $\lambda_1$'' which gives the same signal cross section (with $N_r = 1$) as the natural value of $\lambda_i$ for $N_r > 1$:
\begin{eqnarray}
\nonumber \sigma &=& \sigma_0 (\lambda_1^\mathrm{eff})^2 =   \sigma_0 \lambda_i^2 N_r
\\
&\to& \lambda_1^\mathrm{eff} = \frac{12 y_t}{\sqrt{N_r}}
\label{e.lambdaieff}
\end{eqnarray}
Note that the number of direct production events scales as $1/N_r$.

A similar rescaling approach works for fermionic top partners, since the signal is nearly identical, but in this case we have to compute the scalar vs fermion cross section ratio in MadGraph~\cite{Alwall:2011uj}. Note that this represents a slight \emph{underestimate} of the fermion signal in the top partner search. Since the fermion pair production cross section scales with the parton-level center-of-mass energy $\hat s$ instead of $\lambda_{HS}^2 v^2$ as in the scalar case, fermion pair production is slightly more biased towards  boosted final states, which manifests as more missing energy. Even so, the simple signal rate rescaling using cross sections only, assuming identical kinematics, is sufficient for the purposes of our simple estimate, especially considering the relatively simple nature of these exploratory collider studies.

\fref{xsec} (a) compares the 100  TeV production cross section for a 12 real scalar partners to 4 fermion partners, obeying cancellation conditions \eref{naturalreal} and \eref{naturalfermion} respectively.  \fref{xsec}(b) shows the cross section ratio between natural fermions and scalars for different $N_f$ (with $N_r = 4 N_f$). This allows $S/\sqrt{B}$ in \fref{soversqrtb} to be rescaled for the fermion case.  The resulting mass reach for scalar and fermion top partners is shown in \fref{directproduction}.

\subsection{Higgs Self-Coupling}

The lowest-order Higgs self-coupling is the cubic $\mathcal{L} \supset \lambda_3 h^3$ interaction. In the SM at tree-level, $\lambda_3^\mathrm{SM} = \frac{1}{6} \partial^3 V / \partial h^3|_{h=v} = m_h^2/(2 v) \approx 32 \gev$. $\lambda_3$ can be measured through double-Higgs production. Define the self coupling deviation
\begin{equation}
\delta \lambda_3 = \frac{\lambda_3 - \lambda_3^\mathrm{SM}}{\lambda_3^\mathrm{SM}}.
\end{equation}
Precise measurements of $\delta \lambda_3$ are very challenging, partially due to SM interference effects.
The projected sensitivities are shown in \tref{sensitivities} (bottom). 
(Where only 1 sigma sensitivities were quoted, we assumed that twice that deviation could be excluded at two sigma.)
 Reasonably precise determinations require a future 100 TeV collider, which has been studied most recently in~\cite{Barr:2014sga, Azatov:2015oxa, He:2015spf}.

\begin{figure}
\begin{center}
\includegraphics[width=6cm]{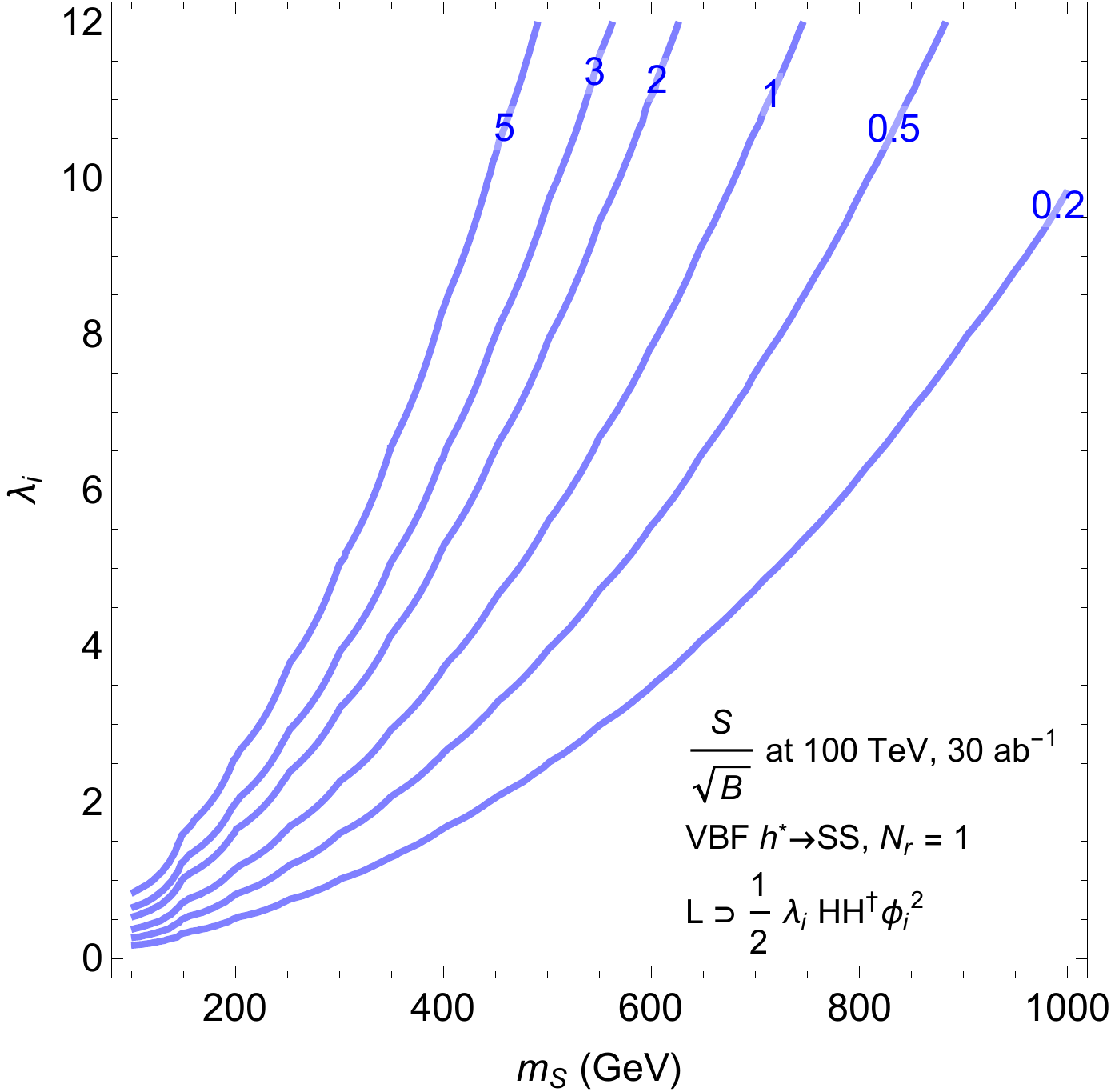}
\end{center}
\caption{
$S/\sqrt{B}$ for VBF $h^* \to \phi_i \phi_i$ production of a single real scalar with the Lagrangian \eref{Lscalar}, at a 100 TeV $pp$ collider with $30\iab$. Adapted from~\cite{Curtin:2014jma}.
}
\label{f.soversqrtb}
\end{figure}

\begin{figure}
\begin{center}
\hspace*{-14mm}
\begin{tabular}{c}
\includegraphics[height=5cm]{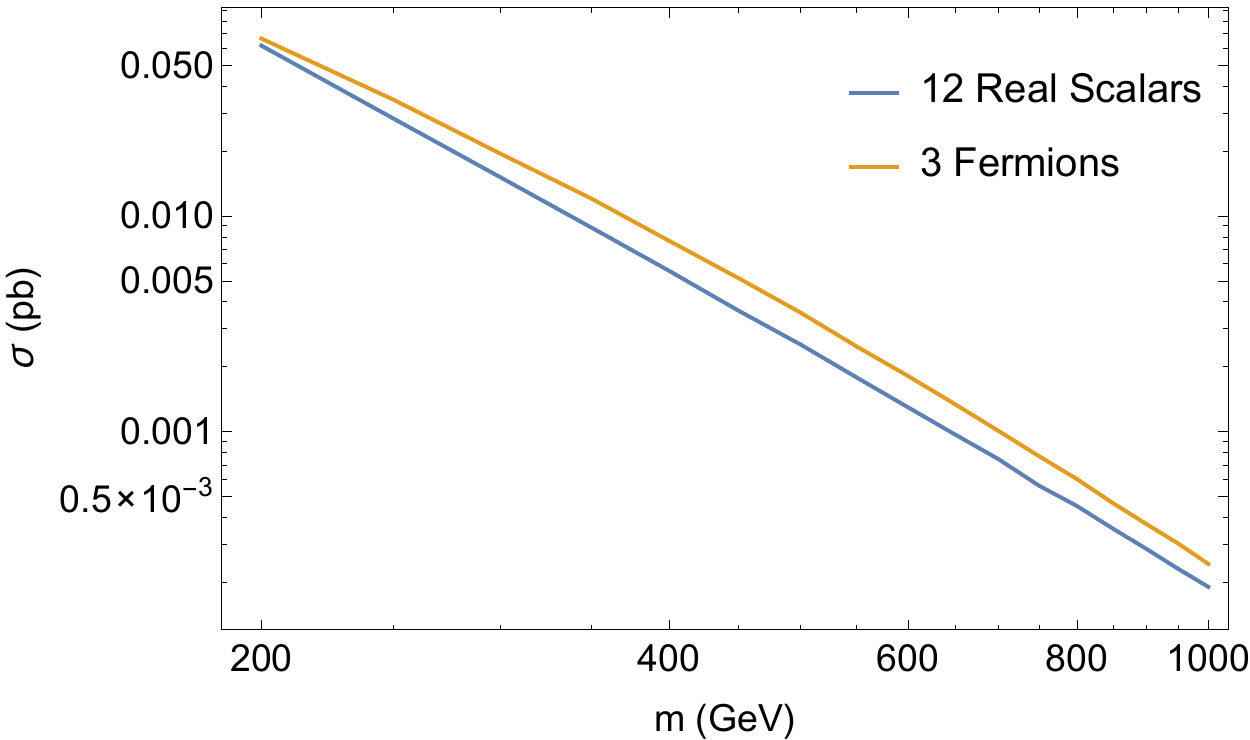}
\\
\hspace{10mm} 
(a)
\\ \\
\hspace*{6mm}
\includegraphics[height=4.9cm]{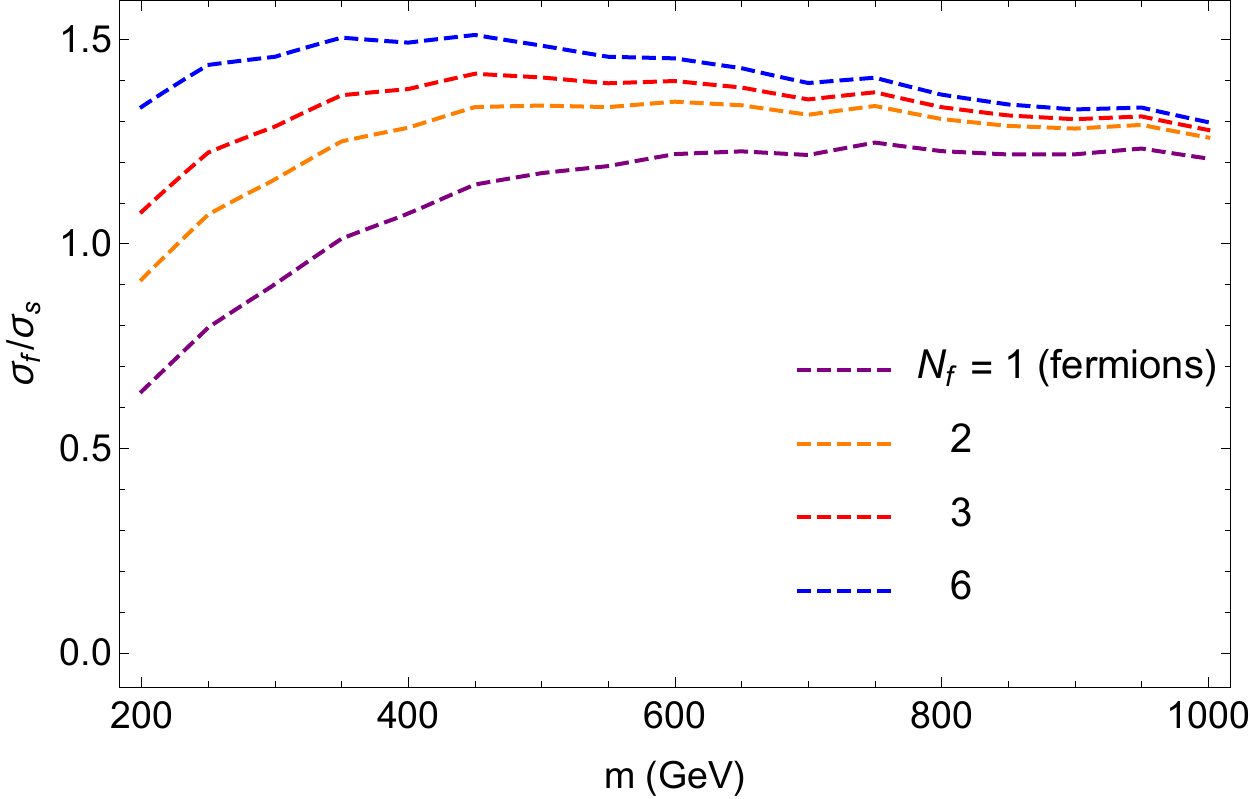}
\\
\hspace{10mm} 
(b)
\end{tabular}
\end{center}
\caption{
(a) Cross section (without any cuts) for VBF $h^* \to XX$ production, where $X$ are either 12 scalar or 3 fermion top partners obeying Eqns.~(\ref{e.naturalreal}) and~(\ref{e.naturalfermion}), at a 100 TeV collider. 
(b) 
Cross section ratio between fermion and scalar top partners, for different $N_f$ (with $N_r = 4 N_f$).
}
\label{f.xsec}
\end{figure}

 It should be pointed out that the 100 TeV studies of  $\delta \lambda_3$ reach are conservative, in that they assume present-day detector capabilities and understanding of QCD fake rates. It would be very useful to know a somewhat \emph{optimistic} estimate of $\delta \lambda_3$ reach, to understand what might be possible. We find that in particular scalar neutral top partners provide ample motivation to study the triple Higgs coupling at high precision, see \fref{scalarexclusionlambda3}.

\section{Further exploration of the Sacrificial Scalar Mechanism}
\label{a.sacrificialscalar}

It is helpful to study the Sacrificial Scalar Mechanism in a concrete example. A well-motivated case is that of a scalar mediator which obeys a $\mathbb{Z}_2$ symmetry $S \to -S$. This restricts the Lagrangian in a way reminiscient of the canonical Twin Higgs in a linear sigma model formulation:
\begin{eqnarray}
\label{e.LscalarmediatorZ2}
\mathcal{L}_\mathrm{int} &=& y_t H Q \bar U +   y_{STT} S \bar T T - V_0(H, S)  
\\ \nonumber \\
\nonumber
V_0(H,S) &=& - \mu^2 |H|^2 + \lambda |H|^4  
\\ \nonumber
&&
- \frac{1}{2} \mu_S^2 S^2 + \frac{1}{4} \lambda_S S^4 + \lambda_{HS} |H|^2 S^2 .
\end{eqnarray}
We write the singlet mass term with negative sign since it will need to acquire a vev to realize our scenario.
First, we compute explicitly the quadratically divergent one-loop corrections in the above two-scalar potential, arising from finite momentum cutoff $\Lambda$: 
\begin{equation}
\delta V_1 = - \delta \mu^2 |H|^2 - \frac{1}{2} \delta \mu_S^2 S^2
\end{equation}
where
\begin{equation}
\delta \mu^2 = \frac{3 y_t^2 \Lambda^2}{8 \pi^2}
\ \ \ , \ \ \ \ \
\delta \mu_S^2 = \frac{N_f y_{STT}^2 \Lambda^2}{4 \pi^2} \ \ .
\end{equation}
For clarity we drop logarithmically divergent and finite terms. We can write the full potential $V = V_0 + \delta V_1$ by replacing $\mu^2, \mu_S^2$ in \eref{LscalarmediatorZ2} by
\begin{equation}
\tilde \mu^2 = \mu_2 + \delta \mu^2 \ \ , \ \ \  \
\tilde \mu_S^2 = \mu_S^2 + \delta \mu_S^2
\ \ .
\end{equation}
Integrating out the scalar by solving $\partial \mathcal{L}/\partial S = 0$ (with the full potential) gives the classical solution
\begin{eqnarray*}
&& S(H) =
\\ \\
&& 
\frac{\tilde \mu_S}{\sqrt{\lambda_S}}\left[ 1 - \left( \frac{\lambda_{HS} |H|^2}{\tilde \mu_S^2} \right) - \frac{1}{2} \left( \frac{\lambda_{HS} |H|^2}{\tilde \mu_S^2} \right)^2 + \mathcal{O}\left(|H|^6\right) \right]
\\ \\
&& + \ 
\frac{y_{STT} \bar T T}{2 \tilde \mu_S^2}
\left[ 1 + \left( \frac{2 \lambda_{HS} |H|^2}{\tilde \mu_S^2} \right)   + \mathcal{O}\left(|H|^4\right)
\right]
\\ \\ 
 && +  \  \mathcal{O}\left( (\bar T T)^2 \right)
\end{eqnarray*}
Substituting this into \eref{LscalarmediatorZ2} yields the effective low-energy Lagrangian for the Higgs, top and top partner:
\begin{eqnarray}
\nonumber
\mathcal{L}_\mathrm{eff} &=& y_t H Q \bar U 
\\ \nonumber
&& + y_{STT} \bar T T \frac{\tilde \mu_S}{\lambda_S} \left( 1 - \frac{\lambda_{HS} |H|^2}{\tilde \mu_S^2} + \mathcal{O}(|H|^4) \right) 
\\ \nonumber 
&& - V_\mathrm{eff}(H)
\\ \nonumber \\ \label{e.LscalarmediatorZ2eff}
V_\mathrm{eff}(H) &=& \mu_\mathrm{eff}^2 |H|^2 + \lambda_\mathrm{eff} |H|^4 + \mathcal{O}(|H|^6)
\end{eqnarray}
where
\begin{equation}
\mu_\mathrm{eff}^2 = \tilde \mu^2 - \frac{\lambda_{HS}}{\lambda_S} \mu_S^2 
\ \ , \ \ \ \  \ \ 
\lambda_\mathrm{eff} = \lambda - \frac{\lambda_{HS}^2}{\lambda_S} \ \ .
\end{equation}
We can now ask the question whether the Higgs mass is protected in this low-energy theory. The quadratically divergent correction to $\mu_\mathrm{eff}^2$ is simply
\begin{equation}
\delta \mu_\mathrm{eff}^2 = \delta \mu^2 - \frac{\lambda_{HS}}{\lambda_S} \delta \mu_S^2
\end{equation}
Immediately we can see that the the quadratically divergent top loop contribution to the Higgs mass (in the full theory with the extra scalar) is canceled in the low-energy theory by the quadratically divergent $T$-loop contribution to the $S$-mass. This is the essence of the Sacrificial Scalar Mechanism: the lack of protection for the singlet mediator is what protects the Higgs. If there were some additional states in the theory which stabilize the $S$-mass against quadratic corrections to the $T$-loops, then $\delta \mu_S = 0$ and the Higgs mass could not be stabilized.

As a consistency check we can derive the explicit cancellation condition for $\delta \mu_\mathrm{eff}^2 = 0$:
\begin{equation}
\frac{3}{N_f} y_t^2 = 2 \frac{\lambda_{HS}}{\lambda_S} y_{STT}^2.
\end{equation}
This is the same condition we derive by matching \eref{LscalarmediatorZ2eff} to the general low-energy fermionic top partner Lagrangian \eref{Lfermion} and applying the naturalness condition \eref{naturalfermion} to the resulting $M_T, M'$.

The above derivation can be repeated for more general scenarios with more scalars, without the  $\mathbb{Z}_2$ symmetry, etc. However, with the above notation, and the understanding that in the EFT formulation, quadratically divergent  contributions are transmitted to the Higgs essentially via the vev of the $S$, we can formulate a compact general argument that $S$ should be heavy if there is no tree-vs-loop tuning in the singlet sector when expanded around some natural origin of field space.

We consider a version of the general scalar mediator Lagrangian:
\begin{eqnarray}
\nonumber
\mathcal{L}_\mathrm{int} &=& y_t H Q \bar U +  (M + y_{STT} S) \bar T T -  V(S,H)
\\ \label{e.LscalarmediatorGeneral}
\\ \nonumber
V(S,H) &=& \left[ - \mu^2 |H|^2 + \lambda |H|^4 + \mu_{HHS} |H|^2 S + \ldots \right].
\end{eqnarray}
In \eref{Lscalarmediator} we made a choice of convenience that $\langle S \rangle = 0$ for the purpose of deriving physical observables. Doing so at one-loop order requires a field re-definition which obfuscates quadratic corrections to the $S$ mass and (if there is no $\mathbb{Z}_2$) tadpole from $T$-loops. Therefore, for this argument we consider the general case of $S$ having some vev, $\langle S \rangle = w$ at tree-level, with respect to whatever is the natural origin of field space. Furthermore, this vev will have quadratically divergent contributions from $T$-loops, which we call $\delta w$. Therefore, the full  S vev is $\langle S \rangle = \tilde w = w + \delta w$. Matching this to the low-energy Lagrangian \eref{Lfermion} gives
\begin{equation}
\label{e.deltaMT}
M_T = M + y_{STT} w + \delta M_T \ ,
\end{equation}
where
\begin{equation}
\delta M_T = y_{STT} \delta w, 
\end{equation}
The effective Higgs mass parameter is
\begin{equation}
\mu_\mathrm{eff}^2 = \mu^2 - \mu_{HHS} w + \delta \mu^2 - \delta \mu_T^2  \ ,
\end{equation}
where
\begin{equation}
\delta\mu^2_T = \mu_{HHS} \delta w,
\end{equation}
and $\delta \mu^2$ is the quadratically divergent Higgs mass contribution from top loops. 
Protecting the Higgs mass requires
\begin{equation}
\mu_{HHS} = \frac{\delta \mu^2}{\delta w}
\end{equation}
The naturalness cancellation condition \eref{naturalfermion} can be written as
\begin{equation}
\frac{3}{N_f} \frac{y_t^2}{2 M_T} = \frac{1}{2 M'} = \frac{y_{STT} \mu_{HHS}}{m_S^2} = \frac{y_{STT} \delta \mu^2 }{\delta w \ m_S^2} \ ,
\end{equation}
where again $m_S$ is the \emph{physical} mass of the singlet. Solving for $m_S^2$ and using \eref{deltaMT} gives
\begin{equation}
\frac{m_S^2 }{m_h^2}
\ = \ 
\frac{2 N_f}{6} \ 
 \frac{y_{STT}^2}{y_t^2} \ 
 \frac{M_T}{\delta M_T} \
  \frac{\delta \mu^2}{\mu^2_\mathrm{eff}}
\end{equation}
The only way $m_S$ can be weak scale is by satisfying one or more of the following requirements:
\begin{itemize}
\item $y_{STT}$ is small. This indicates large $\mu_{HHS}$ and hence large Higgs-singlet mixing, which is detectable at lepton colliders;
\item $\delta \mu^2/\mu^2_\mathrm{eff}$ is not large, which indicates a low momentum cutoff and new states at scales accessible by the HL-LHC or 100 TeV collider;
\item  $\delta M_T \gg M_T$, which is an unnatural tree vs loop tuning. 
\end{itemize}
Therefore, the Sacrificial Scalar Mechanism guarantees that in a natural theory, $m_S^2 \gg m_h^2$.

\bibliography{neutral_naturalness_pheno}

\end{document}